# Geometry and Response of Lindbladians

Victor V. Albert,[1,*] Barry Bradlyn,[2] Martin Fraas,[3] and Liang Jiang[1]

[1]*Departments of Applied Physics and Physics, Yale University, New Haven, Connecticut 06511, USA*
[2]*Princeton Center for Theoretical Science, Princeton University, Princeton, New Jersey 08540, USA*
[3]*Mathematisches Institut der Universität München, University of Munich, Munich 80333, Germany*

Markovian reservoir engineering, in which time evolution of a quantum system is governed by a Lindblad master equation, is a powerful technique in studies of quantum phases of matter and quantum information. It can be used to drive a quantum system to a desired (unique) steady state, which can be an exotic phase of matter difficult to stabilize in nature. It can also be used to drive a system to a unitarily evolving subspace, which can be used to store, protect, and process quantum information. In this paper, we derive a formula for the map corresponding to asymptotic (infinite-time) Lindbladian evolution and use it to study several important features of the unique state and subspace cases. We quantify how subspaces retain information about initial states and show how to use Lindbladians to simulate any quantum channels. We show that the quantum information in all subspaces can be successfully manipulated by small Hamiltonian perturbations, jump operator perturbations, or adiabatic deformations. We provide a Lindblad-induced notion of distance between adiabatically connected subspaces. We derive a Kubo formula governing linear response of subspaces to time-dependent Hamiltonian perturbations and determine cases in which this formula reduces to a Hamiltonian-based Kubo formula. As an application, we show that (for gapped systems) the zero-frequency Hall conductivity is unaffected by many types of Markovian dissipation. Finally, we show that the energy scale governing leakage out of the subspaces, resulting from either Hamiltonian or jump-operator perturbations or corrections to adiabatic evolution, is different from the conventional Lindbladian dissipative gap and, in certain cases, is equivalent to the excitation gap of a related Hamiltonian.

## I. MOTIVATION AND OUTLINE

Consider coupling a quantum mechanical system to a Markovian reservoir which evolves initial states of the system into multiple nonequilibrium (i.e., nonthermal) asymptotic states in the limit of infinite time. After tracing out the degrees of freedom of the reservoir, the time evolution of the system is governed by a Lindbladian $\mathcal{L}$ [1,2] (see also Refs. [3–6]), and its various asymptotic states $\rho_\infty$ are elements of an asymptotic subspace $\text{As}(\mathsf{H})$—a subspace of $\text{Op}(\mathsf{H})$, the space of operators on the system Hilbert space $\mathsf{H}$. The asymptotic subspace attracts all initial states $\rho_{\text{in}} \in \text{Op}(\mathsf{H})$, is free from the decoherence effects of $\mathcal{L}$, and any remaining time evolution within $\text{As}(\mathsf{H})$ is exclusively unitary. If $\text{As}(\mathsf{H})$ has no time evolution, all $\rho_\infty$ are stationary or steady. This work provides a thorough investigation into the response and geometrical properties of the various asymptotic subspaces.

On one hand, $\text{As}(\mathsf{H})$ that support quantum information [7–10] are promising candidates for storing, preserving, and manipulating such information, particularly when their states can be engineered to possess favorable features (e.g., topological protection [11–13]). They have been subject to intense experimental investigation in quantum optics [14,15], liquid-state NMR [16,17], trapped ions [18–23], and (most recently) circuit QED [24]. With many current experimental efforts aimed at engineering Markovian environments admitting asymptotic subspaces, it is important to gain a comprehensive understanding of any differences between the properties of these subspaces and analogous subspaces of Hamiltonian systems (e.g., subspaces spanned by degenerate energy eigenstates).

On the other hand, response properties of $\text{As}(\mathsf{H})$, which do not necessarily support quantum information, can help model experimental probes into exotic nonequilibrium phases of matter resulting from engineered Markovian reservoirs [25–35] (realized in, e.g., optical lattices [36–46] or microwave cavity arrays [47–49]). Because of, e.g., symmetry [50,51] or topology [11], the asymptotic subspace can be degenerate yet not support a qubit [e.g., an $\text{As}(\mathsf{H})$ spanned by two projections $|\psi\rangle\langle\psi|$ and $|\psi'\rangle\langle\psi'|$]. For these and similar cases, standard thermodynamical concepts [52–55] may not apply and steady states may no longer be thermal or even full

*valbert4@gmail.com

rank. (We remind the reader that the rank of a diagonalizable matrix is the number of its not necessarily distinct nonzero eigenvalues.) Our approach is directly tailored to such systems, i.e., those possessing one or more nonequilibrium steady states whose rank is less than the dimension of the system Hilbert space.

Unlike Hamiltonians, Lindbladians have the capacity to model decay. As a result, Lindbladians are often used to describe commonplace non-Hamiltonian processes (e.g., cooling to a ground state). In general Lindbladian-based time evolution, all parts of an initial state $\rho_{in}$ that are outside of As(H) will decay as $\rho_{in}$ evolves toward an asymptotic state $\rho_\infty \in$ As(H). Since As(H) may be multidimensional, the resulting asymptotic state may depend on $\rho_{in}$. The decay of parts of $\rho_{in}$ and the nontrivial dependence of $\rho_\infty$ on $\rho_{in}$ stand out as two distinct features of Lindbladian-based evolution. Nonetheless, $\rho_\infty$ is a collection of states whose behavior is otherwise familiar from Hamiltonian-based quantum mechanics. An asymptotic subspace can thus be thought of as a Hamiltonian-evolving subspace embedded in a larger Lindbladian-evolving space. The aim of this paper is to determine the effects of Lindbladian evolution on the properties of $\rho_\infty$. Namely, we prove a formula for the effect of $\mathcal{L}$ in the limit of infinite time (Proposition 2 in Sec. III) and apply it to the following physically motivated questions, noting that (4)–(6) contain results relevant also to $\mathcal{L}$ with a unique steady state.

(1) *What is the dependence of $\rho_\infty$ on $\rho_{in}$?* Building on previous results [50], in Sec. III we show that $\rho_\infty$ does not depend on any initial coherences between As(H) and subspaces outside of As(H) and that the presence of unitary evolution within As(H) can actually suppress the purity of $\rho_\infty$. We provide a recipe for using infinite-time Lindbladian evolution to implement arbitrary quantum channels, i.e., completely positive trace-preserving maps [56]. This recipe should prove useful in experimental quantum channel simulation [57] and autonomous or passive quantum error correction [58].

(2) *What is the effect of time-independent Hamiltonian perturbations on $\rho_\infty$ within As(H)?* It was recently shown [59,60] that Hamiltonian perturbations and perturbations to the jump operators of $\mathcal{L}$ generate unitary evolution within some As(H) to linear order. In Sec. IV, we prove that such perturbations induce unitary evolution within *all* As(H) to linear order, extending the capabilities of environment-assisted quantum computation and quantum Zeno dynamics [61–66].

(3) *What is the geometric "phase" acquired by $\rho_\infty$ after cyclic adiabatic deformations of $\mathcal{L}$?* In Sec. V, we extend previous results [67–71] to show that cyclic Lindbladian-based [72] adiabatic evolution of states in As(H) is always unitary, extending the capabilities of holonomic quantum computation [73] via reservoir engineering.

(4) *What is the natural metric governing distances between various $\rho_\infty$?* We introduce in Sec. VI a Lindbladian version of the quantum geometric tensor (QGT) [74,75], which encodes both the curvature associated with adiabatic deformations and a metric associated with distances between adiabatically connected steady states.

(5) *What is the energy scale governing leakage out of the asymptotic subspace?* Extending Ref. [69], in Secs. IV C and V C we determine the energy scale governing leakage out of As(H) due to both Hamiltonian perturbations and adiabatic evolution. Contrary to popular belief, this scale is not always the dissipative gap of $\mathcal{L}$ (the nonzero eigenvalue with the smallest real part). We demonstrate this with an example from coherent state quantum information processing [61].

(6) *What is the linear response of $\rho_\infty$ to time-dependent Hamiltonian perturbations?* In. Sec. IV, we derive a Lindbladian-based Kubo formula for response of $\rho_\infty$ and determine when it reduces to the familiar Hamiltonian-based Kubo formula [76]. As an application, we show that the zero-frequency Hall conductivity [77] remains quantized under various kinds of Markovian dissipation.

## II. STATEMENT OF KEY RESULTS

In this section, we introduce necessary notation, state our key result, summarize its ramifications in the form of two properties, the no-leak and clean-leak properties Eqs. (2.9) and (2.11), and apply it to various types of As(H). We conclude with a summary of earlier work and outline the rest of the paper. Readers unfamiliar with Lindbladian evolution are welcome to browse Appendix A.

### A. Four-corners decomposition

Since decay of states is an unavoidable feature of Lindbladian evolution, it is important to make a clear distinction between the decaying and nondecaying parts of the $N$-dimensional system Hilbert space H. Let us group all nondecaying parts of Op(H) into the upper left corner [of the matrix representation of Op(H)] and denote them by the "upper-left" block ◧. Thereby, any completely decaying parts will be in the complementary ◨ block, and coherences between the two will be in the "off-diagonal" blocks ◪. We can discuss such a decomposition in the familiar language of NMR: the ◧ block consists of a degenerate ground state subspace immune to decay and dephasing, the ◨ block contains the set of populations decaying with the well-known rate $1/T_1$, and the ◪ block is the set of coherences dephasing with rate $1/T_2$. More generally, there can be further dephasing *within* ◧ without population decay, so As(H) [gray region in Fig. 1(a)] is in general a subspace of ◧.

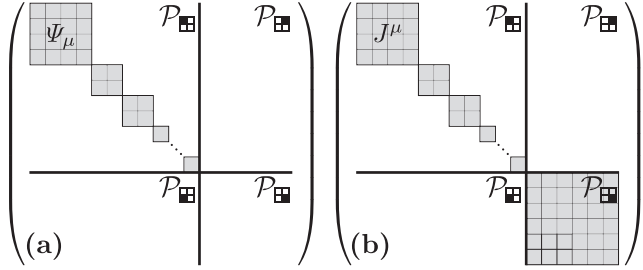

FIG. 1. Decompositions of the space of matrices Op(H) acting on a Hilbert space H using the projections $\{P, Q\}$ defined in Eq. (2.1) and their corresponding superoperator projections $\{\mathcal{P}_\blacksquare, \mathcal{P}_\blacksquare, \mathcal{P}_\blacksquare, \mathcal{P}_\blacksquare\}$ defined in Eq. (2.2). Panel (a) depicts the block diagonal structure of the asymptotic subspace As(H), which is located in $\blacksquare$ and spanned by steady-state basis elements $\Psi_\mu$. Panel (b) depicts the subspace of Op(H), spanned by conserved quantities $J_\mu$, that may leave a footprint on states in As(H) in cases when there are multiple steady states.

Let us now define the superoperator projections on the blocks. Let $P$ be the orthogonal operator projection ($P = P^2 = P^\dagger$) on *and only on* the nondecaying subspace of H. This projection is uniquely defined by the following conditions: for all $\rho_\infty \in \text{As}(\text{H})$,

$$\rho_\infty = P\rho_\infty P,$$
$$\text{Tr}\{P\} = \max_{\rho_\infty}\{\text{rank}(\rho_\infty)\}. \quad (2.1)$$

The first condition makes sure that $P$ projects onto all nondecaying subspaces, while the second guarantees that $P$ does not project onto any decaying subspace. Naturally, the orthogonal projection onto the maximal decaying subspace of H is $Q \equiv I - P$ [with $PQ = QP = 0$ and $Q\rho(t)Q \to 0$ as $t \to \infty$].

We define the four-corners projections acting on $A \in \text{Op}(\text{H})$ as follows:

$$A_\blacksquare \equiv \mathcal{P}_\blacksquare(A) \equiv PAP,$$
$$A_\blacksquare \equiv \mathcal{P}_\blacksquare(A) \equiv PAQ,$$
$$A_\blacksquare \equiv \mathcal{P}_\blacksquare(A) \equiv QAP,$$
$$A_\blacksquare \equiv \mathcal{P}_\blacksquare(A) \equiv QAQ. \quad (2.2)$$

By our convention, taking the conjugate transpose of the upper-right part places it in the lower-left subspace (projection acts *before* adjoint): $A^\dagger_\blacksquare \equiv (A_\blacksquare)^\dagger = (A^\dagger)_\blacksquare$. The operators $\mathcal{P}_\boxplus$ (with $\boxplus \in \{\blacksquare, \blacksquare, \blacksquare, \blacksquare\}$) are projections ($\mathcal{P}_\boxplus = \mathcal{P}_\boxplus^2$), which partition the identity $\mathcal{I}$ on Op(H),

$$\mathcal{P}_\blacksquare + \mathcal{P}_\blacksquare + \mathcal{P}_\blacksquare + \mathcal{P}_\blacksquare = \mathcal{I}, \quad (2.3)$$

analogous to $P + Q = I$. They conveniently add, e.g.,

$$\mathcal{P}_\blacksquare \equiv \mathcal{P}_\blacksquare + \mathcal{P}_\blacksquare \quad \text{and} \quad \mathcal{P}_\blacksquare \equiv \mathcal{P}_\blacksquare + \mathcal{P}_\blacksquare. \quad (2.4)$$

The subspace $\blacksquare \equiv \mathcal{P}_\blacksquare\text{Op}(\text{H})$ consists of all coherences between $P\text{H}$ and $Q\text{H}$, and the "diagonal" subspace $\blacksquare \equiv \mathcal{P}_\blacksquare\text{Op}(\text{H})$ consists of all operators that do not contain any such coherences.

Nontrivial decaying subspaces $\blacksquare$ are ubiquitous in actively researched quantum information schemes (see, e.g., Refs. [61,62]). For instance, consider a bosonic Lindbladian whose steady states are the two coherent states $|\alpha\rangle$ and $|-\alpha\rangle$ (recently realized experimentally [24] and discussed in more detail in Sec. IV C 1). All states orthogonal to $|\pm\alpha\rangle$ constitute the decaying subspace, and our results apply. We thoroughly discuss how our work applies to various As(H) in Sec. II C. Here, before summarizing our key results, we mention two cases without decaying subspaces for which our work reduces to known results.

*Hamiltonian case.*—If $\mathcal{L} = -i[H, \cdot]$ for some Hamiltonian, any state written in terms of the $N$ eigenstate projections $|E_k\rangle\langle E_k|$ of $H$ ($H|E_k\rangle = E_k|E_k\rangle$) is a steady state. Therefore, there is no decaying subspace in Hamiltonian evolution ($P = I$).

*Unique state case (full rank).*—In the case of a one-dimensional As(H), $P$ is the projection on the rank of the unique steady state $\rho_\infty \equiv \varrho$. If the state's spectral decomposition is $\varrho = \sum_{k=0}^{d_\varrho - 1} \lambda_k |\psi_k\rangle\langle\psi_k|$ (with $d_\varrho$ being the number of nonzero eigenvalues $\lambda_k$ of $\varrho$), then $P = \sum_{k=0}^{d_\varrho - 1} |\psi_k\rangle\langle\psi_k|$. If all $N$ eigenvalues are nonzero, then $\varrho$ is full rank (e.g., in a Gibbs state) and there is no decaying subspace ($P = I$).

### B. Key results

States undergoing Lindbladian evolution evolve into asymptotic states for sufficiently long times [78]:

$$\rho_{\text{in}} \xrightarrow{t \to \infty} \rho_\infty \equiv e^{-iH_\infty t}\mathcal{P}_\infty(\rho_{\text{in}})e^{iH_\infty t}. \quad (2.5)$$

The nonunitary effect of Lindbladian time evolution is encapsulated in the *asymptotic projection superoperator* $\mathcal{P}_\infty$ (with $\mathcal{P}_\infty^2 = \mathcal{P}_\infty$). The extra Hamiltonian $H_\infty$ quantifies any further unitary evolution within As(H), which of course does not cause any decoherence. For simplicity, we state our result for the $H_\infty = 0$ case and outline the nontrivial consequences of $H_\infty \neq 0$ later. The asymptotic projection is a trace-preserving quantum process taking a density matrix $\rho_{\text{in}} \in \text{Op}(\text{H})$ into an asymptotic density matrix in $\rho_\infty \in \text{As}(\text{H})$. We determine the following formula for $\mathcal{P}_\infty$ (Proposition 2):

$$\mathcal{P}_\infty = \mathcal{P}_\Psi(\mathcal{P}_\blacksquare - \mathcal{P}_\blacksquare\mathcal{L}\mathcal{P}_\blacksquare\mathcal{L}^{-1}\mathcal{P}_\blacksquare), \quad (2.6)$$

where the *minimal projection* $\mathcal{P}_\Psi$ further maps $\blacksquare$ onto As(H). The form of $\mathcal{P}_\Psi$, a projection onto As(H) of $\mathcal{L}$ which do not possess a decaying subspace, depends on the details of As(H) and is already known [78,79]. Therefore, our work extends previous Lindbladian results to cases when a decaying subspace is present.

The above formula allows us to determine which parts of $\rho_{\text{in}}$ are preserved in the large-time limit [question (1); see Sec. III C]. For example, since the projection $\mathcal{P}_{\blacksquare}$ is not present in the above formula, we can immediately read off that no coherences between the nondecaying subspace and its counterpart are preserved. Moreover, the piece $\mathcal{P}_\infty \mathcal{P}_{\boxplus}$ can be used to simulate an arbitrary quantum channel (Sec. III D). Finally, the formula describes how states that are already in As(H) respond to perturbations. We now apply the formula to show why $\mathcal{P}_\Psi$ is the only part relevant to answering questions (2)–(4).

We sketch the effect of small perturbations $\mathcal{O}$ on a state $\rho_\infty$ already in As(H). The perturbations of interest are either Hamiltonian perturbations $\mathcal{V} \equiv -i[V, \cdot]$ (with Hamiltonian $V$ and small parameter $\epsilon$) or derivatives $\partial_\alpha \equiv \partial/\partial \mathbf{x}_\alpha$ (with parameters $\mathbf{x}_\alpha$ and adiabatic evolution time $T$) of the now parameter-dependent $\rho_\infty(\mathbf{x}_\alpha)$ and $\mathcal{L}(\mathbf{x}_\alpha)$:

$$\mathcal{O} \in \left\{ \epsilon \mathcal{V}, \frac{1}{T} \partial_\alpha \right\}. \quad (2.7)$$

We show that both of these can be used to induce unitary operations on As(H). We show later in the paper that this analysis holds for jump operator perturbations as well, but omit discussing those perturbations for now to keep things simple. The perturbations $\partial_\alpha$ determine adiabatic connection(s) and thus help with defining parallel transport [i.e., adiabatic evolution of As(H)]. Within first order for the case of perturbation theory ($\epsilon \to 0$) and approaching the adiabatic limit for the case of parallel transport ($T \to \infty$), two relevant perturbative processes after the action of $\mathcal{O}$ on an asymptotic state are subsequent projection onto As(H) and leakage out of As(H) via the perturbation and $\mathcal{L}^{-1}$:

$$\rho_\infty \to \mathcal{P}_\infty \mathcal{O}(\rho_\infty) - \mathcal{L}^{-1} \mathcal{O}(\rho_\infty). \quad (2.8)$$

We show below that these two terms occur both in the Kubo formula and in adiabatic response.

We first observe that $\mathcal{O}$ is limited in its effect on $\rho_\infty$. Acting with $\mathcal{O}$ once does not connect $\boxplus$ with $\boxplus$ because $\mathcal{O}$ does not act nontrivially on $\rho_\infty$ from both sides simultaneously. This *no-leak property* can be understood if one observes that Hamiltonian superoperator perturbations $\mathcal{V}$ act nontrivially on $\rho_\infty$ only from one side at a time due to their commutator form. Likewise, derivatives $\partial_\alpha$ act nontrivially on either the "ket" or "bra" parts of all basis elements used to write $\rho_\infty$ due to the product rule. Therefore, acting with $\mathcal{O}$ once only connects $\boxplus$ to itself and nearest-neighbor squares ($\blacksquare$) and does not cause "transitions" into $\boxplus$:

$$\mathcal{O}(\rho_\infty) = \mathcal{P}_{\blacksquare} \mathcal{O}(\rho_\infty), \quad (2.9)$$

where $\mathcal{P}_{\blacksquare} \equiv \mathcal{I} - \mathcal{P}_{\boxplus}$. Moreover, despite two actions of $\mathcal{O}$ connecting $\boxplus$ to $\boxplus$, Eq. (2.9) still provides some insight into second-order effects within As(H) (Sec. IV C).

The no-leak property (2.9) is important in determining the energy scale governing leakage out of As(H) [question (5); see Secs. IV C and V C]. Let us apply this property to the second term in Eq. (2.8):

$$\mathcal{L}^{-1}\mathcal{O}(\rho_\infty) = \mathcal{L}^{-1}\mathcal{P}_{\blacksquare}\mathcal{O}(\rho_\infty) = \mathcal{L}_{\boxplus}^{-1}\mathcal{O}(\rho_\infty), \quad (2.10)$$

where $\mathcal{L}_{\boxplus}^{-1} \equiv (\mathcal{P}_{\boxplus} \mathcal{L} \mathcal{P}_{\boxplus})^{-1}$ and $\boxplus$ is any block. Note that the last step in Eq. (2.10) also uses a property of $\mathcal{L}$, $\mathcal{P}_{\boxplus} \mathcal{L} \mathcal{P}_{\blacksquare} = 0$ [see Eq. (3.3)], which can be understood by remembering that evolution under $\mathcal{L}$ draws states to $\boxplus$. Since the restriction to studying $\mathcal{L}$ on $\blacksquare$ in linear response has previously gone unnoticed, it is conventionally believed that the leakage energy scale is determined by the dissipative (also, dissipation or damping) gap $\Delta_{\text{dg}}$—the nonzero eigenvalue of $\mathcal{L}$ with smallest real part. As shown in Eq. (2.10), that energy scale is actually governed by the effective dissipative gap $\Delta_{\text{edg}} \geq \Delta_{\text{dg}}$—the nonzero eigenvalue of $\mathcal{L}_{\blacksquare}$ with smallest real part. In Hamiltonian systems ($\mathcal{L} = -i[H, \cdot]$), a special case of the no-leak property states that the energy denominator in the first-order perturbative correction to the $k$th eigenstate of $H$ contains only energy differences involving the energy $E_k$ of that eigenstate (and not, e.g., $E_{k-1} - E_{k+1}$).

We now project $\mathcal{O}(\rho_\infty)$ back to As(H) to examine the first term in Eq. (2.8). Applying $\mathcal{P}_\infty$ to Eq. (2.9) and using $\mathcal{P}_\infty \mathcal{P}_{\blacksquare} = 0$ from Eq. (2.6) removes two more squares:

$$\mathcal{P}_\infty \mathcal{O}(\rho_\infty) = \mathcal{P}_\infty \mathcal{P}_{\boxplus} \mathcal{O}(\rho_\infty) = \mathcal{P}_\Psi \mathcal{O} \mathcal{P}_\Psi(\rho_\infty). \quad (2.11)$$

The clean-leak property shows that any leakage of the perturbed $\rho_\infty$ into $\blacksquare$ does not contribute to the first-order effect of $\mathcal{O}$ within As(H). Essentially, the *clean-leak property* (2.11) makes As(H) resistant to the nonunitary effects of Lindbladian evolution and allows for a closer analogue between As(H) and subspaces of unitary systems. The clean-leak property simplifies calculations of both Hamiltonian perturbations [question (2); see Sec. IV] and adiabatic or Berry connections [question (3); see Sec. V]. It can be used to show that $\mathcal{P}_\Psi$ (instead of the full $\mathcal{P}_\infty$) fully governs adiabatic evolution, so the Lindbladian generalization of the QGT [question (4); see Sec. VI] is

$$\mathcal{Q}_{\alpha\beta} \equiv \mathcal{P}_\Psi \partial_\alpha \mathcal{P}_\Psi \partial_\beta \mathcal{P}_\Psi \mathcal{P}_\Psi. \quad (2.12)$$

The part of the QGT antisymmetric in $\alpha, \beta$ corresponds to the adiabatic curvature $\mathcal{F}_{\alpha\beta}$ (determined from the Berry connections). The part of the QGT symmetric in $\alpha, \beta$ corresponds to a metric $\mathcal{M}_{\alpha\beta}$ on the parameter space.

### C. Examples

We now apply the four-corners decomposition and leak conditions to various types of As(H) and summarize some of our main results.

#### 1. Unique state case

In this case, As(H) is one-dimensional (with unique steady state $\varrho$) and the asymptotic projection preserves only the trace of the initial state:

$$\mathcal{P}_\infty(\rho_{\text{in}}) = \varrho \text{Tr}\{\rho_{\text{in}}\} \quad \text{and} \quad \mathcal{P}_\Psi(\rho_{\text{in}}) = \varrho \text{Tr}\{P\rho_{\text{in}}\}. \quad (2.13)$$

Note that we use $\varrho$ for states which are determined only by $\mathcal{L}$ (meaning they are independent of $\rho_{\text{in}}$). Since there is only one steady state, there is nowhere to move within As(H). Indeed, it is easy to show that

$$\mathcal{P}_\infty \mathcal{O} \mathcal{P}_\infty = \text{Tr}\{\mathcal{O}(\varrho)\}\mathcal{P}_\infty = 0 \quad (2.14)$$

for both types of perturbations $\mathcal{O}$ (2.7). Thus, the only novel application of our results to this case is the metric arising from the QGT,

$$\mathcal{M}_{\alpha\beta} = \text{Tr}\{\partial_{(\alpha} P \partial_{\beta)} \varrho\}, \quad (2.15)$$

where $A_{(\alpha}B_{\beta)} = A_\alpha B_\beta + A_\beta B_\alpha$. This metric is distinct from the Hilbert-Schmidt metric $\text{Tr}\{\partial_{(\alpha} \varrho \partial_{\beta)} \varrho\}$ for mixed $\varrho$ and is nonzero only when $\varrho$ is not full rank. For pure steady states, both metrics reduce to the Fubini-Study metric [74].

#### 2. Decoherence-free subspace (DFS) case

The simplest multidimensional As(H) which stores quantum information is a decoherence-free subspace (DFS) [8]. A $d^2$-dimensional DFS block,

$$\boxplus = \text{As(H)}, \quad (2.16)$$

is spanned by matrices $\{|\psi_k\rangle\langle\psi_l|\}_{k,l=0}^{d-1}$, where $\{|\psi_k\rangle\}_{k=0}^{d-1}$ is a basis for a subspace of the $d \leq N$-dimensional system space. The decaying block $\boxminus$ is then spanned by $\{|\psi_k\rangle\langle\psi_l|\}_{k,l=d}^{N-1}$. Evolution of the DFS under $\mathcal{L}$ is exclusively unitary,

$$\partial_t(|\psi_k\rangle\langle\psi_l|) = \mathcal{L}(|\psi_k\rangle\langle\psi_l|) = -i[H_\infty, |\psi_k\rangle\langle\psi_l|], \quad (2.17)$$

where $H_\infty$ is the asymptotic Hamiltonian and $k, l \leq d - 1$.

Since the entire upper-left block is preserved,

$$\mathcal{P}_\Psi(\rho_{\text{in}}) = \mathcal{P}_\boxplus(\rho_{\text{in}}) = P\rho_{\text{in}}P \quad (2.18)$$

for a DFS. We can thus deduce from Eq. (2.11) that the effect of Hamiltonian perturbations $V$ within As(H) is $V_\boxplus = PVP$—the Hamiltonian projected onto the DFS. Likewise, if $\mathcal{O} = \partial_\alpha$, then the Lindbladian adiabatic connection can be shown to reduce to $\partial_\alpha P \cdot P$, the adiabatic connection of the DFS. Naturally, the QGT and its corresponding metric also reduce to that of the DFS states. In other words, all such results are the same regardless of whether the states form a DFS of a Lindbladian or a degenerate subspace of a Hamiltonian.

#### 3. Noiseless subsystem (NS) case

This important case is a combination of the DFS and unique steady-state cases. In this case, the nondecaying portion of the system Hilbert space ($P$H) factors into a $d$-dimensional subspace $\mathsf{H}_{\text{DFS}}$ spanned by DFS states and a $d_{\text{ax}}$-dimensional auxiliary subspace $\mathsf{H}_{\text{ax}}$, which is the range of some unique steady state $\varrho_{\text{ax}}$ [$d_{\text{ax}} = \text{rank}(\varrho_{\text{ax}})$]. This combination of a DFS tensored with the auxiliary state $\varrho_{\text{ax}}$ is called a noiseless subsystem (NS) [9]. For one NS block, H decomposes as

$$\mathsf{H} = P\mathsf{H} \oplus Q\mathsf{H} = (\mathsf{H}_{\text{DFS}} \otimes \mathsf{H}_{\text{ax}}) \oplus Q\mathsf{H}. \quad (2.19)$$

A NS block is possible if $\mathcal{L}$ respects this decomposition and does not cause any decoherence within the DFS part. The DFS basis elements $|\psi_k\rangle\langle\psi_l|$ from Eq. (2.17) generalize to $|\psi_k\rangle\langle\psi_l| \otimes \varrho_{\text{ax}}$. For this case, states in $\boxplus$ are not perfectly preserved, but are instead partially traced over the auxiliary subspace:

$$\mathcal{P}_\Psi(\rho_{\text{in}}) = \text{Tr}_{\text{ax}}\{P\rho_{\text{in}}P\} \otimes \varrho_{\text{ax}}, \quad (2.20)$$

where $P = P_{\text{DFS}} \otimes P_{\text{ax}}$ and $P_{\text{DFS}}$ ($P_{\text{ax}}$) is the identity on $\mathsf{H}_{\text{DFS}}$ ($\mathsf{H}_{\text{ax}}$) and $\text{Tr}_{\text{ax}}$ is a trace over $\mathsf{H}_{\text{ax}}$.

Note that the auxiliary factor becomes trivial when $\varrho_{\text{ax}}$ is a pure state ($d_{\text{ax}} = 1$), reducing the NS to a DFS. This means that the NS case is distinct from the DFS case only when $\varrho_{\text{ax}}$ is mixed ($d_{\text{ax}} \neq 1$). Similarly, if the dimension of the DFS $d^2 = d = 1$, the NS reduces to the unique steady-state case. The NS case thus encapsulates both the DFS and unique state cases.

For this case, the effect of perturbations $\mathcal{V}$ on As(H) is more subtle due to the auxiliary factor, but the induced time evolution on the DFS is nevertheless still unitary. The effective DFS Hamiltonian is

$$W = \text{Tr}_{\text{ax}}\{\varrho_{\text{ax}} V_\boxplus\}. \quad (2.21)$$

Similarly, if we define generators of motion $G_\alpha$ in the $\mathbf{x}_\alpha$ direction in parameter space (i.e., such that $\partial_\alpha \rho_\infty = -i[G_\alpha, \rho_\infty]$), then the corresponding holonomy (Berry phase) after a closed path is the path-ordered integral of the various DFS adiabatic connections

$$\tilde{A}_\alpha^{\text{DFS}} = \text{Tr}_{\text{ax}}\{\varrho_{\text{ax}}(G_\alpha)_\boxplus\}. \quad (2.22)$$

In both cases, the effect of the perturbation on the DFS part depends on $\varrho_{\text{ax}}$, meaning that $\varrho_{\text{ax}}$ can be used to modulate both Hamiltonian-based and holonomic quantum gates. The QGT for this case is rather complicated due to the $\varrho_{\text{ax}}$-assisted adiabatic evolution, but we show that the QGT does endow us with a metric on the parameter space for a NS block.

#### 4. Multiblock case

The noiseless subsystem is the most general form of one block of asymptotic states of $\mathcal{L}$, and the most general As(H) is a direct sum of such NS blocks [78,80,81] [see Fig. 1(a)] with corresponding minimal projection $\mathcal{P}_\Psi$. This important result applies to both Lindbladians and more general quantum channels [79,82–85] (see Ref. [86] for a technical introduction). Throughout the paper, we explicitly calculate properties of one NS block $\{|\psi_k\rangle\langle\psi_l| \otimes \varrho_{ax}\}_{k,l=0}^{d-1}$ and sketch any straightforward generalizations to the multiblock case.

Both Eqs. (2.21) and (2.22) extend straightforwardly to the multiblock case, provided that the blocks maintain their shape during adiabatic evolution. We do not derive a metric for this case, so taking into account any potential interaction of the blocks during adiabatic evolution remains an open problem.

### D. Earlier work

We review efforts related to our work, including studies of the structure, stability, and control of Lindbladian steady-state subspaces.

Regarding the formula for $\mathcal{P}_\infty$ (Proposition 2), we have mentioned that the piece $\mathcal{P}_\infty \mathcal{P}_\blacksquare$ has already been determined in two seminal works, Baumgartner and Narnhofer [78] and Blume-Kohout et al. [79] (see also Ticozzi and Viola [80]). Our four-corners partition of $\mathcal{L}$ produces constraints on the Hamiltonian and jump operators of $\mathcal{L}$ (Proposition 1), which are already known from Refs. [78,80,87]. There exist related formulas for the parts of $\mathcal{P}_\infty \mathcal{P}_\blacksquare$ corresponding to fixed points of discrete-time quantum channels in Lemma 5.8 of Ref. [79] and Proposition 7 of Ref. [88] and of Markov chains in Theorem 3.3 of Ref. [89]. In addition, previous results assume no residual unitary evolution within As(H) (i.e., $H_\infty = 0$).

Regarding question (1), Jakob and Stenholm [90] mentioned the importance of conserved quantities in determining $\rho_\infty$ from $\rho_{in}$, but did not generalize to all As(H). This generalization was done by two of us [50], showing that $\rho_\infty$ does not depend on dynamics at any intermediate times. Here, we provide an analytical formula for the conserved quantities for multidimensional As(H). In contrast, current applications of the Keldysh formalism to Lindbladians [91] do not tackle such cases. Regarding channel simulation, theoretical efforts have focused on minimizing the ancillary resources required to simulate channels on a system [92–95]. To our knowledge, previous efforts did not consider constructing a more general quantum channel out of less general Markovian ones.

Regarding Hamiltonian control of As(H) [question (2)], there are two questions: (a) Is the dominant term generating evolution within As(H) or causing leakage out of it? and (b) does the term acting within As(H) generate unitary evolution? Regarding the first question, it has been widely believed (and often numerically verified, e.g., in Ref. [61]) that the term governing evolution within As(H), $\mathcal{P}_\infty \mathcal{V} \mathcal{P}_\infty$, dominates over the term governing leakage out of As(H) (provided that $\mathcal{V}$ is turned on for some finite time). Several works [59,62,96] have formally justified this claim and provided the necessary constraints on the time scale of the perturbation, interpreting As(H) as a quantum Zeno subspace [63,64,66] (see also Refs. [97,98]). Regarding the second question, Zanardi and Campos Venuti [59] recently proved that if $\mathcal{P}_\infty \mathcal{P}_\blacksquare = 0$, then $\mathcal{P}_\infty \mathcal{V} \mathcal{P}_\infty$ generates unitary evolution for the DFS case. They also showed [60] that Lindbladian jump operator perturbations induce unitary evolution on Lindbladians without decaying subspaces. We generalize both of these results (by showing that $\mathcal{P}_\infty \mathcal{P}_\blacksquare$ is always zero) to all As(H).

Regarding reservoir-engineered holonomic quantum computation [73] on As(H) [question (3)], we are faced again with two similar questions: (a) Is there an adiabatic limit for open systems? and (b) is the holonomy after a closed adiabatic deformation unitary? Regarding the first question, the adiabatic theorem has indeed been generalized to Lindblad master equations [70,72,99–102] and all orders of corrections to adiabatic evolution have been derived (see, e.g., Ref. [72], Theorem 6). This is the adiabatic limit dominated by steady states of $\mathcal{L}$. Another adiabatic limit exists which is dominated by eigenstates of the Hamiltonian part of $\mathcal{L}$ [103–105], which we do not address further here. Regarding question (b), Sarandy and Lidar [68] were the first to make contact between adiabatic or Berry connections and Lindbladians. Avron et al. (Ref. [72], Proposition 3) showed that the corresponding holonomy is trace preserving and completely positive. Carollo, Santos, and Vedral [67] showed that the holonomy is unitary for Lindbladians possessing one DFS block. Oreshkov and Calsamiglia [69] proposed a theory of adiabaticity which extended that result to the multiblock case and arrived at Eq. (2.22). They showed that corrections to their result were $O(1/\sqrt{T})$ (with $T$ being the traversal time), as opposed to $O(1/T)$ as in a proper adiabatic limit. By explicitly calculating the adiabatic connections, we connect the result of Ref. [69] with the formulation of Ref. [68], showing that nonadiabatic corrections are actually $O(1/T)$. We also extend Ref. [69] to NS cases where the dimension of the auxiliary subspace (i.e., the rank of $\varrho_{ax}$) can change. Finally, Zanardi and Campos Venuti (Ref. [60], Proposition 1) showed that first-order Hamiltonian evolution within As(H) can be thought of as a holonomy. We develop this connection further by showing that, for both processes, evolution within As(H) is generated by the same type of effective Hamiltonian [Eqs. (2.21) and (2.22)], and leakage out of As(H) is governed by the same energy scale. We make the same connection between ordinary and adiabatic perturbations to jump operators of $\mathcal{L}$; the latter were first studied in Avron *et al.* [70].

Next, we review the QGT, introduced for Hamiltonian systems in Ref. [74] (the term "QGT" was coined by Berry [75]). It encodes both a metric for measuring distances [106] and the adiabatic curvature. The QGT is experimentally probable (e.g., via current noise measurements [107]). The Berry curvature can be obtained from adiabatic transport in Hamiltonian [108–110] and Lindbladian [70,111] systems and even ordinary linear response ([112] and Appxendix C of [113]). Singularities and scaling behavior of the metric are in correspondence with quantum phase transitions [114–116]. Conversely, flatness of the metric and curvature may be used to quantify stability of a given phase [117–120], a topic of particular interest due to its applications in engineering exotic topological phases. Regarding generalization of the QGT [question (4)], to our knowledge there has been no introduction of a tensor including both the adiabatic curvature and a metric for As(H). However, Refs. [121,122] did apply various known metrics to study distinguishability within families of Gaussian fermionic and spin-chain steady states, respectively.

Regarding leakage out of As(H) [question (5)], the idea that ⊞ is not relevant to first-order nonadiabatic corrections was mentioned in the Supplemental Material of Ref. [69]. We extend that result to ordinary first-order perturbation theory. Regarding response (6), both ordinary [123–125] and adiabatic [72,126] time-dependent perturbation theory for Lindbladians have been developed earlier. In parallel to this work, Campos Venuti and Zanardi [127] further developed the Kubo formula for response to Lindladian perturbations to specific Lindbladians, most of which do not possess a decaying subspace.

Lastly, regarding Hall conductivity, Avron *et al.* [70] used adiabatic perturbation theory to show that the zero-frequency Hall conductivity is unaffected by a Lindbladian whose jump operators are the Landau level lowering (raising) operators $b$ ($b^\dagger$). We confirm their result using linear response (calculated for all frequencies) and extend it to jump operators that are powers of $b$. Still other jump operators are considered in Refs. [111,124].

### E. Structure of the paper

In Sec. III, we prove Eq. (2.6) for $\mathcal{P}_\infty$ by applying the four-corners decomposition to $\mathcal{L}$. We also study the dependence of $\rho_\infty$ on $\rho_{\text{in}}$ and show how $\mathcal{P}_\infty$ can be used to generate any quantum channel. The strategy of the rest of the paper is to apply the four-corners decomposition to leading-order response formulas from ordinary and adiabatic perturbation theory. In Sec. IV, we study the Kubo formula for Lindbladians and state conditions under which it reduces to a Hamiltonian-based formula. We also prove that the evolution within As(H) is unitary, study the effective dissipative gap $\Delta_{\text{edg}}$, and touch upon second-order perturbative effects. In a similar fashion, we study the adiabatic response formula for Lindbladians in Sec. V. There, we prove that adiabatic evolution within As(H) is unitary and link $\Delta_{\text{edg}}$ to nonadiabatic corrections. In Sec. VI, we introduce the Lindbladian QGT and calculate it for most of the examples discussed above. We discuss future directions in Sec. VII. Examples and links to the appendixes are placed throughout the paper when physical concreteness or extra pedagogy are desired.

## III. ASYMPTOTIC PROJECTION

In this section, we apply the four-corners partition to Lindbladian superoperators and derive a formula for the asymptotic projection $\mathcal{P}_\infty$ for nonsteady As(H) ($H_\infty \neq 0$). We also show how the presence of $H_\infty$ can influence the dependence of $\rho_\infty$ on $\rho_{\text{in}}$ and demonstrate how one can embed any quantum channel in $\mathcal{P}_\infty$.

### A. Four-corners partition of Lindbladians

As we introduce in Sec. II, the four-corners projections Eq. (2.2) partition every operator $A \in \text{Op}(\mathsf{H})$ into four independent parts. Combining this notation with the vectorized or double-ket notation for matrices in $\text{Op}(\mathsf{H})$ (see Appendix A), we can express any $A$ as a vector whose components are the respective parts. The following are, therefore, equivalent,

$$A = \begin{pmatrix} A_\boxminus & A_\boxslash \\ A_\boxbslash & A_\boxplus \end{pmatrix} \longleftrightarrow |A\rangle\!\rangle = \begin{bmatrix} |A_\boxminus\rangle\!\rangle \\ |A_\boxplus\rangle\!\rangle \\ |A_\boxdot\rangle\!\rangle \end{bmatrix}, \quad (3.1)$$

and $A_\boxdot = A_\boxslash + A_\boxbslash$. With $A$ written as a block vector, superoperators can now be represented as 3-by-3 block matrices acting on said vector. Note that we use square brackets for partitioning superoperators and parentheses for operators in $\text{Op}(\mathsf{H})$ [as in Fig. 1 and Eq. (3.1)]. We do so as well with the Lindbladian $\mathcal{L}$. Recall that

$$\mathcal{L}(\rho) = -i[H,\rho] + \frac{1}{2}\sum_\ell \kappa_\ell (2F^\ell \rho F^{\ell\dagger} - F^{\ell\dagger}F^\ell \rho - \rho F^{\ell\dagger}F^\ell), \quad (3.2)$$

with Hamiltonian $H$, jump operators $F^\ell \in \text{Op}(\mathsf{H})$, and positive rates $\kappa_\ell$. By writing $\mathcal{L} = \mathcal{I}\mathcal{L}\mathcal{I}$ using Eqs. (2.3) and (2.4) (see Appendixes B and C), we find that

$$\mathcal{L} = \begin{bmatrix} \mathcal{L}_\boxminus & \mathcal{P}_\boxminus \mathcal{L} \mathcal{P}_\boxdot & \mathcal{P}_\boxminus \mathcal{L} \mathcal{P}_\boxplus \\ 0 & \mathcal{L}_\boxdot & \mathcal{P}_\boxdot \mathcal{L} \mathcal{P}_\boxplus \\ 0 & 0 & \mathcal{L}_\boxplus \end{bmatrix}, \quad (3.3)$$

where $\mathcal{L}_{\boxplus} \equiv \mathcal{P}_{\boxplus}\mathcal{L}\mathcal{P}_{\boxplus}$. Note that $\mathcal{L}_{\boxplus}$ is a bona fide Lindbladian governing evolution within $\boxplus$, and the minimal projection $\mathcal{P}_\Psi$ is exactly the asymptotic projection of $\mathcal{L}_{\boxplus}$. The reason for the zeros in the first column is the inability of $\mathcal{L}$ to take anything out of $\boxplus$ (stemming from the definition of the four-corners projections). This turns out to be sufficient for $\mathcal{P}_{\boxplus}\mathcal{L}\mathcal{P}_{\boxplus}$ to also be zero, leading to the block upper-triangular form above. These constraints on $\mathcal{L}$ translate to well-known [78,80,87] constraints on the Hamiltonian and jump operators as follows (see Appendix B).

*Proposition 1.*—Let $\{P, Q\}$ be projections on $\mathsf{H}$ and $\{\mathcal{P}_{\boxplus}, \mathcal{P}_{\boxplus}, \mathcal{P}_{\boxplus}, \mathcal{P}_{\boxplus}\}$ be their corresponding projections on $\mathrm{Op}(\mathsf{H})$. Then

$$\forall \ell: F_{\boxplus}^\ell = 0, \quad (3.4)$$

$$H_{\boxplus} = -\frac{i}{2}\sum_\ell \kappa_\ell F_{\boxplus}^{\ell\dagger} F_{\boxplus}^\ell. \quad (3.5)$$

These constraints on $H_{\boxplus}$ and $F_{\boxplus}^\ell$ (due to Hermiticity, $H_{\boxplus} = H_{\boxplus}^\dagger$) leave only their complements as degrees of freedom. The four-corners decomposition provides simple expressions for the surviving matrix elements of Eq. (3.3) in terms of $H_{\boxplus}, F_{\boxplus}^\ell$; these are shown in Appendix C.

*DFS case.*—Recall that, in this case, $\mathrm{As}(\mathsf{H}) = \boxplus$ and $P = \sum_{k=0}^{d-1} |\psi_k\rangle\langle\psi_k|$ is the DFS projection. In the case of a nonsteady DFS, evolution within $\boxplus$ is exclusively unitary for all times and generated by a Hamiltonian superoperator $\mathcal{H}_\infty \equiv \mathcal{L}_{\boxplus}$. The jump operators in $\mathcal{L}_{\boxplus}$, Eq. (C1), must then act trivially:

$$F_{\boxplus}^\ell = a_\ell P \quad (3.6)$$

for some complex constants $a_\ell$. This implies that $\mathcal{P}_{\boxplus}\mathcal{L}\mathcal{P}_{\boxplus}$ from Eq. (C5) is zero and the partition Eq. (3.3) becomes

$$\mathcal{L} = \begin{bmatrix} \mathcal{H}_\infty & 0 & \mathcal{P}_{\boxplus}\mathcal{L}\mathcal{P}_{\boxplus} \\ 0 & \mathcal{L}_{\boxplus} & \mathcal{P}_{\boxplus}\mathcal{L}\mathcal{P}_{\boxplus} \\ 0 & 0 & \mathcal{L}_{\boxplus} \end{bmatrix}. \quad (3.7)$$

If we assume that $|\psi_k\rangle$ are eigenstates of $H_\infty$ (with $\mathcal{H}_\infty \equiv -i[H_\infty, \cdot]$) and remember condition Eq. (3.5), we reduce to well-known conditions guaranteeing $\mathcal{L}(|\psi_k\rangle\langle\psi_k|) = 0$ (Ref. [25], Theorem 1).

### B. Nonsteady asymptotic subspaces

Armed with the partition of $\mathcal{L}$ from Eq. (3.3), we study cases where $\mathrm{As}(\mathsf{H})$ contains unitarily evolving states [$H_\infty \neq 0$ from Eq. (2.5)]. The basis for $\mathrm{As}(\mathsf{H})$ consists of right eigenmatrices of $\mathcal{L}$ with pure imaginary eigenvalues. By definition, we can expand $|\rho_\infty\rangle\rangle$ in such a basis since all other eigenmatrices will decay to zero under $e^{t\mathcal{L}}$ for sufficiently large $t$. We call such eigenmatrices right asymptotic eigenmatrices $|\Psi_{\Delta\mu}\rangle\rangle$ with purely imaginary eigenvalue $i\Delta$ (used here as an index) and degeneracy index $\mu$ (that depends on $\Delta$). By definition, $|\Psi_{\Delta\mu}\rangle\rangle \in \boxplus$ and the eigenvalue equation is

$$\mathcal{L}|\Psi_{\Delta\mu}\rangle\rangle = i\Delta|\Psi_{\Delta\mu}\rangle\rangle. \quad (3.8)$$

Since $\mathcal{L}$ is not always diagonalizable, any degeneracy may induce a nontrivial Jordan block structure for a given $\Delta$. However, it can be shown (see, e.g., Ref. [50], Appendix C) that all Jordan blocks corresponding to asymptotic eigenmatrices are diagonal. Therefore, there exists a dual set of left asymptotic eigenmatrices $\langle\langle J^{\Delta\mu}|$ such that

$$\langle\langle J^{\Delta\mu}|\mathcal{L} = i\Delta\langle\langle J^{\Delta\mu}|. \quad (3.9)$$

The $J$ are either conserved or oscillating indefinitely:

$$\langle\langle J^{\Delta\mu}|\rho(t)\rangle\rangle = \langle\langle J^{\Delta\mu}|e^{t\mathcal{L}}|\rho_{\mathrm{in}}\rangle\rangle = e^{i\Delta t}\langle\langle J^{\Delta\mu}|\rho_{\mathrm{in}}\rangle\rangle \quad (3.10)$$

by trivial integration of the equations of motion [Eq. (2.17)]. For $\Delta = 0$, such $J$ are conserved quantities, so a natural question is whether they always commute with the Hamiltonian and the jump operators. It turns out that they do not always commute [50,78], and so various generalizations of Noether's theorem have to be considered [70,128]. Using the following analysis, we can say that $J$'s always commute with both the Hamiltonian and jump operators of $\mathcal{L}$ when there is no decaying subspace ($P = I$). If there is decay, then conserved quantities still commute with jump operators and the Hamiltonian in the non-decaying subspace ($[J_{\boxplus}, F_{\boxplus}^\ell] = 0$; see Appendix B), but no longer have to commute in general ($[J, F^\ell] \neq 0$).

The left and right eigenmatrices are dual in the sense that they can be made biorthogonal (while still maintaining the orthonormality of the right ones):

$$\langle\langle J^{\Delta\mu}|\Psi_{\Theta\nu}\rangle\rangle = \delta_{\Delta\Theta}\delta_{\mu\nu},$$
$$\langle\langle \Psi_{\Delta\mu}|\Psi_{\Theta\nu}\rangle\rangle = \delta_{\Delta\Theta}\delta_{\mu\nu}. \quad (3.11)$$

Outer products of such eigenmatrices can then be used to express the asymptotic projection

$$\mathcal{P}_\infty = \sum_{\Delta,\mu} |\Psi_{\Delta\mu}\rangle\rangle\langle\langle J^{\Delta\mu}|. \quad (3.12)$$

This is indeed a projection ($\mathcal{P}_\infty^2 = \mathcal{P}_\infty$) due to Eq. (3.11).

Since it was shown that evolution of asymptotic states is exclusively unitary (Ref. [78], Theorem 2), it must be that the eigenvalue set $\{\Delta\}$ is that of a Hamiltonian superoperator, which we define to be $\mathcal{H}_\infty \equiv -i[H_\infty, \cdot]$. In other words, we use the set $\{\Delta\}$ to construct a Hamiltonian $H_\infty \in \mathcal{P}_{\boxplus}\mathrm{Op}(\mathsf{H})$ (defined up to a constant energy shift) such that each $\Delta$ is a difference of the energies of $H_\infty$ and $|\Psi_{\Delta\mu}\rangle\rangle$ are eigenmatrices of $\mathcal{H}_\infty$. (Note that $\mathcal{H}_\infty$ shares the same eigenvalues as $\mathcal{P}_\infty\mathcal{L}\mathcal{P}_\infty$, but $\mathcal{H}_\infty \neq \mathcal{P}_\infty\mathcal{L}\mathcal{P}_\infty$ because the latter is not anti-Hermitian.) Because of this, the eigenmatrices $\{\Psi, J\}$ must come in complex conjugate pairs: $\Psi_{-\Delta\mu} = \Psi_{\Delta\mu}^{\dagger}$ (which obstructs us from constructing a Hermitian basis for $\{\Psi_{\Delta\neq 0,\mu}\}$) and the same for $J^{\Delta\mu}$. The explicit form of $H_\infty$ depends on the block diagonal structure of $\mathcal{P}_\infty$. Combining $\mathcal{P}_\infty$ with the definition of $\mathcal{H}_\infty$ yields

$$\lim_{t\to\infty} e^{t\mathcal{L}} = \sum_{\Delta,\mu} e^{i\Delta t} |\Psi_{\Delta\mu}\rangle\!\rangle\langle\!\langle J^{\Delta\mu}| \equiv e^{t\mathcal{H}_\infty}\mathcal{P}_\infty. \quad (3.13)$$

The asymptotic state is then expressible as

$$|\rho_\infty(t)\rangle\!\rangle = e^{t\mathcal{H}_\infty}\mathcal{P}_\infty|\rho_{\text{in}}\rangle\!\rangle \quad (3.14a)$$

$$= \sum_{\Delta,\mu} c_{\Delta\mu} e^{i\Delta t}|\Psi_{\Delta\mu}\rangle\!\rangle, \quad (3.14b)$$

with complex coefficients

$$c_{\Delta\mu} \equiv \langle\!\langle J^{\Delta\mu}|\rho_{\text{in}}\rangle\!\rangle = \text{Tr}\{(J^{\Delta\mu})^{\dagger}\rho_{\text{in}}\}. \quad (3.15)$$

These coefficients determine the footprint that $\rho_{\text{in}}$ leaves on $\rho_\infty$. In general, any part of $|\rho_{\text{in}}\rangle\!\rangle$ not in the kernel of $\mathcal{P}_\infty$ imprints on the asymptotic state since, by definition, that part overlaps with some $J^{\Delta\mu}$.

We proceed to determine $|J^{\Delta\mu}\rangle\!\rangle$ by plugging in the partition of $\mathcal{L}$ from Eq. (3.3) into the eigenvalue equation (3.9). The block upper-triangular structure of $\mathcal{L}$ readily implies that $|J^{\Delta\mu}_{\boxplus}\rangle\!\rangle$ are left eigenmatrices of $\mathcal{L}_{\boxplus}$:

$$\langle\!\langle J^{\Delta\mu}_{\boxplus}|\mathcal{L}_{\boxplus} = i\Delta\langle\!\langle J^{\Delta\mu}_{\boxplus}|. \quad (3.16)$$

Writing out the conditions on the remaining components $|J^{\Delta\mu}_{\boxminus}\rangle\!\rangle$ yields an analytic expression for $|J^{\Delta\mu}\rangle\!\rangle$. We state this formula below, noting that $[\mathcal{L}_{\boxminus}, \mathcal{P}_{\boxminus}] = 0$; the proof is given in Appendix B.

*Proposition 2.*—The left eigenmatrices of $\mathcal{L}$ corresponding to pure imaginary eigenvalues $i\Delta$ are

$$\langle\!\langle J^{\Delta\mu}| = \langle\!\langle J^{\Delta\mu}_{\boxplus}|\left(\mathcal{P}_{\boxplus} - \mathcal{L}\frac{\mathcal{P}_{\boxminus}}{\mathcal{L}_{\boxminus} - i\Delta\mathcal{P}_{\boxminus}}\right), \quad (3.17)$$

where $\langle\!\langle J^{\Delta\mu}_{\boxplus}|$ are left eigenmatrices of $\mathcal{L}_{\boxplus}$.

Plugging this result into Eq. (3.12) and setting $\Delta = 0$ yields the formula for $\mathcal{P}_\infty$ from Sec. II for the case when $H_\infty = 0$. We now go through the relevant special cases, introducing notation used throughout the rest of the paper.

*Unique state case.*—Here, As(H) is stationary because there is only one state $\varrho$. The corresponding conserved quantity is the identity $I$ (since $e^{t\mathcal{L}}$ preserves the trace). In the double-ket notation, the asymptotic projection Eq. (2.13) can be written as $\mathcal{P}_\infty = |\varrho\rangle\!\rangle\langle\!\langle I|$. Note that $P$ is the conserved quantity of $\mathcal{L}_{\boxplus}$.

*DFS case.*—In this case, all states in $\boxplus$ are asymptotic. Therefore, steady-state basis elements and conserved quantities of $\mathcal{L}_{\boxplus} = \mathcal{H}_\infty$ are equal: $|J^{\Delta\mu}_{\boxplus}\rangle\!\rangle = |\Psi_{\Delta\mu}\rangle\!\rangle$. Splitting the degeneracy index $\mu$ into two indices $k$, $l$ for convenience, one can express the right asymptotic eigenvectors as $\Psi_{\Delta,kl} = |\psi_k\rangle\langle\psi_l|$, where $\{|\psi_k\rangle\}$ is a basis for the DFS consisting of eigenstates of $H_\infty$ with energies $\{E_k\}$. The eigenvalue equation for $\Psi_{\Delta,kl}$ becomes

$$\mathcal{H}_\infty(\Psi_{\Delta,kl}) = -i[H_\infty, \Psi_{\Delta,kl}] = i(E_l - E_k)\Psi_{\Delta,kl}, \quad (3.18)$$

where $\Delta \equiv E_k - E_l$ is a difference of the energies of $H_\infty$.

*NS case.*—Let us now focus on a stationary As(H) ($H_\infty = 0$), meaning that all $\Delta = 0$, and we denote the respective As(H) basis elements and conserved quantities as $|\Psi_\mu\rangle\!\rangle \equiv |\Psi_{\Delta=0,\mu}\rangle\!\rangle$ and $|J^\mu\rangle\!\rangle \equiv |J^{\Delta=0,\mu}\rangle\!\rangle$. Since As(H) is stationary, we can construct a Hermitian matrix basis for both As(H) and the corresponding conserved quantities that uses one index and is orthonormal (under the trace). For the DFS part of the NS, we define the matrix basis $\{|\Psi_\mu^{\text{DFS}}\rangle\!\rangle\}_{\mu=0}^{d^2-1}$. Each $\Psi_\mu^{\text{DFS}}$ consists of Hermitian linear superpositions of the outer products $|\psi_k\rangle\langle\psi_l|$ and is *not* a density matrix. In this new notation, the basis elements for one NS block are then

$$|\Psi_\mu\rangle\!\rangle = \frac{1}{n_{\text{ax}}}\begin{pmatrix} |\Psi_\mu^{\text{DFS}}\rangle\!\rangle \otimes |\varrho_{\text{ax}}\rangle\!\rangle & 0 \\ 0 & 0 \end{pmatrix}. \quad (3.19)$$

We normalize the states using the auxiliary state norm (purity),

$$n_{\text{ax}} \equiv \sqrt{\langle\!\langle \varrho_{\text{ax}}|\varrho_{\text{ax}}\rangle\!\rangle} = \sqrt{\text{Tr}\{\varrho_{\text{ax}}^2\}}, \quad (3.20)$$

to ensure that $\langle\!\langle \Psi_\mu|\Psi_\nu\rangle\!\rangle = \delta_{\mu\nu}$. Since a NS block is a combination of the unique and DFS cases, the conserved quantities of $\boxplus$ (i.e., of $\mathcal{L}_{\boxplus}$) are direct products of the DFS and auxiliary conserved quantities [78,79]. The unique auxiliary conserved quantity is $P_{\text{ax}}$, the identity on the auxiliary subspace $\mathsf{H}_{\text{ax}}$. Combining this with the result above and multiplying by $n_{\text{ax}}$ so that $\Psi_\mu$ and $J^\mu$ are biorthogonal [see Eq. (3.11)], we see that

$$\langle\!\langle J^\mu| = n_{\text{ax}}\begin{pmatrix} \langle\!\langle \Psi_\mu^{\text{DFS}}| \otimes \langle\!\langle P_{\text{ax}}| & 0 \\ 0 & \langle\!\langle J^\mu_{\boxminus}| \end{pmatrix}. \quad (3.21)$$

We use the NS block basis of the above form throughout the paper. The asymptotic projection $\mathcal{P}_\infty$ is then

$$\mathcal{P}_\infty \equiv \sum_\mu |\Psi_\mu\rangle\!\rangle\langle\!\langle J^\mu| = \mathcal{P}_\Psi - \mathcal{P}_\Psi \mathcal{L}\mathcal{L}_{\boxminus}^{-1}, \quad (3.22)$$

where the minimal projection is

$$\mathcal{P}_\Psi \equiv \sum_\mu |\Psi_\mu\rangle\!\rangle\langle\!\langle J^\mu_{\boxplus}| \equiv \mathcal{P}_{\text{DFS}} \otimes |\varrho_{\text{ax}}\rangle\!\rangle\langle\!\langle P_{\text{ax}}| \quad (3.23)$$

and $\mathcal{P}_{\text{DFS}}(\cdot) = \sum_\mu |\Psi_\mu^{\text{DFS}}\rangle\!\rangle\langle\!\langle \Psi_\mu^{\text{DFS}}| \cdot \rangle\!\rangle = P_{\text{DFS}} \cdot P_{\text{DFS}}$ is the superoperator projection onto the DFS part. Applying this

to a state and remembering that $P = P_{\text{DFS}} \otimes P_{\text{ax}}$ yields the NS projection formula (2.20).

*Multiblock case.*—If there are two NS blocks (characterized by projections $P_{\text{DFS}}^{(\varkappa)} \otimes P_{\text{ax}}^{(\varkappa)}$ with $\varkappa \in \{1, 2\}$) and no decaying subspace, then the conserved quantities $J^{\varkappa,\mu} = \Psi_{\varkappa,\mu}^{\text{DFS}} \otimes P_{\text{ax}}^{(\varkappa)}$ do not have presence in the subspace of coherences between the blocks. Since the most general As(H) is a direct sum of such NS blocks [78,80,81], we can shade gray the blocks in which $J^\mu$ may not be zero [Fig. 1(b)], dual to $\Psi_\mu$ [Fig. 1(a)].

### C. Dependence of $\rho_\infty$ on $\rho_{\text{in}}$ and $H_\infty$

Here, we examine how $\rho_\infty$ depends on $\rho_{\text{in}}$, showing how $H_\infty$ can suppress purity of $\rho_\infty$. The coefficients Eq. (3.15) determining the dependence of $|\rho_\infty\rangle\rangle$ on $|\rho_{\text{in}}\rangle\rangle$ can be split into two parts,

$$c_{\Delta\mu} = \langle\langle J_{\boxminus}^{\Delta\mu}|\rho_{\text{in}}\rangle\rangle + \langle\langle J_{\boxminus}^{\Delta\mu}|\rho_{\text{in}}\rangle\rangle, \qquad (3.24)$$

with each part representing the footprint left by $\mathcal{P}_{\boxminus}|\rho_{\text{in}}\rangle\rangle$ and $\mathcal{P}_{\boxminus}|\rho_{\text{in}}\rangle\rangle$, respectively. We can readily see that coherences $\mathcal{P}_{\boxminus}|\rho_{\text{in}}\rangle\rangle$ decay and cannot imprint in $|\rho_\infty\rangle\rangle$. The second term can be expressed using Proposition 2:

$$\langle\langle J_{\boxminus}^{\Delta\mu}|\rho_{\text{in}}\rangle\rangle = -\langle\langle J_{\boxminus}^{\Delta\mu}|\mathcal{P}_{\boxminus}\mathcal{L}\mathcal{P}_{\boxminus}(\mathcal{L} - i\Delta)_{\boxminus}^{-1}|\rho_{\text{in}}\rangle\rangle. \quad (3.25)$$

Reading from right to left, this part first "scrambles" $\mathcal{P}_{\boxminus}|\rho_{\text{in}}\rangle\rangle$ via the inverse term, then "transfers" the result $\rho^\star$ into $\boxminus$ via the channel [Eq. (C7)]

$$\mathcal{P}_{\boxminus}\mathcal{L}\mathcal{P}_{\boxminus}(\rho^\star) = \sum_\ell \kappa_\ell F_{\boxminus}^\ell \rho^\star F_{\boxminus}^{\ell\dagger}, \qquad (3.26)$$

and, finally, "catches" that result with $\langle\langle J_{\boxminus}^{\Delta\mu}|$. The footprint thus depends on all three actions. The transfer channel in Eq. (3.26) is completely positive (Ref. [56], Theorem 8.1). One can see that this map has to be nonzero for $J_{\boxminus} \neq 0$, i.e., for any footprint to be left at all. This is indeed true when one remembers that all populations in $\boxminus$ are transferred since Lindbladian evolution is trace preserving (see Appendix C).

Now observe the scrambling term $(\mathcal{L} - i\Delta)_{\boxminus}^{-1}$. Since $\Delta$ is an energy difference from $H_\infty$, this tells us that unitary evolution in As(H) affects the dependence of $|\rho_\infty\rangle\rangle$ on $\mathcal{P}_{\boxminus}|\rho_{\text{in}}\rangle\rangle$. This effect cannot be removed by transforming into a rotating frame via $e^{t\mathcal{H}_\infty}$. In such a frame, $|\rho_\infty\rangle\rangle$ becomes a steady state, but the $\Delta$ dependence of $J_{\boxminus}^{\Delta\mu}$ (and therefore the expression for $c_{\Delta\mu}$) remains. This is because the evolution caused by $e^{t\mathcal{H}_\infty}$ is happening in conjunction with the nonunitary decay of $\mathcal{P}_{\boxminus}|\rho_{\text{in}}\rangle\rangle$, which can be interpreted as $\mathcal{H}_\infty$ affecting the "flow" of parts of $\mathcal{P}_{\boxminus}|\rho_{\text{in}}\rangle\rangle$ into As(H). One can thus see that the energy denominator (due to $H_\infty \neq 0$) may dampen the purity of the asymptotic state. We highlight this with a specific example.

#### 1. Example: Four-level system

Let H be four dimensional, with the first two levels $\{|\psi_0\rangle, |\psi_1\rangle\}$ being a DFS and the latter two $\{|\psi_0^\perp\rangle, |\psi_1^\perp\rangle\}$ decaying into the DFS. Let $H = 0$ and

$$F = \sum_{k=0}^{1} |\psi_k\rangle\langle\psi_k^\perp| + \alpha \sum_{k=0}^{1} (-1)^k |\psi_k^\perp\rangle\langle\psi_k^\perp|, \qquad (3.27)$$

where the first term $F_{\boxminus}$ makes sure that everything flows into the DFS and the last term $F_{\boxminus}$ dephases the non-DFS Bloch vector (with $\alpha \in \mathbb{R}$). The steady-state basis elements are $\Psi_{kl} = |\psi_k\rangle\langle\psi_l|$ since $F_{\boxminus} = 0$. We can then use Eq. (3.17): acting on $\Psi_{kl}$ with the adjoint of $\mathcal{L}$ (see Appendix A) and then the adjoint of $\mathcal{L}_{\boxminus}^{-1}$ [Eq. (C4)] yields the corresponding conserved quantities

$$J^{kl} = |\psi_k\rangle\langle\psi_l| + \frac{|\psi_k^\perp\rangle\langle\psi_l^\perp|}{1 + 2\alpha^2(1 - \delta_{kl})}. \qquad (3.28)$$

One can see that $J_{\boxminus}^{kl} = \Psi_{kl}$, a feature of the DFS case, and the absence of $|\psi_k\rangle\langle\psi_l^\perp|$ terms in $J^{kl}$, a key result of the paper. The only nontrivial feature of the steady state is due to $F_{\boxminus}$ and the resulting "scrambling term" $\mathcal{L}_{\boxminus}^{-1}$ in Eq. (3.25). Namely, an initial nonzero coherence $\langle\psi_0^\perp|\rho_{\text{in}}|\psi_1^\perp\rangle$ leads necessarily to a mixed steady state due to coherence suppression of order $O(\alpha^{-2})$.

Letting $\alpha = 0$, a similar effect can be achieved by adding the Hamiltonian $H = \frac{1}{2}\beta(|\psi_0\rangle\langle\psi_0| - |\psi_1\rangle\langle\psi_1|)$ (with $\beta \in \mathbb{R}$). Now the DFS is nonstationary (with $H_\infty = H$) and the off-diagonal DFS elements $\Psi_{k\neq l}$ rotate. Abusing notation by omitting the corresponding eigenvalue $\Delta = \beta$, the left asymptotic eigenvectors become

$$J^{kl} = |\psi_k\rangle\langle\psi_l| + \frac{|\psi_k^\perp\rangle\langle\psi_l^\perp|}{1 + i\beta(-)^l(1 - \delta_{kl})}. \qquad (3.29)$$

Despite $F_{\boxminus} = 0$, the scrambling term still inflicts damage to the initial state due to $H_\infty$ (for nonzero $\beta$), but now the coherence suppression is of order $O(\beta^{-1})$.

### D. Quantum channel simulation

Here, we show how to embed any quantum channel into $\mathcal{P}_\infty$. Recall that a quantum channel $\mathcal{E}$ taking a state $\rho$ from a $d_{\text{in}}$-dimensional input space $H_{\text{in}}$ to a $d_{\text{out}}$-dimensional output space $H_{\text{out}}$ acts as

$$\mathcal{E}(\rho) \equiv \sum_\ell E_\ell \rho E_\ell^\dagger, \qquad (3.30)$$

where $E_\ell$ are $d_{\text{out}}$-by-$d_{\text{in}}$-dimensional matrices and $\sum_\ell E_\ell^\dagger E_\ell$ is the identity on $d_{\text{in}}$. We construct a corresponding $\mathcal{L}$ such that $\mathcal{E} = \mathcal{P}_\infty \mathcal{P}_{\boxminus}$, with the input space matched to $\boxminus$ and output space to $\boxminus$. First, set all rates $\kappa_\ell$ of the Lindbladian equal to one rate $\kappa_{\text{eff}}$, which quantifies convergence to $\text{As}(\mathsf{H})$. Let $H=0$ and pad $E_\ell$ with zeros to obtain jump operators of dimension $d_{\text{in}} + d_{\text{out}}$,

$$F^\ell = F^\ell_{\boxplus} = \begin{pmatrix} 0 & E_\ell \\ 0 & 0 \end{pmatrix}. \tag{3.31}$$

This DFS case greatly simplifies the matrix elements of $\mathcal{L}$ in Appendix C. The decay-generating terms reduce to $\mathcal{L}_{\blacksquare} = -\frac{1}{2}\kappa_{\text{eff}}\mathcal{P}_{\blacksquare}$ and $\mathcal{L}_{\boxminus} = -\kappa_{\text{eff}}\mathcal{P}_{\boxminus}$, so one can think of $\kappa_{\text{eff}}$ as the inverse of a relaxation time $T_1$ for $\boxminus$. Using the Kraus form for the transfer term of $\mathcal{P}_\infty$ from Eq. (3.26) and simplifying yields

$$\mathcal{P}_\infty \mathcal{P}_{\boxminus} = -\mathcal{P}_{\blacksquare}\mathcal{L}\mathcal{L}_{\boxminus}^{-1} = \frac{1}{\kappa_{\text{eff}}}\mathcal{P}_{\blacksquare}\mathcal{L}\mathcal{P}_{\boxminus} = \mathcal{E}. \tag{3.32}$$

In other words, while not all quantum channels can be expressed as $e^{t\mathcal{L}}$ for any finite $t$, all can be embedded in some $\mathcal{P}_\infty = \lim_{t\to\infty} e^{t\mathcal{L}}$.

## IV. LINEAR RESPONSE

In this section, we apply the four-corners decomposition to the Kubo formula. For both Hamiltonian and jump operator perturbations, we show that evolution within $\text{As}(\mathsf{H})$ is of Hamiltonian form and that leakage out of $\text{As}(\mathsf{H})$ is governed by the effective dissipative gap.

### A. Decomposing the Kubo formula

Let us assume that time evolution is governed by a Lindbladian $\mathcal{L}$ and the initial state $\rho_\infty$ is steady; i.e., $\mathcal{L}(\rho_\infty) = 0$. The system is then perturbed as

$$\mathcal{L} \to \mathcal{L} + g(t)\delta\mathcal{L}, \tag{4.1}$$

where the perturbation superoperator $\delta\mathcal{L}$ is multiplied by a time-dependent factor $g(t)$. The Lindbladian-based Kubo formula [123,125,127,129,130] is derived analogously to the Hamiltonian formula; i.e., it is a leading-order Dyson expansion of the full evolution. The main difference is that the derivation is performed in the superoperator formalism. We study the difference between the perturbed and unperturbed expectation values, $\langle\!\langle \delta A(t) \rangle\!\rangle \equiv \langle\!\langle A|\rho(t) - \rho_\infty\rangle\!\rangle$ for some observable $A$. We remind the reader that we use vectorized notation for matrices and the Hilbert-Schmidt inner product $\langle\!\langle A|\rho(t)\rangle\!\rangle \equiv \text{Tr}\{A^\dagger\rho(t)\}$ (see Appendix A). Within first order in $g$, the Kubo formula is

$$\langle\!\langle \delta A(t) \rangle\!\rangle = \int_{-\infty}^{t} d\tau g(\tau) \langle\!\langle A| e^{(t-\tau)\mathcal{L}} \delta\mathcal{L}|\rho_\infty\rangle\!\rangle. \tag{4.2}$$

While this superoperator form looks very different from the usual time-ordered commutator expression, it offers an intuitive interpretation if one thinks of the system as evolving from the right side of the expression to the left. Reading the integrand from right to left, the steady state is perturbed by $\delta\mathcal{L}$ at a time $\tau$, then evolved under the unperturbed Lindbladian $\mathcal{L}$, and finally evaluated using the observable $A$ at a time $t \geq \tau$. The integral represents a sum over different times $\tau$ of the perturbation acting on the steady state. Removing $\langle\!\langle A|$ produces the first-order term in the Dyson series for $|\rho(t)\rangle\!\rangle$.

We dissect Hamiltonian and jump operator perturbations of $\mathcal{L}$ [Eq. (3.2)], respectively,

$$H \to H + g(t)V, \tag{4.3a}$$

$$F^\ell \to F^\ell + g(t)f^\ell \tag{4.3b}$$

[for $V, f^\ell \in \text{Op}(\mathsf{H})$ and $V^\dagger = V$], showing that both generate unitary evolution within all $\text{As}(\mathsf{H})$ and leakage caused by both does not take states into $\boxminus$. We handle the Hamiltonian case first for simplicity,

$$\delta\mathcal{L} = -i[V, \cdot] \equiv \mathcal{V}, \tag{4.4}$$

returning to the general case in Sec. IV D.

*Hamiltonian case.*—As a sanity check, we let $\mathcal{L} = \mathcal{H} = -i[H, \cdot]$ and massage Eq. (4.2) into standard form. For that, let $O(t) \equiv e^{iHt}Oe^{-iHt} = e^{-t\mathcal{H}}(O)$ and recall that $[H, \rho_\infty] = 0$, since $\rho_\infty$ is generically a superposition of projections on eigenstates of $H$. We can then commute $e^{iHt}$ with $\rho_\infty$ and cyclically permute under the trace to obtain

$$\langle\!\langle \delta A(t) \rangle\!\rangle = \frac{1}{i} \int_{-\infty}^{t} d\tau g(\tau) \text{Tr}\{[A(t-\tau), V]\rho_\infty\}. \tag{4.5}$$

We now use four-corners projections $\mathcal{P}_{\boxplus}$ to partition Eq. (4.2). Because of the no-leak property [Eq. (2.9)], $\mathcal{P}_{\boxminus}\mathcal{V}\mathcal{P}_{\blacksquare} = 0$. Remembering that the Lindbladian is block upper-triangular in the four-corners partition [see Eq. (3.3)], it follows that $e^{\mathcal{L}t}$ is also block upper-triangular. We do not make any assumptions on $A$: $\langle\!\langle A| = [\langle\!\langle A_{\blacksquare}| \langle\!\langle A_{\boxminus}| \langle\!\langle A_{\boxminus}|]$. Further decomposing the first term using the asymptotic projection $\mathcal{P}_\infty$ from Eq. (3.22) and its complement $\mathcal{Q}_\infty \equiv \mathcal{I} - \mathcal{P}_\infty$ yields

$$\langle\!\langle \delta A(t) \rangle\!\rangle = \int_{-\infty}^{t} d\tau g(\tau) \langle\!\langle A_{\blacksquare}| e^{(t-\tau)\mathcal{H}_\infty} \mathcal{W}|\rho_\infty\rangle\!\rangle \tag{4.6a}$$

$$+ \int_{-\infty}^{t} d\tau g(\tau) \langle\!\langle A_{\blacksquare}| e^{(t-\tau)\mathcal{L}} \mathcal{Q}_\infty \mathcal{P}_{\blacksquare}\mathcal{V}|\rho_\infty\rangle\!\rangle \tag{4.6b}$$

$$+ \int_{-\infty}^{t} d\tau g(\tau) \langle\!\langle A_{\boxminus}| e^{(t-\tau)\mathcal{L}} \mathcal{P}_{\blacksquare}\mathcal{V}|\rho_\infty\rangle\!\rangle. \tag{4.6c}$$

The terms differ by which parts of $\mathcal{V}$ perturb $\rho_\infty$ and also which parts of $A$ "capture" the evolved result. The three relevant parts of $A$ correspond to the three labels in Fig. 2. One can readily see that $A_\square$ is irrelevant to this order due to Eq. (2.9). Equation (4.6a) consists of perturbing and evolving *within* the asymptotic subspace **A**, shaded gray in the Fig 2. The effect of the perturbation within As(H) is $\mathcal{W} \equiv \mathcal{P}_\infty \mathcal{V} \mathcal{P}_\infty$ (shown in Sec. IV B to be of Hamiltonian form), and $\mathcal{H}_\infty$ is the part of the unperturbed $\mathcal{L}$ that generates unitary evolution within As(H). Equation (4.6a) therefore most closely resembles the traditional Hamiltonian-based Kubo formula. The remaining two terms quantify leakage out of As(H) and contain non-Hamiltonian contributions. Equation (4.6b) consists of perturbing into regions **B** and **C** in Fig. 2, but then evolving under $\mathcal{P}_\square e^{t\mathcal{L}} \mathcal{P}_\square$ strictly into region **B** (since $\mathcal{P}_\infty e^{t\mathcal{L}} \mathcal{Q}_\infty = 0$). Equation (4.6c) consists of perturbing into region **C** and remaining there after evolution due to $\mathcal{P}_\square e^{t\mathcal{L}} \mathcal{P}_\square$. This term is eliminated if $A_\square = 0$, i.e., if the observable is strictly in $\square$.

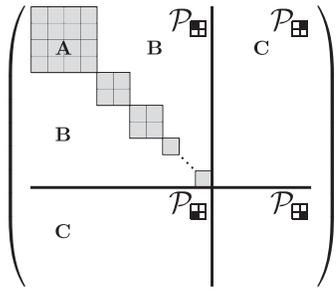

FIG. 2. Sketch of regions of linear response of the asymptotic subspace As(H) (gray) to a Hamiltonian perturbation. Each of three regions **A**, **B**, and **C** corresponds to the respective response term Eqs. (4.6a), (4.6b), and (4.6c) in the text.

*DFS case.*—Recall that in this case $\square$ is a DFS ($\mathcal{P}_\infty \mathcal{P}_\square = \mathcal{P}_\square$), and we do not assume it is stationary ($H_\infty \neq 0$). From Eq. (3.7), we can see that $\mathcal{L}$ cannot take any coherences in $\square$ back into the DFS ($\mathcal{P}_\square \mathcal{L} \mathcal{P}_\square = 0$). Therefore, the interference term [Eq. (4.6b)] is eliminated and the response formula reduces to

$$\langle\langle \delta A(t) \rangle\rangle = \int_{-\infty}^{t} d\tau g(\tau) \langle\langle A_\square | e^{(t-\tau)\mathcal{H}_\infty} \mathcal{W} | \rho_\infty \rangle\rangle \quad (4.7a)$$

$$+ \int_{-\infty}^{t} d\tau g(\tau) \langle\langle A_\square | e^{(t-\tau)\mathcal{L}} \mathcal{P}_\square \mathcal{V} | \rho_\infty \rangle\rangle. \quad (4.7b)$$

If, furthermore, $A_\square = 0$, there are no interference terms coming from outside of the DFS and the Lindbladian linear response reduces to the purely Hamiltonian-based term [Eq. (4.7a)]. Such a simplification can also be achieved when $\mathcal{V}_\square = 0$, which implies that the Hamiltonian perturbation does not take $\rho_\infty$ out of the DFS to begin with ($\mathcal{P}_\square \mathcal{V} \mathcal{P}_\square = 0$).

In the next section, we use the no-leak and clean-leak properties to determine that evolution within As(H) is of Hamiltonian form and to quantify the leakage scale of the remaining two terms [Eqs. (4.6b) and (4.6c)]. Before doing that, however, let us first show how and when the above decomposition is useful with an important example.

### 1. Example: Hall conductivity with dissipation

As an application of the Lindblad Kubo formula, let us consider a quantum Hall system with Markovian dissipation. We do not aim to represent physically sensible environments of electronic systems; such environments have already been thoroughly studied (see, e.g., Ref. [131]). Rather, we aim to describe artificial quantum Hall systems induced by light-matter interactions and/or photonic reservoir engineering. Such systems are being extensively studied both theoretically [38–41] and experimentally [43–45,132].

Consider a two-dimensional system of $N$ particles of mass $m$, charge $e = +1$, position **r**, and momentum **p**, in an area $A = L^2$ and external magnetic field $B$ (with $\hbar = 1$). The Hamiltonian is

$$H = \frac{1}{2m} \sum_i \pi_\varsigma^i \pi_\varsigma^i + \frac{1}{2} \sum_{i \neq j} U_{ij}$$

$$= \omega_c \sum_i \left( b_i^\dagger b_i + \frac{1}{2} \right) + \frac{1}{2} \sum_{i \neq j} U_{ij}, \quad (4.8)$$

where $i, j \in \{1, ..., N\}$ are particle indices and $\varsigma, \tau \in \{x, y\}$ index the spatial direction (with repeated indices summed). Above, we define the kinetic momentum $\boldsymbol{\pi}^i = \mathbf{p}^i - \mathbf{A}$ (with **A** the magnetic vector potential, $[\pi_\varsigma^i, \pi_\tau^j] = iB\epsilon_{\varsigma\tau}\delta_{ij}$, and $\epsilon_{\varsigma\tau}$ the antisymmetric Levi-Cività symbol), the Landau level lowering operators

$$b_j = \frac{1}{\sqrt{2B}} (\pi_x^j + i\pi_y^j), \quad (4.9)$$

a two-electron interaction potential $U_{ij}$, and the cyclotron frequency $\omega_c = B/m$. For simplicity, we assume

$$[U_{ij}, b_k] = [U_{ij}, b_k^\dagger] = 0. \quad (4.10)$$

Let us take the number of electrons $N$ to satisfy $\nu \equiv 2\pi N/BA = p/q \leq 1$ for $p, q \in \mathbb{Z}$, and let us assume the interaction potential $U_{ij}$ is chosen such that there is a gap above the ground state $|0\rangle$ [133] in the absence of dissipation. We take for our perturbation the electric potential corresponding to a uniform electric field,

$$\mathcal{V}_\tau \equiv -i[V_\tau, \cdot] \quad \text{with} \quad V_\tau \equiv -E_\tau \sum_i r_\tau^i, \quad (4.11)$$

and we measure the total current in the $\varsigma$ direction. The frequency-dependent conductivity tensor $\sigma_{\varsigma\tau}^0$ for the Hamiltonian system can be extracted from Eq. (4.2) and is given by [77]

$$\sigma^0_{\varsigma\tau}(\omega) = \frac{1}{A}\lim_{\epsilon\downarrow 0}\int_0^\infty dt\, e^{i(\omega+i\epsilon)t}\langle\!\langle J^0_\varsigma|e^{t\mathcal{H}}\mathcal{V}_\tau|\rho_\infty\rangle\!\rangle \quad (4.12\text{a})$$

$$= \frac{\nu\omega_c}{2\pi(\omega^2-\omega_c^2)}(i\omega\delta_{\varsigma\tau} - \omega_c\epsilon_{\varsigma\tau}), \quad (4.12\text{b})$$

where $\rho_\infty = |0\rangle\langle 0|$ is the ground state, the total current

$$J^0_\varsigma = \frac{1}{m}\sum_i \pi^i_\varsigma \quad (4.13)$$

is obtained from the Hamiltonian-based continuity equation, and $\mathcal{H} = -i[H,\cdot]$. We can further extract the quantized zero-frequency Hall conductivity:

$$\sigma_H \equiv \sigma^0_{xy}(\omega\to 0) = \frac{\nu}{2\pi}. \quad (4.14)$$

We now examine the fate of the conductivity in the presence of dissipation. Let us subject the system to Lindblad evolution [Eq. (A3)] with rates $\kappa_i = 1$ and single-particle jump operators

$$F_i = \sqrt{2\gamma_1}b_i + \sqrt{2\gamma_2}b_i^2 + \cdots. \quad (4.15)$$

Note that the coefficients $\gamma_J$ must be independent of particle index $i$ for identical particles. One has to be careful about defining the current operator $J_\varsigma$. The current density $j_\varsigma(\mathbf{r})$ now obeys the Lindbladian-based continuity equation,

$$\partial_t n + \partial_\varsigma j_\varsigma = \mathcal{L}^\ddagger(n) + \partial_\varsigma j_\varsigma = 0, \quad (4.16)$$

where $n = \sum_i \delta(\mathbf{r}-\mathbf{r}_i)$ is the particle density operator (see Appendix A for a formal definition of $\ddagger$). The total current is then expressed as

$$J_\varsigma = \frac{1}{L}\int d^2r\, j_\varsigma(\mathbf{r}). \quad (4.17)$$

This is the sensible and measurable definition of current in a dissipative system (Ref. [70], Sec. 5.2) since it represents the time-rate change of charge density in a region. Taking the Fourier transform of Eq. (4.16) and expanding to lowest order in wave vector yields

$$J_\varsigma = \sum_i \partial_t r^i_\varsigma = \sum_i \mathcal{L}^\ddagger(r^i_\varsigma), \quad (4.18)$$

and the Kubo formula (4.12a) generalizes to

$$\sigma_{\varsigma\tau}(\omega) = \frac{1}{A}\lim_{\epsilon\downarrow 0}\int_0^\infty dt\, e^{i(\omega+i\epsilon)t}\langle\!\langle J_\varsigma|e^{t\mathcal{L}}\mathcal{V}_\tau|\rho_\infty\rangle\!\rangle. \quad (4.19)$$

*Unique state case.*—We first consider the case when $\gamma \equiv \gamma_1 \neq 0$, $\gamma_{J>1} = 0$, so that $F_i \propto b_i$. The key observation is that the current operator is given by

$$J_\varsigma = J^0_\varsigma + \frac{1}{m}\sum_i \epsilon_{\varsigma\tau}\frac{\gamma}{\omega_c}\pi^i_\tau. \quad (4.20)$$

With this form of the current operator and our choice of $F_i$ and $U_{ij}$, Eq. (4.19) can be evaluated for all frequencies:

$$\sigma_{\varsigma\tau}(\omega) = \left(\delta_{\varsigma\lambda} + \frac{\gamma}{\omega_c}\epsilon_{\varsigma\lambda}\right)\sigma^0_{\lambda\tau}(\tilde{\omega})$$

$$= \frac{\nu\omega_c}{2\pi(\tilde{\omega}^2-\omega_c^2)}\left[i\omega\delta_{\varsigma\tau} - \omega_c\left(1 - i\frac{\gamma\tilde{\omega}}{\omega_c^2}\right)\epsilon_{\varsigma\tau}\right], \quad (4.21)$$

with complex frequency $\tilde{\omega} \equiv \omega + i\gamma$. Quite surprisingly, the Hall conductivity at zero frequency is *still* given by its quantized value,

$$\sigma_{xy}(\omega\to 0) = \sigma_H = \frac{\nu}{2\pi}, \quad (4.22)$$

due to an interesting interplay between the Lindbladian time evolution and the modification to the current operator. This effect can also be observed when calculating the quantized Hall conductivity using adiabatic perturbation theory (Ref. [70], Sec. 7). It is even present when we extend this case to the case of a low (but nonzero) temperature thermal bath, up to exponential corrections due to leakage out of the lowest Landau level (see Appendix D). Additionally, we see that the usual cyclotron pole at $\omega = \omega_c$—guaranteed to be present in the Hamiltonian case by Kohn's theorem [134]—is broadened into a Lorentzian due to the presence of dissipation. This shows that while the cyclotron resonance is independent of the details of interactions, it is in fact sensitive to dissipation.

*DFS case.*—Here, we look at the case when $\gamma_1 = 0$ and $\gamma_{J>1} \neq 0$. Now the asymptotic subspace consists of all states in the lowest two Landau levels—a DFS case. Therefore, it is useful to consider the DFS Kubo formula [Eq. (4.7)]. The key point now is that the perturbation $\mathcal{V}_\tau$ leaves $\rho_\infty$ in the steady-state subspace, and, hence, the second term [Eq. (4.7b)] in the Kubo formula vanishes. Although the current operators $J_\varsigma$, determined by Eq. (4.18), depend on the jump operators $F_i$, the projection $(J_\varsigma)_\boxplus$, which appears in the first term [Eq. (4.7a)], is independent of $F_i$ and equivalent to the Hamiltonian-based current Eq. (4.13): $(J_\varsigma)_\boxplus = J^0_\varsigma$. These two observations conspire to ensure that the conductivity at all frequencies is unaffected by dissipation and is still given by $\sigma^0_{\varsigma\tau}(\omega)$ from Eq. (4.12b).

### B. Evolution within As(H)

Let us now focus on the term $\mathcal{W} \equiv \mathcal{P}_\infty \mathcal{V}\mathcal{P}_\infty$ [Eq. (4.6a)] quantifying the effect of the perturbation within As(H). Becayse of a lack of a formula for $\mathcal{P}_\infty$, it was previously unclear whether $\mathcal{W}$ is capable of causing any decoherence within As(H). We now show that it is not. Therefore, the first-order effect of the perturbation within As(H) will always be of Hamiltonian form.

A swift application of the no-leak and clean-leak properties, Eqs. (2.9) and (2.11), allows us to substitute $\mathcal{P}_\Psi \equiv \mathcal{P}_\infty \mathcal{P}_\blacksquare$ for $\mathcal{P}_\infty$. Recall that $\mathcal{P}_\blacksquare \mathcal{V} \mathcal{P}_\blacksquare = 0$ and that $\mathcal{W}$ is acting on a steady state $\rho_\infty \in \text{As}(H)$, yielding

$$\mathcal{W}|\rho_\infty\rangle\!\rangle = \mathcal{P}_\infty \mathcal{P}_\blacksquare \mathcal{V} \mathcal{P}_\infty |\rho_\infty\rangle\!\rangle = \mathcal{P}_\Psi \mathcal{V} \mathcal{P}_\Psi |\rho_\infty\rangle\!\rangle. \quad (4.23)$$

As seen from the full Kubo formula, this term is of the same order in the perturbation as the two leakage terms [Eqs. (4.6b) and (4.6c)]. However, if $H_\infty = 0$ and if the perturbation is turned on for a finite time $T$ and rescaled by $1/T$, it can be shown [59,62,96] that $\mathcal{W}$ is the only leading-order effect. Therefore, the *entire* state undergoes quantum Zeno dynamics according to $\mathcal{W}$ (Refs. [63,64,66]; see also Refs. [97,98]). We show below that such dynamics is unitary for all As(H).

*DFS case.*—We immediately read off the effective Hamiltonian for the DFS case. Since $\mathcal{P}_\Psi = \mathcal{P}_\blacksquare$,

$$\mathcal{W} = -i[V_\blacksquare, \cdot], \quad (4.24)$$

with $V_\blacksquare$ the perturbation projected onto the DFS. Applications of this formula to circuit and waveguide QED quantum computation schemes can, respectively, be found in Refs. [61,62].

*NS case.*—In this case, we have to use the formula for $\mathcal{P}_\Psi$ from Eq. (3.23), restated below:

$$\mathcal{P}_\Psi = \mathcal{P}_{\text{DFS}} \otimes |\varrho_{\text{ax}}\rangle\!\rangle\!\langle\!\langle P_{\text{ax}}|, \quad (4.25)$$

with $\mathcal{P}_{\text{DFS}}(\cdot) = P_{\text{DFS}} \cdot P_{\text{DFS}}$ being the superoperator projection on the DFS part, $P_{\text{ax}}$ being the operator projection on the auxiliary part, and $P = P_{\text{DFS}} \otimes P_{\text{ax}}$. Direct multiplication yields

$$\mathcal{W} = \mathcal{P}_\Psi \mathcal{V} \mathcal{P}_\Psi = \langle\!\langle P_{\text{ax}}|\mathcal{V}|\varrho_{\text{ax}}\rangle\!\rangle \otimes |\varrho_{\text{ax}}\rangle\!\rangle\!\langle\!\langle P_{\text{ax}}|, \quad (4.26)$$

where the evolution within the auxiliary part is trivial and evolution within the DFS part is generated by the effective DFS Hamiltonian $W$:

$$\langle\!\langle P_{\text{ax}}|\mathcal{V}|\varrho_{\text{ax}}\rangle\!\rangle = -i[\text{Tr}_{\text{ax}}\{\varrho_{\text{ax}} V_\blacksquare\}, \cdot] \equiv -i[W, \cdot]. \quad (4.27)$$

To better reveal the effect of $\varrho_{\text{ax}}$, it is worthwhile to express $V_\blacksquare$ as a sum of tensor products of various DFS and auxiliary Hamiltonians: $V_\blacksquare = \sum_\iota V^\iota \otimes V^\iota_{\text{ax}}$. The effective Hamiltonian then becomes

$$W = \sum_\iota \text{Tr}_{\text{ax}}\{\varrho_{\text{ax}} V^\iota_{\text{ax}}\} V^\iota. \quad (4.28)$$

In words, $\mathcal{W}$ is a linear combination of Hamiltonian perturbations $V^\iota$ on the DFS, with each perturbation weighed by the expectation value of the corresponding auxiliary operator $V^\iota_{\text{ax}}$ in the state $\varrho_{\text{ax}}$.

### C. Leakage out of As(H)

Now, let us set $H_\infty = 0$ and focus on the two leakage terms [Eqs. (4.6b) and (4.6c)] from the Kubo formula. For simplicity, let us slowly ramp up the perturbation $g(t)\mathcal{V}$ to a constant, so $g(t) = \lim_{\eta \to 0} e^{\eta t \Theta(-t)}$, with $\Theta(t)$ the Heaviside step function. This simplifies the leakage part of the Kubo formula using the Drazin inverse of $\mathcal{L}$:

$$\int_{-\infty}^{t} d\tau g(\tau) e^{(t-\tau)\mathcal{L}} \mathcal{Q}_\infty = \int_0^\infty dt e^{t\mathcal{L}} \mathcal{Q}_\infty \equiv -\mathcal{L}^{-1}. \quad (4.29)$$

This pseudoinverse ($\mathcal{L}^{-1}\mathcal{L} = \mathcal{L}\mathcal{L}^{-1} = \mathcal{Q}_\infty$) is also the inverse of all invertible parts in the Jordan normal form of $\mathcal{L}$ (Ref. [59], Appendix D). Plugging this in and omitting $\langle\!\langle A|$, the leakage terms Eqs. (4.6b) and (4.6c) reduce to

$$\mathcal{Q}_\infty |\rho(t)\rangle\!\rangle = -\mathcal{L}^{-1}\mathcal{V}|\rho_\infty\rangle\!\rangle. \quad (4.30)$$

Now we can apply the clean-leak property Eq. (2.11) to narrow down those eigenvalues of $\mathcal{L}$ that are relevant in characterizing the scale of the leakage. By definition Eq. (4.29), $\mathcal{L}^{-1}$ has the same block upper-triangular structure as $\mathcal{L}$ from Eq. (3.3). This fact conspires with $\mathcal{P}_\blacksquare \mathcal{V} \mathcal{P}_\blacksquare = 0$ to allow us to ignore $\mathcal{L}_\blacksquare$ and write

$$\mathcal{Q}_\infty |\rho(t)\rangle\!\rangle = -\mathcal{L}_\blacksquare^{-1} \mathcal{V}|\rho_\infty\rangle\!\rangle. \quad (4.31)$$

Therefore, the relevant gap is the nonzero eigenvalue of $\mathcal{L}_\blacksquare$ with the smallest absolute value. However, we now show how the spectrum of $\mathcal{L}_\blacksquare$ is actually contained in the spectrum of $\mathcal{L}_\blacksquare + \mathcal{L}_\blacksquare$. Recalling the block upper-triangular structure of $\mathcal{L}$ from Eq. (3.3), one can establish that its eigenvalues must consist of eigenvalues of $\mathcal{L}_\blacksquare$, $\mathcal{L}_\blacksquare$, and $\mathcal{L}_\blacksquare$. However, evolution of the two coherence blocks is decoupled, $\mathcal{L}_\blacksquare = \mathcal{L}_\blacksquare + \mathcal{L}_\blacksquare$ (see Appendix C), and eigenvalues of $\mathcal{L}_\blacksquare$ come in pairs. Therefore, one can then define the effective dissipative gap $\Delta_{\text{edg}}$ to be the nonzero eigenvalue of $\mathcal{L}_\blacksquare + \mathcal{L}_\blacksquare$ with the smallest absolute value.

As a brief aside, we mention that the piece $\mathcal{L}_\blacksquare$ is also not relevant in a term $\mathcal{P}_\infty \mathcal{V} \mathcal{L}^{-1} \mathcal{V} \mathcal{P}_\infty$ [135–137] that acts on As(H) and is second order in the perturbation. Since $\mathcal{P}_\blacksquare \mathcal{V} \mathcal{P}_\blacksquare = 0$, one can reduce this term to $\mathcal{P}_\infty \mathcal{V} \mathcal{L}_\blacksquare^{-1} \mathcal{V} \mathcal{P}_\Psi$. However, we cannot replace the remaining $\mathcal{P}_\infty$ with $\mathcal{P}_\Psi$ since two actions of $\mathcal{V}$ can take the state from $\blacksquare$ to $\blacksquare$.

*DFS case.*—Recall that now all of $\blacksquare$ is stationary (provided that $H_\infty = 0$). We show that for certain DFS cases, $\Delta_{\text{edg}}$ is the excitation gap of a related Hamiltonian. Such DFS cases are those where $\mathcal{L}$ [Eq. (A3)] can be written without a Hamiltonian part,

$$\mathcal{L}(\rho) = \frac{1}{2}\sum_{\ell}\kappa_{\ell}(2F^{\ell}\rho F^{\ell\dagger} - F^{\ell\dagger}F^{\ell}\rho - \rho F^{\ell\dagger}F^{\ell}), \quad (4.32)$$

and where DFS states are annihilated by the jump operators, $F^{\ell}|\psi_k\rangle = 0$. This implies that $F^{\ell}_\boxplus = PF^{\ell}P = 0$ (with $P = \sum_{k=0}^{d-1}|\psi_k\rangle\langle\psi_k|$). We now determine $\Delta_{\text{edg}}$ for such systems. Since there is no evolution in $\boxplus$, $\mathcal{L}_\boxplus = 0$. Borrowing from Appendix C and using the above assumptions,

$$\mathcal{L}_{\boxminus}(\rho) = -\frac{1}{2}\sum_{\ell}\kappa_{\ell}P\rho(F^{\ell\dagger}F^{\ell})_{\boxminus}. \quad (4.33)$$

From this, we can extract the *decoherence* [138] or *parent* [31] Hamiltonian:

$$H_{\text{edg}} \equiv \frac{1}{2}\sum_{\ell}\kappa_{\ell}F^{\ell\dagger}F^{\ell}. \quad (4.34)$$

The (zero-energy) ground states of $H_{\text{edg}}$ are exactly the DFS states $|\psi_k\rangle$ [31,138] and the excitation gap of $H_{\text{edg}}$ is $\Delta_{\text{edg}}$.

#### 1. Example: Driven two-photon absorption

As an example of the above DFS simplification, consider the bosonic Lindbladian [61,71,139,140] with one jump operator $F = a^2 - \alpha^2$ and rate $\kappa = 1$, where $\alpha \in \mathbb{R}$, $[a, a^\dagger] = I$ and $n \equiv a^\dagger a$. For sufficiently large $\alpha$, this Lindbladian possesses a DFS spanned by the bosonic coherent states $|\alpha\rangle$ and $|-\alpha\rangle$. All states orthogonal to $|\pm\alpha\rangle$ constitute the decaying subspace $\boxminus$. The decoherence Hamiltonian is readily calculated to be

$$H_{\text{edg}} = \frac{1}{2}[n(n-1) - \alpha^2(a^2 + a^{\dagger 2}) + \alpha^4]. \quad (4.35)$$

The excitation gap of $H_{\text{edg}}$ ($\Delta_{\text{edg}}$) is plotted in Fig. 3 versus $\alpha$, along with $\Delta_{\text{dg}}$ and the eigenvalue of $\mathcal{L}_{\boxminus}$ with smallest real part. One can see that for $\alpha > 1.5$, the dissipative gap of $\mathcal{L}$ is smaller and does not coincide with the energy scale governing leakage.

### D. Jump operator perturbations

Having covered Hamiltonian perturbations, let us return to jump operator perturbations of the Lindbladian Eq. (3.2). Recall from Eq. (4.3b) that

$$F \to F + g(t)f, \quad (4.36)$$

with $f \in \text{Op}(\mathsf{H})$, not necessarily Hermitian. It was first shown in Ref. [60] that such perturbations actually induce unitary evolution on NS blocks of those Lindbladians that do not possess a nontrivial decaying subspace ($P = I$). Here, we extend this interesting result to cases where $P \neq I$, thereby covering all $\mathcal{L}$. Namely, just like Hamiltonian perturbations $\mathcal{V}$, jump operator perturbations induce unitary evolution within As(H) and the leakage scale associated with them is still $\Delta_{\text{edg}} \geq \Delta_{\text{dg}}$.

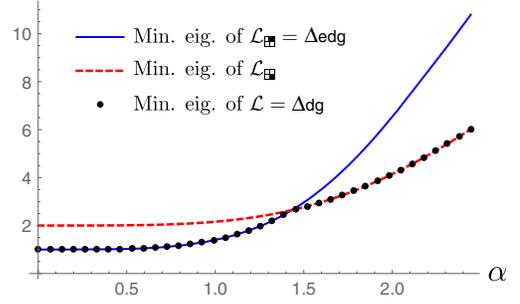

FIG. 3. Plot of the effective dissipative gap $\Delta_{\text{edg}}$, the nonzero eigenvalue of $\mathcal{L}_{\boxminus}$ with smallest real part, and the dissipative gap $\Delta_{\text{dg}}$ versus $\alpha$ for the Lindbladian with jump operator $F = a^2 - \alpha^2$. One can see that $\Delta_{\text{edg}} \geq \Delta_{\text{dg}}$.

Returning to Eq. (4.4), the action of the perturbation to first order in $g$ is characterized by

$$\delta\mathcal{L}(\rho) \equiv \mathcal{Y}(\rho) = \kappa\left(F\rho f^\dagger + \text{H.c.} - \frac{1}{2}\{f^\dagger F + F^\dagger f, \rho\}\right), \quad (4.37)$$

with $\kappa$ being the rate corresponding to the jump operator $F$ (we ignore the index $\ell$ for clarity). We hope to invoke the clean-leak property [Eq. (2.11)] once again, but the first term on the right-hand side of the above equation acts simultaneously and nontrivially on *both* sides of $\rho$. There is thus a possibility that one can reach $\boxminus$ when acting with $\mathcal{Y}$ on a steady state $\rho_\infty$. However, the condition $F_\boxminus = 0$ from Proposition 1 implies that $\mathcal{P}_\boxminus(F\rho_\infty f^\dagger)$ is zero for all $f$, so one can still substitute $\mathcal{P}_\Psi$ for $\mathcal{P}_\infty$:

$$\mathcal{P}_\infty \mathcal{Y} \mathcal{P}_\infty |\rho_\infty\rangle\!\rangle = \mathcal{P}_\Psi \mathcal{Y} \mathcal{P}_\Psi |\rho_\infty\rangle\!\rangle. \quad (4.38)$$

Furthermore, the fact that $\mathcal{P}_\boxminus \mathcal{Y} \mathcal{P}_\boxminus = 0$ allows us to ignore $\boxminus$ in determining the leakage energy scale associated with these jump operator perturbations. We finish with calculating the corresponding effective Hamiltonian for the most general cases.

*NS case.*—Having eliminated the influence of the decaying subspace $\boxminus$, we can now repeat the calculation done for Hamiltonian perturbations using the NS projection Eq. (4.25), yielding

$$\mathcal{P}_\Psi \mathcal{Y} \mathcal{P}_\Psi = \langle\!\langle P_{\text{ax}}|\mathcal{Y}|\varrho_{\text{ax}}\rangle\!\rangle \otimes |\varrho_{\text{ax}}\rangle\!\rangle\langle\!\langle P_{\text{ax}}|. \quad (4.39)$$

After some algebra, the DFS part reduces to Hamiltonian form [60]: $\langle\!\langle P_{\text{ax}}|\mathcal{Y}|\varrho_{\text{ax}}\rangle\!\rangle = -i[Y,\cdot]$, where

$$Y \equiv \frac{i}{2}\kappa \mathrm{Tr}_{\mathrm{ax}}\{\varrho_{\mathrm{ax}}(F^\dagger_\blacksquare f_\blacksquare - f^\dagger_\blacksquare F_\blacksquare)\}. \quad (4.40)$$

*Multiblock case.*—We now sketch the calculation of both Hamiltonian and jump operator perturbations, $\delta\mathcal{L} = \mathcal{V} + \mathcal{Y}$, for the most general case of $\blacksquare$ housing multiple NS blocks. Once again, we can get rid of the decaying subspace and substitute $\mathcal{P}_\Psi$ for $\mathcal{P}_\infty$. In addition, since $\mathcal{P}_\Psi$ does not have any presence except within the (gray) NS blocks of $\blacksquare$ [(see Fig. 1), $\mathcal{P}_\Psi$ will not project onto any coherences between the NS blocks. The contributing part of $\mathcal{P}_\infty \delta\mathcal{L} \mathcal{P}_\infty$ thus consists of the Hamiltonian and jump operator perturbations projected onto each NS block. Combining the effective Hamiltonians arising from $\mathcal{V}$ and $\mathcal{Y}$ [respectively, Eqs. (4.27) and (4.40)], the effective evolution within the DFS part of each NS block (indexed by $\varkappa$) is generated by the Hamiltonian

$$X^{(\varkappa)} \equiv \mathrm{Tr}_{\mathrm{ax}}^{(\varkappa)}\left\{\varrho_{\mathrm{ax}}^{(\varkappa)}\left(V_\blacksquare + \frac{i}{2}\kappa(F^\dagger_\blacksquare f_\blacksquare - f^\dagger_\blacksquare F_\blacksquare)\right)\right\}. \quad (4.41)$$

In fact, the unprojected Hamiltonian,

$$X \equiv V + \frac{i}{2}\kappa(F^\dagger f - f^\dagger F), \quad (4.42)$$

has previously been introduced (Ref. [70], Theorem 5) as the operator resulting from joint adiabatic variation of the Hamiltonian and jump operators of $\mathcal{L}$. It is thus not surprising that the effect of perturbations to the Hamiltonian and jump operators on $\rho_\infty$ is $X$ projected onto $\mathrm{As}(\mathsf{H})$.

## V. ADIABATIC RESPONSE

We now apply the four-corners decomposition to adiabatic perturbation theory. Here, the leading-order term governs adiabatic evolution within $\mathrm{As}(\mathsf{H})$ while all other terms are nonadiabatic corrections. We show that for a cyclic adiabatic deformation of steady $\mathrm{As}(\mathsf{H})$, the holonomy is unitary. We also determine that the energy scale governing nonadiabatic corrections is once again governed by the effective dissipative gap $\Delta_{\mathrm{edg}}$.

### A. Decomposing the adiabatic formula

First, let us briefly recall the setup of the standard adiabatic limit for Lindbladians (see Sec. II D for a reference list). Readers who are unfamiliar are encouraged to read about the closely related Hamiltonian-based adiabatic limit in Appendix E 1. Unlike adiabatic evolution of "non-Hermitian Hamiltonian" systems, Lindbladian adiabatic evolution always obeys the rules of quantum mechanics (i.e., is completely-positive and trace-preserving). Throughout this entire section, we assume that $\mathrm{As}(\mathsf{H})$ is steady ($H_\infty = 0$) but note that this analysis can be extended to non-steady $\mathrm{As}(\mathsf{H})$ by carefully including a "dynamical phase" contribution from $H_\infty$. Recall that a system evolves in a rescaled time $s \equiv t/T \in [0, 1]$ according to a time-dependent Lindbladian $\mathcal{L}(s)$, where the end time $T$ is infinite in the adiabatic limit. For all $s$, we define a continuous and differentiable family of instantaneous asymptotic subspaces with corresponding asymptotic projections

$$\mathcal{P}_\infty^{(s)} = \sum_\mu |\Psi_\mu^{(s)}\rangle\!\rangle\langle\!\langle J^\mu(s)|, \quad (5.1)$$

steady-state basis elements $\Psi_\mu^{(s)}$ [such that $\mathcal{L}(s)|\Psi_\mu^{(s)}\rangle\!\rangle = 0$], and conserved quantities $J^\mu(s)$ [such that $\langle\!\langle J^\mu(s)|\mathcal{L}(s) = 0$]. The dimension of the instantaneous subspaces (i.e., the rank of $\mathcal{P}_\infty^{(s)}$) is assumed to stay constant during this evolution. In other words, the zero eigenvalue of $\mathcal{L}(s)$ is isolated from all other eigenvalues at all points $s$ by the dissipative gap $\Delta_{\mathrm{dg}}$ (analogous to the excitation gap in Hamiltonian systems). We further assume that $s \in [0, 1]$ parametrizes a path in a space of control parameters $\mathsf{M}$, whose coordinate basis is $\{\mathbf{x}_\alpha\}$. In other words, we can parametrize

$$\partial_t = \frac{1}{T}\partial_s = \frac{1}{T}\sum_\alpha \dot{\mathbf{x}}_\alpha \partial_\alpha, \quad (5.2)$$

where $\partial_s$ is the derivative along the path, $\partial_\alpha \equiv \partial/\partial\mathbf{x}_\alpha$ are derivatives in various directions in parameter space, and $\dot{\mathbf{x}}_\alpha \equiv \frac{d\mathbf{x}_\alpha}{ds}$ are (unitless) parameter velocities.

Following Ref. [72], starting with an initially steady state $|\rho(0)\rangle\!\rangle \in \mathrm{As}(\mathsf{H})$, adiabatic perturbation theory is an expansion of the equation of motion

$$\frac{1}{T}\partial_s|\rho(s)\rangle\!\rangle = \mathcal{L}(s)|\rho(s)\rangle\!\rangle \quad (5.3)$$

in a series in $1/T$. Each term in the expansion is further divided using the decomposition $\mathcal{I} = \mathcal{P}_\infty + \mathcal{Q}_\infty$ into terms inside and outside the instantaneous $\mathrm{As}(\mathsf{H})$. This allows one to derive both the adiabatic limit (when $T \to \infty$) and all corrections. The $O(1/T)$ expansion for the final state from Theorem 6 of Ref. [72] reads

$$|\rho(s)\rangle\!\rangle = \mathcal{U}^{(s,0)}|\rho(0)\rangle\!\rangle + \frac{1}{T}\mathcal{L}^{-1}(s)\dot{\mathcal{P}}_\infty^{(s)}\mathcal{U}^{(s,0)}|\rho(0)\rangle\!\rangle$$
$$+ \frac{1}{T}\int_0^s dr \mathcal{U}^{(s,r)}\{\dot{\mathcal{P}}_\infty\mathcal{L}^{-1}\dot{\mathcal{P}}_\infty\}^{(r)}\mathcal{U}^{(r,0)}|\rho(0)\rangle\!\rangle, \quad (5.4)$$

where all quantities in curly brackets are functions of $r$, $\dot{\mathcal{P}}_\infty \equiv \partial_s \mathcal{P}_\infty$, $\mathcal{Q}_\infty \equiv \mathcal{I} - \mathcal{P}_\infty$, and $\mathcal{L}^{-1}$ is the instantaneous inverse [Eq. (4.29)]. The superoperator

$$\mathcal{U}^{(s,s')} = \mathbb{P}\exp\left(\int_{s'}^s \dot{\mathcal{P}}_\infty^{(r)}\mathcal{P}_\infty^{(r)}dr\right) \quad (5.5)$$

parallel transports states in $\mathcal{P}_\infty^{(s')}\mathrm{Op}(\mathsf{H})$ to states in $\mathcal{P}_\infty^{(s)}\mathrm{Op}(\mathsf{H})$ and is a path-ordered product of exponentials of the adiabatic connection $\dot{\mathcal{P}}_\infty \mathcal{P}_\infty$, the generator of adiabatic evolution (see Appendix E).

Like the Kubo formula, all terms can be interpreted when read from right to left. The first term in Eq. (5.4) represents adiabatic evolution of As(H), the (second) leakage term quantifies leakage of $|\rho(0)\rangle\rangle$ out of As(H), and the (last) tunneling term represents interference coming back into As(H) from outside. This term is a continuous sum of adiabatically evolved steady states that are perturbed by $\dot{\mathcal{P}}_\infty \mathcal{L}^{-1} \mathcal{P}_\infty$ at all points $r \in [0, s]$ during evolution. Because of its dependence on the spectrum of $\mathcal{L}$, this term needs to be minimized to determine the optimal adiabatic path through As(H) [141]. Notice also the similarity between the leakage term and the leakage term Eq. (4.31) of the Kubo formula. Motivated by this, we proceed to apply the four-corners decomposition to all three terms.

### B. Evolution within As(H)

Let us now assume a closed path $[\mathcal{L}(s) = \mathcal{L}(0)]$. However, due to the geometry of the parameter space M, the state may be changed (e.g., acquire a Berry phase). In the adiabatic limit [according to Eq. (5.4)], an initial steady state evolves in closed path $C$ as $|\rho(0)\rangle\rangle \to \mathcal{U}|\rho(0)\rangle\rangle$, acquiring a holonomy

$$\mathcal{U} \equiv \mathcal{U}^{(1,0)} = \mathbb{P} \exp\left( \oint_C \dot{\mathcal{P}}_\infty \mathcal{P}_\infty ds \right). \quad (5.6)$$

The above expression acts on the steady-state basis elements $\Psi_\mu^{(s=0)}$ used to express

$$|\rho(0)\rangle\rangle = \sum_\mu c_\mu |\Psi_\mu^{(0)}\rangle\rangle, \quad (5.7)$$

so we deem it the *operator representation* of the holonomy $\mathcal{U}$ and connection $\dot{\mathcal{P}}_\infty \mathcal{P}_\infty$.

Instead of looking at how the basis elements evolve, let us instead express the effect of the holonomy on the coordinates $c_\mu$ of the state above. This can be done by generalizing the Hamiltonian analysis of Appendix E to Lindbladians [68,72], which produces a parallel transport condition

$$\mathcal{P}_\infty \partial_s |\rho\rangle\rangle = 0 \quad (5.8)$$

characterizing the Lindbladian adiabatic limit. After expressing $\partial_s$ in terms of the various $\partial_\alpha$'s [Eq. (5.2)], this condition provides an equation of motion for the coordinate vector $c_\mu$. Solving this equation yields (what we call) the *coordinate representation* of the holonomy,

$$\mathcal{B} = \mathbb{P} \exp\left( -\sum_\alpha \oint_C \mathcal{A}_\alpha d\mathbf{x}_\alpha \right), \quad (5.9)$$

and corresponding adiabatic connection

$$\mathcal{A}_{\alpha,\mu\nu} \equiv \langle\langle J^\mu | \partial_\alpha \Psi_\nu \rangle\rangle. \quad (5.10)$$

Note that $\mathcal{A}_\alpha$ is a real matrix since $\{J^\mu, \Psi_\nu\}$ are Hermitian. The connection transforms as a gauge potential under $|\Psi_\mu\rangle\rangle \to |\Psi_\nu\rangle\rangle \mathcal{R}_{\nu\mu}$ and $\langle\langle J^\mu| \to \mathcal{R}^{-1}_{\mu\nu}\langle\langle J^\nu|$ for any $\mathcal{R} \in GL[\dim \text{As}(H), \mathbb{R}]$:

$$\mathcal{A}_\alpha \to \mathcal{R}^{-1} \mathcal{A}_\alpha \mathcal{R} + \mathcal{R}^{-1} \partial_\alpha \mathcal{R}. \quad (5.11)$$

Upon evolution in the closed path, the density matrix transforms as

$$|\rho(0)\rangle\rangle = \sum_{\mu=0}^{d^2-1} c_\mu |\Psi_\mu^{(0)}\rangle\rangle \to \sum_{\mu,\nu=0}^{d^2-1} \mathcal{B}_{\mu\nu} c_\nu |\Psi_\mu^{(0)}\rangle\rangle, \quad (5.12)$$

equivalent to the operator representation. We study both representations below, showing that the holonomy is unitary for all As(H).

First, let us remove the decaying subspace from both representations of the connection by applying the clean-leak property Eq. (2.11). Simplifying $\mathcal{A}_\alpha$ turns out to be similar to calculating the effective Hamiltonian perturbation $\mathcal{W}$ within As(H) in Sec. IV. By Eq. (2.11),

$$\mathcal{A}_{\alpha,\mu\nu} \equiv \langle\langle J^\mu | \partial_\alpha \Psi_\nu \rangle\rangle = \langle\langle J^\mu_\blacksquare | \partial_\alpha \Psi_\nu \rangle\rangle. \quad (5.13)$$

For the operator representation, one first applies Eq. (2.11) to the parallel transport condition Eq. (5.8):

$$0 = \mathcal{P}_\infty |\partial_s \rho\rangle\rangle = \mathcal{P}_\Psi |\partial_s \rho\rangle\rangle. \quad (5.14)$$

Then, one uses this condition to obtain an equation of motion for $\rho$:

$$|\partial_s \rho\rangle\rangle = (\mathcal{I} - \mathcal{P}_\Psi)|\partial_s \rho\rangle\rangle = \dot{\mathcal{P}}_\Psi \mathcal{P}_\Psi |\rho\rangle\rangle. \quad (5.15)$$

The last equality above can be checked by expressing both sides in terms of the steady-state basis elements $\Psi_\mu$ and conserved quantities $J^\mu$. For a closed path, the solution to this equation of motion is then the same holonomy, but now with the minimal projection $\mathcal{P}_\Psi$ instead of the asymptotic projection $\mathcal{P}_\infty$:

$$\mathcal{U} = \mathbb{P} \exp\left( \oint_C \dot{\mathcal{P}}_\Psi \mathcal{P}_\Psi ds \right). \quad (5.16)$$

The holonomy $\mathcal{U}$ thus does not depend on the piece $\mathcal{P}_\infty \mathcal{P}_\blacksquare$ associated with the decaying subspace.

*DFS case.*—Since $\mathcal{P}_\Psi = \mathcal{P}_\blacksquare$, the operator representation allows us to readily extract the DFS case. The (unitary) holonomy for a set of states $|\psi_k\rangle$ (with $P = \sum_{k=0}^{d-1} |\psi_k\rangle\langle\psi_k|$) is determined by the adiabatic connections of the states themselves, namely, $\dot{P}P$ and its corresponding superoperator form

$$\dot{\mathcal{P}}_{\boxempty}\mathcal{P}_{\boxempty}(\rho) = \dot{P}P\rho P + P\rho P\dot{P}. \qquad (5.17)$$

This result is known [67] and is a cornerstone of reservoir-engineered holonomic quantum computation (see example below). We study this case in the coordinate representation in Appendix E 2.

*Unique state case.*—Now, the only conserved quantity is the identity $J = I$, so it is easy to show that

$$\mathcal{A}_\alpha = \langle\!\langle I|\partial_\alpha \varrho\rangle\!\rangle = \text{Tr}\{\partial_\alpha \varrho\} = 0. \qquad (5.18)$$

The unique steady state can never acquire a Berry phase. While the Hamiltonian formalism does yield a Berry phase for a unique ground state, that (overall) phase disappears when the state is written as a density matrix. Since the Lindbladian formalism deals with density matrices, one never encounters an overall phase.

*NS case.*—For this case, the NS factors into a DFS and an auxiliary part for each $s \in [0, 1]$. The DFS part is mapped into a reference DFS spanned by a (parameter-independent Hermitian matrix) basis $\{|\bar{\Psi}_\mu^{\text{DFS}}\rangle\!\rangle\}_{\mu=0}^{d^2-1}$. [Note that, in general, $|\bar{\Psi}_\mu^{\text{DFS}}\rangle\!\rangle \neq |\Psi_\mu^{\text{DFS}}(s=0)\rangle\!\rangle$ since $s$ parametrizes a particular path in $M$ while $\{|\bar{\Psi}_\mu^{\text{DFS}}\rangle\!\rangle\}$ is fixed.] We let $S(s)$ [with $\mathcal{S}(\rho) \equiv S\rho S^\dagger$] be the unitary operator that simultaneously maps the instantaneous basis elements $|\Psi_\mu^{(s)}\rangle\!\rangle$ into the reference DFS basis and diagonalizes $\varrho_{\text{ax}}^{(s)}$. Similarly, this $S(s)$ will factor the instantaneous conserved quantities $\langle\!\langle J_\boxempty^\mu(s)|$ into a DFS part and the identity $P_{\text{ax}}^{(s)}$ on the auxiliary space. Therefore, we define the family of instantaneous minimal projections as

$$\mathcal{P}_\Psi^{(s)} = \mathcal{S}(s)(\bar{\mathcal{P}}_{\text{DFS}} \otimes |\varrho_{\text{ax}}^{(s)}\rangle\!\rangle\langle\!\langle P_{\text{ax}}^{(s)}|)\mathcal{S}^\ddagger(s), \qquad (5.19)$$

where $\bar{\mathcal{P}}_{\text{DFS}}(\cdot) = \sum_{\mu=0}^{d^2-1} |\bar{\Psi}_\mu^{\text{DFS}}\rangle\!\rangle\langle\!\langle\bar{\Psi}_\mu^{\text{DFS}}|\cdot\rangle\!\rangle = \bar{P}_{\text{DFS}} \cdot \bar{P}_{\text{DFS}}$ is the superoperator projection onto the $\mathbf{x}_\alpha$-independent DFS reference basis. The generators of motion

$$G_\alpha \equiv iS^\dagger \partial_\alpha S \quad \text{and} \quad \mathcal{G}_\alpha \equiv -i[G_\alpha, \cdot] \qquad (5.20)$$

can mix up the DFS with the auxiliary part, generating novel dissipation-assisted adiabatic dynamics.

We note that $\varrho_{\text{ax}}^{(s)}$ (and, therefore, $P_{\text{ax}}^{(s)}$) can change rank ($d_{\text{ax}}^{(s)}$) and purity ($n_{\text{ax}}^{(s)} = \sqrt{\text{Tr}\varrho_{\text{ax}}^2}$), provided that $\mathcal{P}_\Psi^{(s)}$ remains differentiable. For example, one can imagine $\varrho_{\text{ax}}^{(s)}$ to be a thermal state associated with some Hamiltonian on $\mathsf{H}_{\text{ax}}$ whose rank jumps from one to $d_{\text{ax}}$ as the temperature is turned up from zero. This implies that $P^{(s)}$ and thus $\mathcal{P}_\boxempty^{(s)}$ can change rank also.

However, such deformations do not change the dimension $d^2$ of $\text{As}(\mathsf{H})$ and thus do not close the dissipative gap. To account for such deformations in the one NS block case, the path can be partitioned into segments of constant $\text{rank}\{P\}$ and the connection calculation below can be applied to each segment.

Simplifying Eq. (5.13) by invoking the reference basis structure of $\{J, \Psi\}$ from Eq. (5.19) yields

$$\mathcal{A}_\alpha = \tilde{\mathcal{A}}_\alpha^{\text{DFS}} + \mathcal{A}_\alpha^{\text{ax}} = -i[\tilde{A}_\alpha^{\text{DFS}}, \cdot] \otimes |\varrho_{\text{ax}}\rangle\!\rangle\langle\!\langle P_{\text{ax}}| + \mathcal{A}_\alpha^{\text{ax}}, \qquad (5.21)$$

where the DFS effective Hamiltonian is [69]

$$\tilde{A}_\alpha^{\text{DFS}} \equiv \text{Tr}_{\text{ax}}\{(\bar{P}_{\text{DFS}} \otimes \varrho_{\text{ax}}^{(s)})G_\alpha\}, \qquad (5.22)$$

and the second term is the $n_{\text{ax}}$-dependent constant

$$\mathcal{A}_{\alpha,\mu\nu}^{\text{ax}} = -\partial_\alpha \ln n_{\text{ax}}^{(s)} \delta_{\mu\nu}. \qquad (5.23)$$

The first term clearly leaves the auxiliary part invariant and generates unitary evolution within the DFS part of the NS. We can thus see that DFS holonomies can be influenced by $\varrho_{\text{ax}}^{(s)}$. We will see that the second term's only role is to preserve the trace for open paths.

Sticking with the convention that $\Psi_0^{(s)}$ is traceful and the traceless $\Psi_{\mu\neq 0}^{(s)}$ carry the DFS Bloch vector, we notice that $\mathcal{A}_\alpha$ transforms as a gauge potential under orthogonal Bloch vector rotations $\mathcal{R} \in SO(d^2 - 1)$:

$$|\Psi_{\mu\neq 0}\rangle\!\rangle \to |\Psi_{\nu\neq 0}\rangle\!\rangle\mathcal{R}_{\nu\mu} \quad \text{and} \quad |J_\boxempty^{\mu\neq 0}\rangle\!\rangle \to |J_\boxempty^{\nu\neq 0}\rangle\!\rangle\mathcal{R}_{\nu\mu}. \qquad (5.24)$$

In addition, one can internally rotate $\varrho_{\text{ax}}$ without mixing $\Psi_\mu$ with $\Psi_{\nu\neq\mu}$. Under such a transformation $\mathcal{S}_{\text{ax}}$,

$$|\Psi_\mu\rangle\!\rangle \to \mathcal{S}_{\text{ax}}|\Psi_\mu\rangle\!\rangle = \mathcal{S}\left|\bar{\Psi}_\mu^{\text{DFS}} \otimes \frac{R_{\text{ax}}\varrho_{\text{ax}}R_{\text{ax}}^\dagger}{n_{\text{ax}}}\right\rangle\!\rangle \qquad (5.25)$$

for some $R_{\text{ax}} \in U(d_{\text{ax}})$, and the connection transforms as an Abelian gauge potential:

$$\mathcal{A}_{\alpha,\mu\nu} \to \mathcal{A}_{\alpha,\mu\nu} + \langle\!\langle J_\boxempty^\mu|\mathcal{S}_{\text{ax}}^\ddagger \partial_\alpha \mathcal{S}_{\text{ax}}|\Psi_\nu\rangle\!\rangle. \qquad (5.26)$$

Plugging Eq. (5.21) into the Lindblad holonomy Eq. (5.9), we can see that $\mathcal{A}_\alpha^{\text{ax}}$ is proportional to the identity matrix (of the space of coefficients $c_\mu$) and thus can be factored out. Therefore,

$$\mathcal{B} = \exp\left(\sum_\alpha \oint_C \partial_\alpha \ln n_{\text{ax}} d\mathbf{x}_\alpha\right)\mathcal{B}^{\text{DFS}}, \qquad (5.27)$$

where $\mathcal{B}^{\text{DFS}}$ is the unitary $\varrho_{\text{ax}}$-enhanced holonomy associated with $\tilde{A}^{\text{DFS}}$. The first term in the above product for an open path $s \in [0, 1]$ is simply $n_{\text{ax}}^{(1)}/n_{\text{ax}}^{(0)}$, providing the proper rescaling of the coefficients $c_\mu$ to preserve the trace of $|\rho(0)\rangle\!\rangle$ [142]. For a closed path, this term vanishes (since $n_{\text{ax}}$ is real and positive) and $\mathcal{B} = \mathcal{B}^{\text{DFS}}$. Thus, the holonomy after a closed-loop traversal of one NS block is unitary.

*Multiblock case.*—The generalization to multiple NS blocks is straightforward: the reference basis now consists of multiple blocks. Recall that $J^\mu$ do not have presence in the off-diagonal parts neighboring the NS blocks [Fig. 1(b)] and that the only NS block that $\partial_\alpha \Psi_\mu$ has presence in is that of $\Psi_\mu$. Therefore, each NS block is imparted with its own unitary holonomy.

### 1. Adiabatic curvature

The adiabatic connection $\mathcal{A}_\alpha$ [Eq. (5.10)] can be used to define an adiabatic curvature defined on the parameter space induced by the steady states. For simply connected parameter spaces M [145], the adiabatic curvature can be shown to generate the corresponding holonomy. More precisely, the Ambrose-Singer theorem (Ref. [148], Theorem 10.4) implies that the holonomy for an infinitesimal closed path $C$ with base point $\mathbf{x}_\alpha^{(0)}$ is the adiabatic curvature at $\mathbf{x}_\alpha^{(0)}$. One can alternatively use a generalization of Stokes's theorem to non-Abelian connections [149] to express the holonomy in terms of a "surface-ordered" integral of the corresponding adiabatic curvature, generalizing the Abelian case, Eq. (E19). Letting $\partial_{[\alpha} A_{\beta]} = \partial_\alpha A_\beta - \partial_\beta A_\alpha$, the curvature is

$$\mathcal{F}_{\alpha\beta,\mu\nu} \equiv \partial_{[\alpha} \mathcal{A}_{\beta],\mu\nu} + [\mathcal{A}_\alpha, \mathcal{A}_\beta]_{\mu\nu}. \quad (5.28)$$

*NS case.*—Using the NS adiabatic connection Eq. (5.21) and remembering that $\partial_\alpha \mathcal{A}_\beta^{\text{ax}}$ is symmetric in $\alpha, \beta$, the adiabatic curvature for one NS block,

$$\mathcal{F}_{\alpha\beta,\mu\nu} = \partial_{[\alpha} \tilde{\mathcal{A}}_{\beta],\mu\nu}^{\text{DFS}} + [\tilde{\mathcal{A}}_\alpha^{\text{DFS}}, \tilde{\mathcal{A}}_\beta^{\text{DFS}}]_{\mu\nu}, \quad (5.29)$$

is just the curvature associated with the connection $\tilde{A}^{\text{DFS}}$.

### 2. Example: Driven two-photon absorption

A concrete example of Lindbladian-assisted holonomic manipulation of As(H) is a generalized version of the driven two-photon absorption example from Sec. IV C. One can generalize the jump operator to

$$F = (a - \alpha_0)(a - \alpha_1), \quad (5.30)$$

where $\alpha_0$, $\alpha_1$ are complex. For the well-separated case ($|\alpha_0 - \alpha_1| \gg 1$), the DFS is spanned by coherent states $|\alpha_0\rangle$ and $|\alpha_1\rangle$. After adiabatically traversing a closed loop in the parameter space of the two $\alpha$'s, the DFS acquires a holonomy. For example, if $\alpha_0$ is fixed and $\alpha_1$ is varied in a closed loop far away from $\alpha_0$, then $|\alpha_1\rangle \to e^{i\phi}|\alpha_1\rangle$, where $\phi$ is twice the area (in phase space) enclosed by the path. This scheme can be generalized to obtain universal quantum computation on superpositions of coherent states of multiple modes [71].

### C. Leakage out of As(H)

We now return to the adiabatic response formula (5.4) to apply the four-corners decomposition to the $O(1/T)$ nonadiabatic corrections. By Eq. (4.29), $\mathcal{L}^{-1}$ has the same block upper-triangular structure as $\mathcal{L}$ from Eq. (3.3). The derivative of the asymptotic projection has partition

$$\dot{\mathcal{P}}_\infty = \begin{bmatrix} (\dot{\mathcal{P}}_\Psi)_{\blacksquare} & \mathcal{P}_{\blacksquare} \dot{\mathcal{P}}_\infty \mathcal{P}_{\blacksquare} & \mathcal{P}_{\blacksquare} \dot{\mathcal{P}}_\infty \mathcal{P}_{\square} \\ \mathcal{P}_{\square} \dot{\mathcal{P}}_\Psi \mathcal{P}_{\blacksquare} & 0 & \mathcal{P}_{\square} \dot{\mathcal{P}}_\infty \mathcal{P}_{\square} \\ 0 & 0 & 0 \end{bmatrix}. \quad (5.31)$$

One can interpret $\dot{\mathcal{P}}_\infty$ as a perturbation, analogous to $\mathcal{V}$ from Sec. IV, and observe from the above partition that $\dot{\mathcal{P}}_\infty$ does not connect block diagonal spaces: $\mathcal{P}_{\square} \dot{\mathcal{P}}_\infty \mathcal{P}_{\square} = 0$. In addition, whenever $\dot{\mathcal{P}}_\infty^{(r)}$ acts on a parallel transported state living in $\mathcal{P}_{\blacksquare}^{(r)} \text{Op}(H)$, only the first column in the above partition ($\dot{\mathcal{P}}_\infty \mathcal{P}_{\blacksquare}$) is relevant. These observations result in $\mathcal{L}^{-1} \to \mathcal{L}_{\blacksquare}^{-1}$ and the replacement of two factors of $\dot{\mathcal{P}}_\infty$ with $\dot{\mathcal{P}}_\Psi$ in Eq. (5.4). (We cannot replace the remaining $\dot{\mathcal{P}}_\infty$ since $\mathcal{P}_{\blacksquare} \dot{\mathcal{P}}_\infty \mathcal{P}_{\blacksquare}$ contains contributions from $\partial_s J_{\blacksquare}^\mu$.)

$$|\rho(s)\rangle\!\rangle = \mathcal{U}^{(s,0)} |\rho(0)\rangle\!\rangle + \frac{1}{T} \mathcal{L}_{\blacksquare}^{-1}(s) \dot{\mathcal{P}}_\Psi^{(s)} \mathcal{U}^{(s,0)} |\rho(0)\rangle\!\rangle$$
$$+ \frac{1}{T} \int_0^s dr \mathcal{U}^{(s,r)} \{\dot{\mathcal{P}}_\infty \mathcal{L}_{\blacksquare}^{-1} \dot{\mathcal{P}}_\Psi\}^{(r)} \mathcal{U}^{(r,0)} |\rho(0)\rangle\!\rangle. \quad (5.32)$$

Using the results of Sec. IV C, the energy scale governing the leading-order nonadiabatic corrections is once again the adiabatic dissipative gap $\Delta_{\text{edg}}$—the nonzero eigenvalue of $\mathcal{L}_{\blacksquare} + \mathcal{L}_{\square}$ with the smallest real part. A similar result is shown for the leakage term in the Supplemental Material of Ref. [69]. In addition, the tunneling term, which is similar to the second-order perturbative correction $\mathcal{P}_\infty \mathcal{V} \mathcal{L}^{-1} \mathcal{V} \mathcal{P}_\infty$ we discuss in Sec. IV C, does not contain contributions from $\mathcal{L}_{\square}$.

## VI. LINDBLADIAN QUANTUM GEOMETRIC TENSOR

Here, we introduce the Lindbladian QGT $\mathcal{Q}$ and explicitly calculate it for the unique state and NS block cases. The antisymmetric part of the QGT is equal to the curvature $\mathcal{F}$ generated by the connection $\mathcal{A}$ (see Sec. V B 1). The symmetric part of the QGT produces a generalized metric tensor for Lindbladian steady-state subspaces. We review the Hamiltonian QGT and cover in detail the DFS case in Appendix F. Most of the relevant quantities for the Hamiltonian, degenerate Hamiltonian or DFS, and NS cases are summarized in Table I. We introduce other geometric quantities in Appendix G, including an alternative geometric tensor $\mathcal{Q}^{\text{alt}}$ whose curvature is different from the adiabatic curvature from Sec. V B 1, but whose metric appears in the Lindbladian adiabatic path length.

In Sec. V, we show, using the operator representation of the adiabatic connection and the conditions Eqs. (2.9) and (2.11), that the minimal projection $\mathcal{P}_\Psi = \mathcal{P}_\infty \mathcal{P}_\boxminus$ (and not $\mathcal{P}_\infty$) generates adiabatic evolution within $\text{As}(\mathsf{H})$. Following this, we define

$$\mathcal{Q}_{\alpha\beta} \equiv \mathcal{P}_\Psi \partial_\alpha \mathcal{P}_\Psi \partial_\beta \mathcal{P}_\Psi \mathcal{P}_\Psi \quad (6.1)$$

to be the associated QGT. While $\mathcal{P}_\Psi = \sum_\mu |\Psi_\mu\rangle\rangle\langle\langle J_\boxminus^\mu|$ is not always Hermitian due to $J_\boxminus^\mu \neq \Psi_\mu$ (e.g., in the NS case), we show that the QGT nevertheless remains a meaningful geometric quantity. Looking at the matrix elements of $\mathcal{Q}_{\alpha\beta}$ and explicitly plugging in the instantaneous $\mathcal{P}_\Psi$ [Eq. (5.19)] yields the following three forms:

$$\mathcal{Q}_{\alpha\beta,\mu\nu} \equiv \langle\langle J_\boxminus^\mu | \partial_\alpha \mathcal{P}_\Psi \partial_\beta \mathcal{P}_\Psi | \Psi_\nu\rangle\rangle \quad (6.2a)$$

$$= \langle\langle \partial_\alpha J_\boxminus^\mu | (\mathcal{I} - \mathcal{P}_\Psi) | \partial_\beta \Psi_\nu \rangle\rangle \quad (6.2b)$$

$$= \partial_\alpha \mathcal{A}_{\beta,\mu\nu} + (\mathcal{A}_\alpha \mathcal{A}_\beta)_{\mu\nu} - \langle\langle J_\boxminus^\mu | \partial_\alpha \partial_\beta \Psi_\nu\rangle\rangle, \quad (6.2c)$$

with $\mathcal{A}_\alpha$ the Lindblad adiabatic connection Eq. (5.10). Since $\mathcal{A}_{\alpha,\mu\nu}$ are real and $\{J^\mu, \Psi_\nu\}$ are Hermitian, the matrix elements are all real. From its second form, one easily deduces that the QGT transforms as $\mathcal{Q}_{\alpha\beta} \to \mathcal{R}^{-1}\mathcal{Q}_{\alpha\beta}\mathcal{R}$ for any basis transformation $\mathcal{R} \in GL[\dim \text{As}(\mathsf{H}), \mathbb{R}]$ [see Eq. (5.11)]. The QGT $\mathcal{Q}_{\alpha\beta}$ consists of parts symmetric ($\mathcal{Q}_{(\alpha\beta)}$) and antisymmetric ($\mathcal{Q}_{[\alpha\beta]}$) in $\alpha, \beta$. From the third form, it is evident that its antisymmetric part is exactly the adiabatic curvature $\mathcal{F}_{\alpha\beta}$ from Eq. (5.28) (cf. Ref. [70], Proposition 13). The rest of this section is devoted to calculating the symmetric part and its corresponding metric on $\mathsf{M}$, which is defined as the trace TR (i.e., trace in superoperator space) of the QGT's symmetric part,

$$\mathcal{M}_{\alpha\beta} \equiv \text{TR}\{\mathcal{P}_\Psi \partial_{(\alpha} \mathcal{P}_\Psi \partial_{\beta)} \mathcal{P}_\Psi\} = \sum_{\mu=0}^{d^2-1} \mathcal{Q}_{(\alpha\beta),\mu\mu}. \quad (6.3)$$

Before proving this is a metric for some of the relevant cases, let us first reveal how such a structure corresponds to an infinitesimal distance between adiabatically connected Lindbladian steady states by adapting results from non-Hermitian Hamiltonian systems [150–152]. The zero eigenspace of $\mathcal{L}_\boxminus$ is diagonalized by right and left eigenmatrices $|\Psi_\mu\rangle\rangle$ and $\langle\langle J_\boxminus^\mu|$, respectively. In accordance with this duality between $\Psi$ and $J_\boxminus$, we introduce an associated operator $|\hat{\rho}_\infty\rangle\rangle$ [151,152],

$$|\rho_\infty\rangle\rangle = \sum_{\mu=0}^{d^2-1} c_\mu |\Psi_\mu\rangle\rangle \leftrightarrow |\hat{\rho}_\infty\rangle\rangle \equiv \sum_{\mu=0}^{d^2-1} c_\mu |J_\boxminus^\mu\rangle\rangle, \quad (6.4)$$

for every steady-state subspace operator $|\rho_\infty\rangle\rangle$. This allows us to define a modified inner product $\langle\langle\widehat{A}|B\rangle\rangle$ for matrices $A$ and $B$ living in the steady-state subspace. Since $\Psi_\mu$ and $J_\boxminus^\mu$ are biorthogonal ($\langle\langle J_\boxminus^\mu|\Psi_\nu\rangle\rangle = \delta_{\mu\nu}$), this inner product is surprisingly equivalent to the Hilbert-Schmidt inner product $\langle\langle A|B\rangle\rangle$. However, the infinitesimal distance is not the same:

$$\langle\langle \partial_s \hat{\rho}_\infty | \partial_s \rho_\infty \rangle\rangle \neq \langle\langle \partial_s \rho_\infty | \partial_s \rho_\infty \rangle\rangle. \quad (6.5)$$

TABLE I. Summary of quantities defined in Secs. V and VI.

|  | Hamiltonians: Operator notation | Hamiltonians: Superoperator notation | Lindbladians: One NS block |
|---|---|---|---|
| State basis | $\|\psi_k\rangle$ = DFS states | $\Psi_\mu^{\text{DFS}} = (\Psi_\mu^{\text{DFS}})^\dagger \in \text{span}\{\|\psi_k\rangle\langle\psi_l\|\}$ | $\|\Psi_\mu\rangle\rangle = \|\Psi_\mu^{\text{DFS}}\rangle\rangle \otimes \|\frac{\varrho_{\text{ax}}}{n_{\text{ax}}}\rangle\rangle$ |
|  | $P_{\text{DFS}} = \sum_{k=0}^{d-1} \|\psi_k\rangle\langle\psi_k\|$ | $\mathcal{P}_{\text{DFS}} = \sum_{\mu=0}^{d^2-1} \|\Psi_\mu^{\text{DFS}}\rangle\rangle\langle\langle\Psi_\mu^{\text{DFS}}\|$ | $\mathcal{P}_\infty = \sum_{\mu=0}^{d^2-1} \|\Psi_\mu\rangle\rangle\langle\langle J^\mu\|$ |
|  |  |  | $= \mathcal{P}_\Psi + \mathcal{P}_\infty \mathcal{P}_\boxminus$ |
| Connection | $A_{\alpha,kl}^{\text{DFS}} = i\langle\psi_k\|\partial_\alpha\psi_l\rangle$ | $\mathcal{A}_\alpha^{\text{DFS}} = \langle\langle\Psi_\mu^{\text{DFS}}\|\partial_\alpha\Psi_\nu^{\text{DFS}}\rangle\rangle$ | $\mathcal{A}_{\alpha,\mu\nu} = \langle\langle J^\mu\|\partial_\alpha\Psi_\nu\rangle\rangle$ |
|  |  |  | $= \tilde{\mathcal{A}}_{\alpha,\mu\nu}^{\text{DFS}} + \mathcal{A}_{\alpha,\mu\nu}^{\text{ax}}$ |
| Curvature | $F_{\alpha\beta}^{\text{DFS}} = \partial_{[\alpha} A_{\beta]}^{\text{DFS}} - i[A_\alpha^{\text{DFS}}, A_\beta^{\text{DFS}}]$ | $\mathcal{F}_{\alpha\beta}^{\text{DFS}} = \partial_{[\alpha} \mathcal{A}_{\beta]}^{\text{DFS}} + [\mathcal{A}_\alpha^{\text{DFS}}, \mathcal{A}_\beta^{\text{DFS}}]$ | $\mathcal{F}_{\alpha\beta} = \partial_{[\alpha}\mathcal{A}_{\beta]} + [\mathcal{A}_\alpha, \mathcal{A}_\beta]$ |
|  |  |  | $= \partial_{[\alpha}\tilde{\mathcal{A}}_{\beta]}^{\text{DFS}} + [\tilde{\mathcal{A}}_\alpha^{\text{DFS}}, \tilde{\mathcal{A}}_\beta^{\text{DFS}}]$ |
|  | $F_{\alpha\beta,kl}^{\text{DFS}} = \langle\psi_k\|\partial_{[\alpha}P_{\text{DFS}}\partial_{\beta]}P_{\text{DFS}}\|\psi_l\rangle$ | $\mathcal{F}_{\alpha\beta,\mu\nu}^{\text{DFS}} = \langle\langle\Psi_\mu^{\text{DFS}}\|\partial_{[\alpha}\mathcal{P}_{\text{DFS}}\partial_{\beta]}\mathcal{P}_{\text{DFS}}\|\Psi_\nu^{\text{DFS}}\rangle\rangle$ | $\mathcal{F}_{\alpha\beta,\mu\nu} = \langle\langle J_\boxminus^\mu\|\partial_{[\alpha}\mathcal{P}_\Psi\partial_{\beta]}\mathcal{P}_\Psi\|\Psi_\nu\rangle\rangle$ |
| QGT | $Q_{\alpha\beta,kl}^{\text{DFS}} = \langle\psi_k\|\partial_\alpha P_{\text{DFS}}\partial_\beta P_{\text{DFS}}\|\psi_l\rangle$ | $\mathcal{Q}_{\alpha\beta,\mu\nu}^{\text{DFS}} = \langle\langle\Psi_\mu^{\text{DFS}}\|\partial_\alpha \mathcal{P}_{\text{DFS}}\partial_\beta \mathcal{P}_{\text{DFS}}\|\Psi_\nu^{\text{DFS}}\rangle\rangle$ | $\mathcal{Q}_{\alpha\beta,\mu\nu} = \langle\langle J_\boxminus^\mu\|\partial_\alpha \mathcal{P}_\Psi \partial_\beta \mathcal{P}_\Psi\|\Psi_\nu\rangle\rangle$ |
|  | $= -i\partial_\alpha A_{\beta,kl}^{\text{DFS}} - (A_\alpha^{\text{DFS}} A_\beta^{\text{DFS}})_{kl}$ | $= \partial_\alpha \mathcal{A}_{\beta,\mu\nu}^{\text{DFS}} + (\mathcal{A}_\alpha^{\text{DFS}} \mathcal{A}_\beta^{\text{DFS}})_{\mu\nu}$ | $= \partial_\alpha \mathcal{A}_{\beta,\mu\nu} + (\mathcal{A}_\alpha \mathcal{A}_\beta)_{\mu\nu}$ |
|  | $-\langle\psi_k\|\partial_\alpha\partial_\beta\psi_l\rangle$ | $-\langle\langle\Psi_\mu^{\text{DFS}}\|\partial_\alpha\partial_\beta\Psi_\nu^{\text{DFS}}\rangle\rangle$ | $-\langle J_\boxminus^\mu\|\partial_\alpha\partial_\beta\Psi_\nu\rangle$ |
| Metric | $M_{\alpha\beta}^{\text{DFS}} = \text{Tr}\{P_{\text{DFS}}\partial_{(\alpha}P_{\text{DFS}}\partial_{\beta)}P_{\text{DFS}}\}$ | $\mathcal{M}_{\alpha\beta}^{\text{DFS}} = \text{TR}\{\mathcal{P}_{\text{DFS}}\partial_{(\alpha}\mathcal{P}_{\text{DFS}}\partial_{\beta)}\mathcal{P}_{\text{DFS}}\}$ | $\mathcal{M}_{\alpha\beta} = \text{TR}\{\mathcal{P}_\Psi \partial_{(\alpha}\mathcal{P}_\Psi \partial_{\beta)}\mathcal{P}_\Psi\}$ |

The symmetric part $\mathcal{Q}_{(\alpha\beta)}$ shows up in precisely this modified infinitesimal distance. Using Eq. (6.4), the parallel transport condition Eq. (5.8), and parametrizing $\partial_s$ in terms of the $\partial_\alpha$'s [Eq. (5.2)] yields

$$\langle\!\langle\partial_s\hat{\rho}_\infty|\partial_s\rho_\infty\rangle\!\rangle = \frac{1}{2}\sum_{\alpha,\beta}\sum_{\mu,\nu=0}^{d^2-1}\mathcal{Q}_{(\alpha\beta),\mu\nu}\dot{\mathbf{x}}_\alpha\dot{\mathbf{x}}_\beta c_\mu c_\nu, \quad (6.6)$$

as evidenced by the second form Eq. (6.2b) of the Lindblad QGT. Tracing the symmetric part over the steady-state subspace gives the metric $\mathcal{M}_{\alpha\beta}$.

*Unique state case.*—Here, things simplify significantly, yet the obtained metric turns out to be novel nonetheless. The asymptotic projection is $\mathcal{P}_\Psi = |\varrho\rangle\!\rangle\langle\!\langle P|$ and a straightforward calculation using Eq. (6.2b) yields

$$\mathcal{M}_{\alpha\beta} = \langle\!\langle\partial_{(\alpha}P|\partial_{\beta)}\varrho\rangle\!\rangle. \quad (6.7)$$

Using the eigendecomposition $\varrho = \sum_{k=0}^{d_\varrho-1}\lambda_k|\psi_k\rangle\langle\psi_k|$,

$$\mathcal{M}_{\alpha\beta} = 2\sum_{k=0}^{d_\varrho-1}\lambda_k\langle\partial_{(\alpha}\psi_k|Q|\partial_{\beta)}\psi_k\rangle, \quad (6.8)$$

where $Q = I - P$ and $\langle\partial_{(\alpha}\psi_k|Q|\partial_{\beta)}\psi_k\rangle$ is the Fubini-Study metric corresponding to the eigenstate $|\psi_k\rangle$. In words, $\mathcal{M}_{\alpha\beta}$ is the sum of the eigenstate Fubini-Study metrics weighed by their respective eigenvalues or populations. If $\varrho$ is pure, then it is clear that $\mathcal{M}_{\alpha\beta}$ reduces to the Fubini-Study metric. Finally, if $\varrho$ is full rank, then $P = I$ and $\mathcal{M}_{\alpha\beta} = 0$. This means that the metric is nonzero only for those $\varrho$ that are not full rank.

*NS case.*—Recall from Eq. (5.19) that adiabatic evolution on the NS is parametrized by the instantaneous minimal projections

$$\mathcal{P}_\Psi^{(s)} = \mathcal{S}(s)(\bar{\mathcal{P}}_{\mathrm{DFS}}\otimes|\varrho_{\mathrm{ax}}^{(s)}\rangle\!\rangle\langle\!\langle P_{\mathrm{ax}}^{(s)}|)\mathcal{S}^\ddagger(s), \quad (6.9)$$

where $\bar{\mathcal{P}}_{\mathrm{DFS}}(\cdot) = \sum_{\mu=0}^{d^2-1}|\bar{\Psi}_\mu^{\mathrm{DFS}}\rangle\!\rangle\langle\!\langle\bar{\Psi}_\mu^{\mathrm{DFS}}|\cdot\rangle\!\rangle = \bar{P}_{\mathrm{DFS}}\cdot\bar{P}_{\mathrm{DFS}}$ is the superoperator projection onto the $\mathbf{x}_\alpha$-independent DFS reference basis. We remind the reader (see Sec. V B) that the only assumption of such a parametrization is that the state $|\rho_\infty^{(s)}\rangle\!\rangle$ is unitarily equivalent (via unitary $\mathcal{S}$) to a tensor product of a DFS state and auxiliary part for all points $s \in [0, 1]$ in the path.

We can simplify $\mathcal{M}_{\alpha\beta}$ and show that it is indeed a metric. In the reference basis decomposition of $\mathcal{P}_\Psi$ from Eq. (6.9), the operators $G_\alpha \equiv i\mathcal{S}^\dagger\partial_\alpha\mathcal{S}$ [with $\mathcal{S}(s)|\rho\rangle\!\rangle \equiv |S\rho S^\dagger\rangle\!\rangle$] generate motion in parameter space. After significant simplification, one can express $\mathcal{M}_{\alpha\beta}$ in terms of these generators:

$$\mathcal{M}_{\alpha\beta} = \mathcal{M}_{\alpha\beta}^{(1)} + \mathcal{M}_{\alpha\beta}^{(2)},$$
$$\mathcal{M}_{\alpha\beta}^{(1)} = 2d\langle\!\langle\bar{P}_{\mathrm{DFS}}\otimes\varrho_{\mathrm{ax}}|G_{(\alpha}(I-\bar{P}_{\mathrm{DFS}}\otimes P_{\mathrm{ax}})G_{\beta)}\rangle\!\rangle,$$
$$\mathcal{M}_{\alpha\beta}^{(2)} = 2d\langle\!\langle G_{(\alpha}|\bar{\mathcal{P}}_{\mathrm{DFS}}^\star\otimes\mathcal{O}_{\mathrm{ax}}|G_{\beta)}\rangle\!\rangle, \quad (6.10)$$

with projection $\bar{\mathcal{P}}_{\mathrm{DFS}}^\star$ consisting of only traceless DFS generators (we set $\bar{\Psi}_0^{\mathrm{DFS}} = \frac{1}{\sqrt{d}}\bar{P}_{\mathrm{DFS}}$),

$$\bar{\mathcal{P}}_{\mathrm{DFS}}^\star \equiv \sum_{\mu=1}^{d^2-1}|\bar{\Psi}_\mu^{\mathrm{DFS}}\rangle\!\rangle\langle\!\langle\bar{\Psi}_\mu^{\mathrm{DFS}}| = \bar{\mathcal{P}}_{\mathrm{DFS}} - |\bar{\Psi}_0^{\mathrm{DFS}}\rangle\!\rangle\langle\!\langle\bar{\Psi}_0^{\mathrm{DFS}}|, \quad (6.11)$$

and auxiliary superoperator defined (for all auxiliary operators $A$) as $\mathcal{O}_{\mathrm{ax}}(A) \equiv (A - \langle\!\langle\varrho_{\mathrm{ax}}|A\rangle\!\rangle)\varrho_{\mathrm{ax}}$.

The quantity $\mathcal{M}_{\alpha\beta}$ is clearly real and symmetric in $\alpha, \beta$, so to show that it is a (semi)metric, we need to prove positivity [$\mathbf{w}_\alpha\mathcal{M}_{\alpha\beta}\mathbf{w}_\beta \geq 0$, with sum over $\alpha, \beta$ implied, for all vectors $\mathbf{w}$ in the tangent space $\mathsf{T}_\mathsf{M}(\mathbf{x})$ at a point $\mathbf{x} \in \mathsf{M}$ [148]]. Since $\varrho_{\mathrm{ax}}$ is positive definite, one can show that

$$\mathbf{w}_\alpha\mathcal{M}_{\alpha\beta}^{(1)}\mathbf{w}_\beta = 4d\langle\!\langle O|O\rangle\!\rangle \geq 0, \quad (6.12)$$

with $O = (I - \bar{P}_{\mathrm{DFS}}\otimes P_{\mathrm{ax}})(G_\alpha\mathbf{w}_\alpha)(\bar{P}_{\mathrm{DFS}}\otimes\sqrt{\varrho_{\mathrm{ax}}})$. For the second term $\mathcal{M}_{\alpha\beta}^{(2)}$, we can see that $\bar{\mathcal{P}}_{\mathrm{DFS}}^\star$ is positive semidefinite since it is a projection. We show that $\mathcal{O}_{\mathrm{ax}}$ is positive semidefinite by utilizing yet another inner product associated with open systems [52]. First, note that

$$\langle\!\langle A|\mathcal{O}_{\mathrm{ax}}|A\rangle\!\rangle = \mathrm{Tr}\{\varrho_{\mathrm{ax}}A^\dagger A\} - |\mathrm{Tr}\{\varrho_{\mathrm{ax}}A\}|^2. \quad (6.13)$$

Since $\varrho_{\mathrm{ax}}$ is full rank, $\langle\!\langle A|B\rangle\!\rangle_{\varrho_{\mathrm{ax}}} \equiv \mathrm{Tr}\{\varrho_{\mathrm{ax}}A^\dagger B\}$ is a valid inner product [52] and $\langle\!\langle A|\mathcal{O}_{\mathrm{ax}}|A\rangle\!\rangle \geq 0$ is merely a statement of the Cauchy-Schwarz inequality associated with this inner product. For Hermitian $A$, Eq. (6.13) reduces to the variance of $\langle\!\langle A|\varrho_{\mathrm{ax}}\rangle\!\rangle$.

Roughly speaking, the first term $\mathcal{M}_{\alpha\beta}^{(1)}$ describes how much the DFS and auxiliary parts mix and the second term $\mathcal{M}_{\alpha\beta}^{(2)}$ describes how much they leave the ⊞ block while moving in parameter space. For the DFS case, $\mathcal{M}_{\alpha\beta}^{(2)} = 0$ (due to $\mathcal{O}_{\mathrm{ax}} = 0$ for that case) and the metric reduces to the standard DFS metric covered in Appendix F. For the unique state case, $\mathcal{M}_{\alpha\beta}^{(2)}$ is also zero (due to $\bar{\mathcal{P}}_{\mathrm{DFS}}^\star$ not containing any traceful DFS elements and thus reducing to zero when $\bar{P}_{\mathrm{DFS}} = 1$). The mixing term $\mathcal{M}_{\alpha\beta}^{(2)}$ is thus nonzero only in the NS block case.

## VII. OUTLOOK

This work examines the properties of asymptotic (e.g., steady-state) subspaces of Lindbladians, comparing them to analogous subspaces of Hamiltonian systems. We characterize such subspaces as "not very different" from their Hamiltonian cousins in terms of their geometrical and response properties. A quantitative description of our results is found in Sec. II.

While we focus on response to Hamiltonian perturbations within first order and evolution within the adiabatic limit, it would be interesting to apply our results further to Lindbladian perturbations [127], second-order perturbative effects [137,153], and corrections to adiabatic evolution. While several elements of this study consider asymptotic subspaces consisting of only one block of steady states, it is not unreasonable to imagine that the aforementioned second-order and/or nonadiabatic effects could produce transfer of information between two or more blocks. Similar to the first-order case, we anticipate that jump operator perturbations may provide alternative ways to generate second-order effects [137,153], which are currently only producible with Hamiltonian perturbations. Recently developed diagrammatic series aimed at determining perturbed steady states [154] (see also Ref. [155]) may benefit from the four-corners decomposition (whenever the unperturbed steady state is not full rank).

It has recently been postulated [156] that Lindbladian metastable states also possess the same structure as the steady states. This may mean that our results regarding conserved quantities (which are dual to the steady states) also apply to the pseudoconserved quantities (dual to the metastable states).

We obtain a Lindblad generalization of the quantum geometric tensor for Hamiltonian systems [74]. The Lindblad QGT encodes both the adiabatic curvature of the steady-state subspace and also a novel metric which generalizes the Fubini-Study metric for Hamiltonians. This metric will be examined in future work, particularly to see whether it reveals information about bounds on convergence rates [157–160]. It remains to be seen whether the scaling behavior of the metric is correlated with phase stability [117–120] and phase transitions [114–116] for Lindbladian phases with nonequilibrium steady states. It would also be of interest to see whether the adiabatic curvature is related to the Uhlmann phase [161] and various mixed state Chern numbers [162–164].

We show that the dissipative gap of Lindbladians is not always relevant in linear response and in corrections to adiabatic evolution. In fact, another scale, the effective dissipative gap, is the relevant energy scale for those processes. It would be of interest to determine how the effective gap scales with system size in physically relevant dissipative systems [31,135,136,165].

At this point, the only way to find the projection $P$ onto the range of the steady states of a Lindbladian $\mathcal{L}$ is to diagonalize $\mathcal{L}$ [79]. It could be of interest to determine whether diagonalization of $\mathcal{L}$ is necessary for determining $P$. Interestingly, there exists an algorithm [166] (see also [167]) to verify whether a given projection is equal to $P$ that does not rely on diagonalization.

Lastly, the properties of Lindbladian eigenmatrices can be extended to eigenmatrices of more general quantum channels [79,83–85]. Statements similar to Proposition 2 exist for fixed points of quantum channels [79,88], and their extension to rotating points will be a subject of future work. These results may also be useful in determining properties of asymptotic algebras of observables [168,169] and properties of quantum jump trajectories when the Lindbladian is "unraveled" [170,171].

## ACKNOWLEDGMENTS

V. V. A. acknowledges N. Read, L. I. Glazman, S. M. Girvin, F. Ticozzi, A. Rivas, A. del Campo, L. Viola, K.-J. Noh, and C. Shen for stimulating discussions and thanks R. T. Brierley, A. Alexandradinata, Z. K. Minev, J. I. Väyrynen, J. Höller, and A. Dua for useful comments on the manuscript. V. V. A. was supported by the NSF GRFP under Grant No. DGE-1122492 through part of this work. L. J. is supported by the ARO, AFOSR MURI, ARL CDQI program, the Alfred P. Sloan Foundation, and the David and Lucile Packard Foundation.

## APPENDIX A: LINDBLADIANS AND DOUBLE-KET NOTATION

### 1. Introduction to Lindbladians

Lindbladians operate on the space of (linear) operators on $\mathsf{H}$, or $\mathrm{Op}(\mathsf{H}) \equiv \mathsf{H} \otimes \mathsf{H}^\star$ [172,173] (also known as Liouville space [129], von Neumann space, or Hilbert-Schmidt space [174]). This space is also a Hilbert space when endowed with the Hilbert-Schmidt inner product and Frobenius norm (for $N \equiv \dim \mathsf{H} < \infty$). An operator $A$ in quantum mechanics is thus both in the space of operators *acting on* ordinary states and in the space of vectors *acted on* by superoperators. We denote the two respective cases as $A|\psi\rangle$ and $\mathcal{O}|A\rangle\rangle$ (for $|\psi\rangle \in \mathsf{H}$ and for a superoperator $\mathcal{O}$). While (strictly speaking) $|A\rangle\rangle$ is an $N^2$-by-1 vector and $A$ is an $N$-by-$N$ matrix, they are isomorphic, and so we define $\mathcal{O}|A\rangle\rangle$, $\mathcal{O}(A)$, and $|\mathcal{O}(A)\rangle\rangle$ by their context.

For $A, B \in \mathrm{Op}(\mathsf{H})$, the Hilbert-Schmidt inner product and Frobenius norm are, respectively,

$$\langle\langle A|B\rangle\rangle \equiv \mathrm{Tr}\{A^\dagger B\} \quad \text{and} \quad \|A\| \equiv \sqrt{\langle\langle A|A\rangle\rangle}. \tag{A1}$$

The inner product allows one to define an adjoint operation $\ddagger$ that complements the adjoint operation $\dagger$ on matrices in $\mathrm{Op}(\mathsf{H})$:

$$\langle\langle A|\mathcal{O}(B)\rangle\rangle = \langle\langle A|\mathcal{O}|B\rangle\rangle = \langle\langle \mathcal{O}^\ddagger(A)|B\rangle\rangle. \tag{A2}$$

Writing $\mathcal{O}$ as an $N^2$-by-$N^2$ matrix (see Ref. [50], Appendix A, for the explicit form), $\mathcal{O}^{\ddagger}$ is just the conjugate transpose of that matrix. For example, if $\mathcal{O}(\cdot) = A \cdot B$, then one can use Eq. (A1) to verify that $\mathcal{O}^{\ddagger}(\cdot) = A^{\dagger} \cdot B^{\dagger}$. Similar to the Hamiltonian description of quantum mechanics, $\mathcal{O}$ is Hermitian if $\mathcal{O}^{\ddagger} = \mathcal{O}$. For example, all projections $\mathcal{P}_{\boxplus}$ from Eq. (2.2) are Hermitian.

A Lindbladian acts on a density matrix $\rho$ as

$$\mathcal{L}(\rho) = -i[H,\rho] + \frac{1}{2}\sum_{\ell}\kappa_{\ell}(2F^{\ell}\rho F^{\ell\dagger} - F^{\ell\dagger}F^{\ell}\rho - \rho F^{\ell\dagger}F^{\ell}), \tag{A3}$$

with Hamiltonian $H$, jump operators $F^{\ell} \in \text{Op}(\mathsf{H})$, and real nonzero rates $\kappa_{\ell}$. References [170,175,176] describe the conditions on a system and reservoir for which Lindbladian evolution captures the dynamics of the system. The form of the Lindbladian Eq. (A3) is not unique due to the following "gauge" transformation (for complex $g_{\ell}$),

$$H \to H - \frac{i}{2}\sum_{\ell}\kappa_{\ell}(g_{\ell}^{\star}F^{\ell} - g_{\ell}F^{\ell\dagger}),$$
$$F^{\ell} \to F^{\ell} + g_{\ell}I, \tag{A4}$$

that allows parts of the Hamiltonian to be included in the jump operators (and vice versa) while keeping $\mathcal{L}$ invariant. Note that there exists a unique "gauge" in which $F^{\ell}$ are traceless (Ref. [2], Theorem 2.2). It is easy to determine how an observable $A \in \text{Op}(\mathsf{H})$ evolves (in the Heisenberg picture) using the definition of the adjoint Eq. (A2) and cyclic permutations under the trace:

$$\mathcal{L}^{\ddagger}(A) = -\mathcal{H}(A) + \frac{1}{2}\sum_{\ell}\kappa_{\ell}(2F^{\ell\dagger}AF^{\ell} - \{F^{\ell\dagger}F^{\ell},A\}). \tag{A5}$$

The superoperator $\mathcal{H}(\cdot) \equiv -i[H,\cdot]$ corresponding to the Hamiltonian is therefore anti-Hermitian because we have absorbed the "$i$" in its definition.

Time evolution of states is determined by the equation $|\partial_t\rho(t)\rangle\rangle = \mathcal{L}|\rho(t)\rangle\rangle$, so for $t \geq 0$,

$$|\rho(t)\rangle\rangle = e^{t\mathcal{L}}|\rho_{\text{in}}\rangle\rangle, \tag{A6}$$

with $\rho_{\text{in}}$ being the initial state. The norm of a wave function corresponds to the trace of $\rho$ ($\langle\langle I|\rho\rangle\rangle$); it is preserved under both Hamiltonian and Lindbladian evolution. It is easy to check that the exponential of any superoperator of the above form preserves both trace [$\langle\langle I|\mathcal{L}|\rho\rangle\rangle = 0$ with $I$ the identity of Op($\mathsf{H}$)] and Hermiticity [$\mathcal{L}(A^{\dagger}) = [\mathcal{L}(A)]^{\dagger}$, as can be verified from Eq. (A3)]. However, the norm or purity of $\rho$ ($\langle\langle\rho|\rho\rangle\rangle = \text{Tr}\{\rho^2\}$) is not always preserved under Lindbladian evolution.

### 2. Double-bra-ket basis for steady states

We now bring in intuition from Hamiltonian-based quantum mechanics by writing the eigenmatrices as vectors using double-ket notation. First, we introduce some bases for Op($\mathsf{H}$), with which we can build bases for As($\mathsf{H}$). Given any orthonormal basis $\{|\phi_k\rangle\}_{k=0}^{N-1}$ for the system Hilbert space $\mathsf{H}$, one can construct the corresponding orthonormal (under the trace) outer product basis for Op($\mathsf{H}$),

$$\{|\Phi_{kl}\rangle\rangle\}_{k,l=0}^{N-1}, \quad \text{where } \Phi_{kl} \equiv |\phi_k\rangle\langle\phi_l|. \tag{A7}$$

The analogy with quantum mechanics is that the matrices $\Phi_{kl} \leftrightarrow |\Phi_{kl}\rangle\rangle$ and $\Phi_{kl}^{\dagger} \leftrightarrow \langle\langle\Phi_{kl}|$ are vectors in the vector space Op($\mathsf{H}$) and superoperators $\mathcal{O}$ are linear operators on those vectors. Furthermore, one can save an index and use properly normalized Hermitian matrices $\Gamma_{\kappa}^{\dagger} = \Gamma_{\kappa}$ to form an orthonormal basis $\{|\Gamma_{\kappa}\rangle\rangle\}_{\kappa=0}^{N^2-1}$:

$$\langle\langle\Gamma_{\kappa}|\Gamma_{\lambda}\rangle\rangle \equiv \text{Tr}\{\Gamma_{\kappa}^{\dagger}\Gamma_{\lambda}\} = \text{Tr}\{\Gamma_{\kappa}\Gamma_{\lambda}\} = \delta_{\kappa\lambda}. \tag{A8}$$

For example, an orthonormal Hermitian matrix basis for Op($\mathsf{H}$) with $\mathsf{H}$ two dimensional consists of the identity matrix and the three Pauli matrices, all normalized by $1/\sqrt{2}$.

It is easy to see that the coefficients in the expansion of any Hermitian operator in such a matrix basis are real. For example, the coefficients $c_{\kappa}$ in the expansion of a density matrix,

$$|\rho\rangle\rangle = \sum_{\kappa=0}^{N^2-1} c_{\kappa}|\Gamma_{\kappa}\rangle\rangle \quad \text{with} \quad c_{\kappa} = \langle\langle\Gamma_{\kappa}|\rho\rangle\rangle, \tag{A9}$$

are clearly real and represent the components of a generalized Bloch (coherence) vector [177,178]. Furthermore, defining

$$\mathcal{O}_{\kappa\lambda} \equiv \langle\langle\Gamma_{\kappa}|\mathcal{O}|\Gamma_{\lambda}\rangle\rangle \equiv \text{Tr}\{\Gamma_{\kappa}^{\dagger}\mathcal{O}(\Gamma_{\lambda})\} \tag{A10}$$

for any superoperator $\mathcal{O}$, one can write

$$\mathcal{O} = \sum_{\kappa,\lambda=0}^{N^2-1} \mathcal{O}_{\kappa\lambda}|\Gamma_{\kappa}\rangle\rangle\langle\langle\Gamma_{\lambda}|. \tag{A11}$$

There are many physical $\mathcal{O}$ for which the "matrix" elements $\mathcal{O}_{\kappa\lambda}$ are real. For example, it is easy to show that matrix elements $\mathcal{H}_{\kappa\lambda}$, where $\mathcal{H}(\cdot) \equiv -i[H,\cdot]$, are real using cyclic permutations under the trace and Hermiticity of the $\Gamma$'s:

$$\mathcal{H}_{\kappa\lambda}^{\star} = i\text{Tr}\{\Gamma_{\lambda}[H,\Gamma_{\kappa}]\} = -i\text{Tr}\{\Gamma_{\kappa}[H,\Gamma_{\lambda}]\} = \mathcal{H}_{\kappa\lambda}. \tag{A12}$$

This calculation easily extends to all Hermiticity-preserving $\mathcal{O}$, i.e., superoperators such that $\mathcal{O}(A^\dagger) = [\mathcal{O}(A)]^\dagger$ for all operators $A$.

Given a Lindbladian, one can provide necessary and sufficient conditions under which it generates Hamiltonian time evolution. This early key result in open quantum systems [5] can be used to determine whether a perturbation generates unitary evolution (Proposition 3). We now proceed to state and prove it as well as the other two propositions in the main text.

## APPENDIX B: PROOFS OF PROPOSITIONS 1, 2, AND 3

*Proposition 1.*—Let $\{P, Q\}$ be projections on $\mathsf{H}$ and $\{\mathcal{P}_\square, \mathcal{P}_\square, \mathcal{P}_\square, \mathcal{P}_\square\}$ be their corresponding projections on $\mathrm{Op}(\mathsf{H})$. Then,

$$\forall \ell : F^\ell_\square = 0,$$

$$H_\square = -\frac{i}{2}\sum_\ell \kappa_\ell F^{\ell\dagger}_\square F^\ell_\square.$$

*Proof.*—By definition Eq. (2.1), $\mathcal{P}_\square \mathrm{Op}(\mathsf{H})$ is the smallest subspace of $\mathrm{Op}(\mathsf{H})$ containing all asymptotic states. Therefore, all states evolving under $\mathcal{L}$ converge to states in $\mathcal{P}_\square \mathrm{Op}(\mathsf{H})$ as $t \to \infty$ (Ref. [78], Theorem 2-1). This implies invariance, i.e., states $\rho_\square = \mathcal{P}_\square(\rho)$ remain there under application of $\mathcal{L}$:

$$\mathcal{L}(\rho_\square) = \mathcal{L}\mathcal{P}_\square(\rho) = \mathcal{P}_\square \mathcal{L}\mathcal{P}_\square(\rho). \tag{B1}$$

Applying $\mathcal{P}_\square$, we get

$$\mathcal{P}_\square \mathcal{L} \mathcal{P}_\square(\rho) = \sum_\ell \kappa_\ell F^\ell_\square \rho F^{\ell\dagger}_\square = 0,$$

since the projections are mutually orthogonal. Taking the trace,

$$\langle\!\langle I | \mathcal{P}_\square \mathcal{L} \mathcal{P}_\square | \rho \rangle\!\rangle = \sum_\ell \kappa_\ell \mathrm{Tr}\{\rho F^{\ell\dagger}_\square F^\ell_\square\} = 0.$$

If $\rho$ is a full-rank density matrix ($\mathrm{rank}\{\rho\} = \mathrm{Tr}\{P\}$), then each summand above is non-negative (since $\kappa_\ell > 0$ and $F^{\ell\dagger}_\square F^\ell_\square$ are positive semidefinite). Thus, the only way for the above to hold for all $\rho$ is for $F^{\ell\dagger}_\square F^\ell_\square = 0$ for all $\ell$, which implies that $F^\ell_\square = 0$. Applying $\mathcal{P}_\square$ to Eq. (B1) and simplifying using $F^\ell_\square = 0$ gives

$$\mathcal{P}_\square \mathcal{L} \mathcal{P}_\square(\rho) = P\rho\left(iH_\square - \frac{1}{2}\sum_\ell \kappa_\ell F^{\ell\dagger}_\square F^\ell_\square\right) = 0,$$

implying the condition on $H_\square$. ∎

*Proposition 2.*—The left eigenmatrices of $\mathcal{L}$ corresponding to pure imaginary eigenvalues $i\Delta$ are

$$\langle\!\langle J^{\Delta\mu}| = \langle\!\langle J^{\Delta\mu}_\square| \left(\mathcal{P}_\square - \mathcal{L}\frac{\mathcal{P}_\square}{\mathcal{L}_\square - i\Delta\mathcal{P}_\square}\right), \tag{B2}$$

where $\langle\!\langle J^{\Delta\mu}_\square|$ are left eigenmatrices of $\mathcal{L}_\square$.

*Proof.*—For a left eigenmatrix $\langle\!\langle J^{\Delta\mu}|$ with eigenvalue $i\Delta$,

$$\mathcal{L}^\ddagger |J^{\Delta\mu}\rangle\!\rangle = -i\Delta |J^{\Delta\mu}\rangle\!\rangle.$$

Now partition this eigenvalue equation using the projections $\{\mathcal{P}_\square, \mathcal{P}_\square, \mathcal{P}_\square\}$. Taking the $\ddagger$ of the partitioned $\mathcal{L}$ from Eq. (3.3) results in

$$\mathcal{L}^\ddagger |J^{\Delta\mu}\rangle\!\rangle = \begin{bmatrix} \mathcal{L}^\ddagger_\square & 0 & 0 \\ \mathcal{P}_\square \mathcal{L}^\ddagger \mathcal{P}_\square & \mathcal{L}^\ddagger_\square & 0 \\ \mathcal{P}_\square \mathcal{L}^\ddagger \mathcal{P}_\square & \mathcal{P}_\square \mathcal{L}^\ddagger \mathcal{P}_\square & \mathcal{L}^\ddagger_\square \end{bmatrix} \begin{bmatrix} |J^{\Delta\mu}_\square\rangle\!\rangle \\ |J^{\Delta\mu}_\square\rangle\!\rangle \\ |J^{\Delta\mu}_\square\rangle\!\rangle \end{bmatrix}.$$

The eigenvalue equation is then equivalent to the following three conditions on the components of $J^{\Delta\mu}$:

$$-i\Delta J^{\Delta\mu}_\square = \mathcal{L}^\ddagger_\square(J^{\Delta\mu}_\square), \tag{B3a}$$

$$-i\Delta J^{\Delta\mu}_\square = \mathcal{P}_\square \mathcal{L}^\ddagger \mathcal{P}_\square(J^{\Delta\mu}_\square) + \mathcal{L}^\ddagger_\square(J^{\Delta\mu}_\square), \tag{B3b}$$

$$-i\Delta J^{\Delta\mu}_\square = \mathcal{P}_\square \mathcal{L}^\ddagger \mathcal{P}_\square(J^{\Delta\mu}_\square) + \mathcal{L}^\ddagger_\square(J^{\Delta\mu}_\square) + \mathcal{P}_\square \mathcal{L}^\ddagger \mathcal{P}_\square(J^{\Delta\mu}_\square). \tag{B3c}$$

We now examine them in order.

(i) Condition Eq. (B3a) implies that $[F^{\ell\dagger}_\square, J^{\Delta\mu}_\square] = 0$ for all $\ell$. [This part is essentially the Lindblad version of a similar statement for quantum channels (Ref. [79], Lemma 5.2). Another way to prove this is to apply "well-known" algebra decomposition theorems (see, e.g., Ref. [9], Theorem 5)]. To show this, we use the dissipation function $\mathcal{J}$ associated with $\mathcal{L}_\square$ [1]. For some $A \in \mathcal{P}_\square \mathrm{Op}(\mathsf{H})$,

$$\mathcal{J}(A) \equiv \mathcal{L}^\ddagger_\square(A^\dagger A) - \mathcal{L}^\ddagger_\square(A^\dagger)A - A^\dagger \mathcal{L}^\ddagger_\square(A)$$
$$= \sum_\ell \kappa_\ell [F^\ell_\square, A]^\dagger [F^\ell_\square, A].$$

Using Eq. (B3a) and remembering that $J^{\Delta\mu\dagger}_\square = J^{-\Delta\mu}_\square$, the two expressions for $\mathcal{J}(J^{\Delta\mu}_\square)$ imply that

$$\mathcal{L}^\ddagger_\square(J^{\Delta\mu\dagger}_\square J^{\Delta\mu}_\square) = \sum_\ell \kappa_\ell [F^\ell_\square, J^{\Delta\mu}_\square]^\dagger [F^\ell_\square, J^{\Delta\mu}_\square]. \tag{B4}$$

We now take the trace using the full-rank steady-state density matrix:

$$|\rho_{\text{ss}}\rangle\!\rangle = \mathcal{P}_{\blacksquare}|\rho_{\text{ss}}\rangle\!\rangle \equiv \sum_\mu c_\mu |\Psi_{0\mu}\rangle\!\rangle.$$

Such an asymptotic state is simply $|\rho_\infty\rangle\!\rangle$ from Eq. (3.14b) with $c_{\Delta\mu} = \delta_{\Delta 0} c_\mu$ and $c_\mu \neq 0$. It is full rank because it is a linear superposition of projections on eigenstates of $H_\infty$, and such projections provide a basis for all diagonal matrices of $\mathcal{P}_{\blacksquare}\text{Op}(\mathsf{H})$. Taking the trace of the left-hand side of Eq. (B4) yields

$$\langle\!\langle \rho_{\text{ss}} | \mathcal{L}_{\blacksquare}^{\ddagger}(J_{\blacksquare}^{\Delta\mu\dagger} J_{\blacksquare}^{\Delta\mu})\rangle\!\rangle = \langle\!\langle \mathcal{L}_{\blacksquare}(\rho_{\text{ss}}) | J_{\blacksquare}^{\Delta\mu\dagger} J_{\blacksquare}^{\Delta\mu}\rangle\!\rangle = 0,$$

implying that the trace of the right-hand side is zero:

$$\sum_\ell \kappa_\ell \text{Tr}\{\rho_{\text{ss}} [F_{\blacksquare}^\ell, J_{\blacksquare}^{\Delta\mu}]^\dagger [F_{\blacksquare}^\ell, J_{\blacksquare}^{\Delta\mu}]\} = 0.$$

Each summand above is non-negative (since $\kappa_\ell > 0$, the commutator products are positive semidefinite, and $\rho_{\text{ss}}$ is positive definite). Thus, the only way for the above to hold is for $[F_{\blacksquare}^\ell, J_{\blacksquare}^{\Delta\mu}]^\dagger [F_{\blacksquare}^\ell, J_{\blacksquare}^{\Delta\mu}] = 0$, which implies that $F_{\blacksquare}^\ell$ and $J_{\blacksquare}^{\Delta\mu}$ commute for all $\ell, \Delta, \mu$. If we once again remember that $J_{\blacksquare}^{\Delta\mu\dagger} = J_{\blacksquare}^{-\Delta\mu}$ and that the eigenvalues come in pairs $\pm\Delta$, then

$$[F_{\blacksquare}^{\ell\dagger}, J_{\blacksquare}^{\Delta\mu}] = [F_{\blacksquare}^\ell, J_{\blacksquare}^{\Delta\mu}] = 0. \quad (\text{B5})$$

(ii) Now consider condition Eq. (B3b). The first term on the right-hand side can be obtained from Eq. (C5) and is as follows:

$$\mathcal{P}_{\blacksquare}\mathcal{L}^{\ddagger}\mathcal{P}_{\blacksquare}(J_{\blacksquare}^{\Delta\mu}) = \sum_\ell \kappa_\ell (F_{\blacksquare}^{\ell\dagger} J_{\blacksquare}^{\Delta\mu} F_{\blacksquare}^\ell - J_{\blacksquare}^{\Delta\mu} F_{\blacksquare}^{\ell\dagger} F_{\blacksquare}^\ell)$$
$$+ \sum_\ell \kappa_\ell (F_{\blacksquare}^{\ell\dagger} J_{\blacksquare}^{\Delta\mu} F_{\blacksquare}^\ell - F_{\blacksquare}^{\ell\dagger} F_{\blacksquare}^\ell J_{\blacksquare}^{\Delta\mu}).$$

This term is identically zero due to Eq. (B5), reducing condition Eq. (B3b) to $\mathcal{L}_{\blacksquare}^{\ddagger}(J_{\blacksquare}^{\Delta\mu}) = -i\Delta J_{\blacksquare}^{\Delta\mu}$. We now show that this implies

$$\mathcal{P}_{\blacksquare}|J^{\Delta\mu}\rangle\!\rangle = 0 \quad (\text{B6})$$

for all $\Delta$ and $\mu$. By contradiction, assume $J_{\blacksquare}^{\Delta\mu}(\neq 0)$ is a left eigenmatrix of $\mathcal{L}_{\blacksquare}$. Then there must exist a corresponding right eigenmatrix $\Psi'_{\Delta\mu} = \mathcal{P}_{\blacksquare}(\Psi'_{\Delta\mu})$ since the sets of $\Psi$ and $J$ are biorthogonal (see, e.g., Ref. [78], Theorem 18). However, all right eigenmatrices are contained in $\mathcal{P}_{\blacksquare}\text{Op}(\mathsf{H})$ by definition Eq. (2.1), so we have a contradiction and $J_{\blacksquare}^{\Delta\mu} = 0$.

(iii) Finally, consider condition Eq. (B3c). Applying Eq. (B6) removes the last term on the right-hand side of that condition and simplifies it to

$$[\mathcal{L}_{\blacksquare}^{\ddagger} + i\Delta\mathcal{P}_{\blacksquare}](J_{\blacksquare}^{\Delta\mu}) = -\mathcal{P}_{\blacksquare}\mathcal{L}^{\ddagger}(J_{\blacksquare}^{\Delta\mu}).$$

Now, we can show that the operator $\mathcal{L}_{\blacksquare}^{\ddagger} + i\Delta\mathcal{P}_{\blacksquare}$ is invertible when restricted to $\mathcal{P}_{\blacksquare}\text{Op}(\mathsf{H})$ using a proof by contradiction similar to the one used to prove Eq. (B6). Inversion gives a formula for $J_{\blacksquare}^{\Delta\mu}$ which is used along with Eq. (B6) to obtain the statement. ∎

*Proposition 3.*—The matrix $\mathcal{L}_{\kappa\lambda} = \langle\!\langle \Gamma_\kappa | \mathcal{L} | \Gamma_\lambda \rangle\!\rangle$ is real. Moreover,

$$\mathcal{L}_{\lambda\kappa} = -\mathcal{L}_{\kappa\lambda} \Leftrightarrow \mathcal{L} = -i[H, \cdot] \quad \text{with Hamiltonian } H. \quad (\text{B7})$$

*Proof.*—To prove reality, use the definition of the adjoint of $\mathcal{L}$, Hermiticity of $\Gamma_\kappa$, and cyclicity under the trace:

$$\mathcal{L}_{\kappa\lambda}^\star = \langle\!\langle \Gamma_\lambda | \mathcal{L}^\ddagger | \Gamma_\kappa \rangle\!\rangle = \langle\!\langle \mathcal{L}(\Gamma_\lambda) | \Gamma_\kappa \rangle\!\rangle = \langle\!\langle \Gamma_\kappa | \mathcal{L} | \Gamma_\lambda \rangle\!\rangle = \mathcal{L}_{\kappa\lambda}.$$

$\Leftarrow$ Assume $\mathcal{L}$ generates unitary evolution. Then there exists a Hamiltonian $H$ such that $\mathcal{L}|\Gamma_\kappa\rangle\!\rangle = -i|[H, \Gamma_\kappa]\rangle\!\rangle$ and $\mathcal{L}$ is antisymmetric:

$$\mathcal{L}_{\lambda\kappa} = -i\text{Tr}\{\Gamma_\lambda [H, \Gamma_\kappa]\} = i\text{Tr}\{\Gamma_\kappa [H, \Gamma_\lambda]\} = -\mathcal{L}_{\kappa\lambda}.$$

$\Rightarrow$ (An alternative way to prove this part is to observe that all eigenvalues of $\mathcal{L}$ lie on the imaginary axis and use Theorem 18-3 in Ref. [78].) Assume $\mathcal{L}_{\kappa\lambda}$ is antisymmetric, so $\mathcal{L}^\ddagger = -\mathcal{L}$. Then the dynamical semigroup $\{e^{t\mathcal{L}}; t \geq 0\}$ is isometric (norm-preserving): let $t \geq 0$ and $|A\rangle\!\rangle \in \text{Op}(\mathsf{H})$ and observe that

$$\langle\!\langle e^{t\mathcal{L}}(A) | e^{t\mathcal{L}}(A)\rangle\!\rangle = \langle\!\langle A | e^{-t\mathcal{L}} e^{t\mathcal{L}} | A\rangle\!\rangle = \langle\!\langle A | A\rangle\!\rangle.$$

Since it is clearly invertible, $e^{t\mathcal{L}}: \text{Op}(\mathsf{H}) \to \text{Op}(\mathsf{H})$ is a surjective map. All surjective isometric one-parameter dynamical semigroups can be expressed as $e^{t\mathcal{L}}(\rho) = U_t \rho U_t^\dagger$, with $U_t$ belonging to a one-parameter unitary group $\{U_t; t \in \mathbb{R}\}$ acting on $\mathsf{H}$ (Ref. [5], Theorem 6). By Stone's theorem on one-parameter unitary groups, there then exists a Hamiltonian $H$ such that $U_t = e^{-iHt}$ and $\mathcal{L}(\rho) = -i[H, \rho]$. ∎

# APPENDIX C: FOUR-CORNERS DECOMPOSITION OF $\mathcal{L}$

Based on conditions Eq. (2.1) and after simplifications due to Proposition 1, the nonzero elements of Eq. (3.3) acting on a Hermitian matrix $\rho = \rho_{\blacksquare} + \rho_{\blacksquare} + \rho_{\blacksquare}$ are

$$\mathcal{L}_{\square}(\rho) = -i[H_{\square}, \rho_{\square}] + \frac{1}{2}\sum_{\ell}\kappa_{\ell}(2F^{\ell}_{\square}\rho_{\square}F^{\ell\dagger}_{\square} - F^{\ell\dagger}_{\square}F^{\ell}_{\square}\rho_{\square} - \rho_{\square}F^{\ell\dagger}_{\square}F^{\ell}_{\square}), \tag{C1}$$

$$\mathcal{L}_{\square}(\rho) = -i(H_{\square}\rho_{\square} - \rho_{\square}H_{\square}) + \frac{1}{2}\sum_{\ell}\kappa_{\ell}[2F^{\ell}_{\square}\rho_{\square}F^{\ell\dagger}_{\square} - F^{\ell\dagger}_{\square}F^{\ell}_{\square}\rho_{\square} - \rho_{\square}(F^{\ell\dagger}F^{\ell})_{\square}], \tag{C2}$$

$$\mathcal{L}_{\square}(\rho) = [\mathcal{L}_{\square}(\rho)]^{\dagger}, \tag{C3}$$

$$\mathcal{L}_{\square}(\rho) = -i[H_{\square}, \rho_{\square}] + \frac{1}{2}\sum_{\ell}\kappa_{\ell}[2F^{\ell}_{\square}\rho_{\square}F^{\ell\dagger}_{\square} - (F^{\ell\dagger}F^{\ell})_{\square}\rho_{\square} - \rho_{\square}(F^{\ell\dagger}F^{\ell})_{\square}], \tag{C4}$$

$$\mathcal{P}_{\square}\mathcal{L}\mathcal{P}_{\square}(\rho) = \sum_{\ell}\kappa_{\ell}(F^{\ell}_{\square}\rho_{\square}F^{\ell\dagger}_{\square} - \rho_{\square}F^{\ell\dagger}_{\square}F^{\ell}_{\square}) + \text{H.c.}, \tag{C5}$$

$$\mathcal{P}_{\square}\mathcal{L}\mathcal{P}_{\square}(\rho) = \sum_{\ell}\kappa_{\ell}(F^{\ell}_{\square}\rho_{\square}F^{\ell\dagger}_{\square} - F^{\ell\dagger}_{\square}F^{\ell}_{\square}\rho_{\square}) + \text{H.c.}, \tag{C6}$$

$$\mathcal{P}_{\square}\mathcal{L}\mathcal{P}_{\square}(\rho) = \sum_{\ell}\kappa_{\ell}F^{\ell}_{\square}\rho_{\square}F^{\ell\dagger}_{\square}. \tag{C7}$$

Note that $(F^{\ell\dagger}F^{\ell})_{\square} = F^{\ell\dagger}_{\square}F^{\ell}_{\square} + F^{\ell\dagger}_{\square}F^{\ell}_{\square}$ and that evolution of coherences decouples: $\mathcal{P}_{\square}\mathcal{L}\mathcal{P}_{\square} = \mathcal{P}_{\square}\mathcal{L}\mathcal{P}_{\square} = 0$. Since $F^{\ell}_{\square} \neq 0$, $\mathcal{L}_{\square}$ [see Eq. (C4)] is *not* of Lindblad form and instead ensures decay of states in $\square$ (see Ref. [80], Proposition 3, and Ref. [171], Lemma 4).

## APPENDIX D: CONDUCTIVITY FOR A THERMAL LINDBLADIAN

Here, we compute the conductivity for the thermal Landau Lindbladian for noninteracting particles, making contact with Example 9 of Ref. [70]. The thermal Lindbladian consists the Hamiltonian Eq. (4.8) along with two jump operators, $F_i = \sqrt{2\gamma}b_i$ and $\bar{F}_i = \sqrt{2\bar{\gamma}}b_i^{\dagger}$, and the effective temperature is given by

$$\beta \equiv \frac{1}{\omega_c}\log\left(\frac{\gamma}{\bar{\gamma}}\right). \tag{D1}$$

For this portion, we consider noninteracting particles. This is necessary in order to interpret the Lindblad operator as representing coupling to a thermal bath. We work in the grand-canonical ensemble: our initial steady state is $\rho_\infty \propto e^{\beta(H-\mu N)}$, where $N$ is the number operator and $\mu$ is the chemical potential lying in a gap between the Landau levels. Since $\bar{F}_i$ is quite similar to $F_i$, its contribution is calculated in analogous fashion and the temperature-dependent conductivity tensor is

$$\sigma_{\varsigma\tau}(\omega,\beta) = \frac{\langle \nu \rangle \omega_c(i\omega\delta_{\varsigma\tau} - \omega_c[1 - i(\gamma - \bar{\gamma})\frac{\omega_T}{\omega_c^2}]\epsilon_{\varsigma\tau})}{2\pi(\omega_T^2 - \omega_c^2)}, \tag{D2}$$

where we define $\omega_T = \omega + i(\gamma - \bar{\gamma})$, the thermally averaged filling factor is given by

$$\langle \nu \rangle = \sum_m n_F\left(\omega_c\left[m + \frac{1}{2}\right], \mu\right), \tag{D3}$$

$m$ sums over the occupied Landau levels, and $n_F(\epsilon, \mu)$ is the Fermi distribution function.

## APPENDIX E: HAMILTONIAN ADIABATIC THEORY

Here, we review the adiabatic (Berry) connection for a Hamiltonian system and the DFS case.

### 1. Hamiltonian case

First, let us review two important consequences of the (Hamiltonian) quantum mechanical adiabatic theorem. In this work, the adiabatic theorem holds by assumption. We do not make the adiabatic approximation to $H(s)$ [and, later, $\mathcal{L}(s)$ [179]] since it is not sufficient for the adiabatic theorem to hold; see Ref. [180] and references therein. Adiabatic evolution can be thought of as either (1) being generated by an effective operator [181] or (2) generating transport of vectors in parameter space, leading to abelian [182–185] or non-abelian [186] holonomies. We loosely follow the excellent expositions in Chap. 2.1.2 of Ref. [187] and Sec. 9 of Ref. [188]. We conclude with a summary of four different ways [Eqs. (E17)–(E20)] of writing holonomies for the nondegenerate case.

Let $|\psi_0^{(t)}\rangle$ be the instantaneous unique (up to a phase) zero-energy ground state of a Hamiltonian $H(t)$. We assume that the ground state is separated from all other eigenstates of $H(t)$ by a nonzero excitation gap for all times of interest. Let us also rescale time ($s = t/T$) such that the exact state $|\psi(s)\rangle$ evolves according to

$$\frac{1}{T}\partial_s|\psi(s)\rangle = -iH(s)|\psi(s)\rangle. \quad (E1)$$

The adiabatic theorem states that $|\psi(s)\rangle$ [with $|\psi(0)\rangle = |\psi_0^{(s=0)}\rangle$] remains an instantaneous eigenstate of $H(s)$ (up to a phase $\theta$) in the limit as $T \to \infty$, with corrections of order $O(1/T)$. Let $P_0^{(s)} = |\psi_0^{(s)}\rangle\langle\psi_0^{(s)}|$ be the projection onto the instantaneous ground state. In the adiabatic limit,

$$|\psi(s)\rangle = e^{i\theta(s)}|\psi_0^{(s)}\rangle, \quad (E2)$$

and the initial projection $P_0^{(0)}$ evolves into

$$P_0^{(s)} = U_{\text{ad}}(s)P_0^{(0)}U_{\text{ad}}^\dagger(s). \quad (E3)$$

The adiabatic evolution operator $U_{\text{ad}}$ is determined by the Kato equation

$$\partial_s U_{\text{ad}} = -iKU_{\text{ad}}, \quad (E4)$$

with the so-called Kato Hamiltonian [181] ($\dot{P}_0 \equiv \partial_s P_0$)

$$K = i[\dot{P}_0, P_0]. \quad (E5)$$

The adiabatic operator $U_{\text{ad}}$ can be shown to satisfy Eq. (E3) (see Ref. [187], Proposition 2.1.1) using

$$P_0\dot{P}_0P_0 = Q_0\dot{P}_0Q_0 = 0. \quad (E6)$$

The $P_0\dot{P}_0P_0 = 0$ is a key consequence of the idempotence of projections while $Q_0\dot{P}_0Q_0 = 0$ is obtained by application of the no-leak property Eq. (2.9); both are used throughout the text. The adiabatic evolution operator is then a product of exponentials of $-iK$ ordered along the path $s' \in [0, s]$ (with path ordering denoted by $\mathbb{P}$):

$$U_{\text{ad}}(s) = \mathbb{P}\exp\left(\int_0^s [\dot{P}_0, P_0]ds'\right). \quad (E7)$$

Because of the intertwining property Eq. (E3), $U_{\text{ad}}(s)$ simultaneously transfers states in $P_0^{(0)}\mathsf{H}$ to $P_0^{(s)}\mathsf{H}$ and states in $Q_0^{(0)}\mathsf{H}$ to $Q_0^{(s)}\mathsf{H}$ (with $Q_0 \equiv I - P_0$) without mixing the two subspaces during the evolution. The term $\dot{P}_0 P_0$ in Eq. (E5) is responsible for generating the adiabatic evolution of $P_0\mathsf{H}$, while the term $P_0\dot{P}_0$ generates adiabatic evolution of $Q_0\mathsf{H}$. To see this, observe that the adiabatically evolving state $|\psi(s)\rangle = U_{\text{ad}}(s)|\psi_0^{(s=0)}\rangle \in P_0^{(s)}\mathsf{H}$ obeys the Schrödinger equation

$$\partial_s|\psi(s)\rangle = [\dot{P}_0, P_0]|\psi(s)\rangle. \quad (E8)$$

Applying Eq. (E6), the second term in the commutator can be removed without changing the evolution. Since we are interested only in adiabatic evolution of the zero-eigenvalue subspace $P_0\mathsf{H}$ (and not its complement), we can simplify $U_{\text{ad}}$ by removing the second term in the Kato Hamiltonian. This results in the adiabatic Schrödinger equation

$$\partial_s|\psi(s)\rangle = \dot{P}_0 P_0|\psi(s)\rangle \quad (E9)$$

and effective adiabatic evolution operator

$$U^{(s)} = \mathbb{P}\exp\left(\int_0^s \dot{P}_0 P_0 ds'\right). \quad (E10)$$

We now assume that $s$ parametrizes a path in the parameter space $\mathsf{M}$ of some external time-dependent parameters of $H(s)$. For simplicity, we assume that $\mathsf{M}$ is simply connected [145]. By writing $P_0$ and $\dot{P}_0$ in terms of $|\psi_0\rangle$ and explicitly differentiating, the adiabatic Schrödinger equation (E9) becomes

$$\partial_s|\psi\rangle = (I - P_0)\partial_s|\psi\rangle. \quad (E11)$$

This implies a parallel transport condition,

$$0 = P_0\partial_s|\psi\rangle = \langle\psi|\partial_s\psi\rangle|\psi\rangle, \quad (E12)$$

which describes how to move the state vector from one point in $\mathsf{M}$ to another. The particular condition resulting from adiabatic evolution eliminates any first-order deviation from the unit overlap between nearby adiabatically evolving states [189]:

$$\langle\psi(s + \delta s)|\psi(s)\rangle = 1 + O(\delta s^2). \quad (E13)$$

Therefore, we show two interpretations stemming from the adiabatic theorem. The first is that adiabatic evolution of $|\psi(s)\rangle$ [with $|\psi(0)\rangle = |\psi_0^{(s=0)}\rangle$] is generated (in the ordinary quantum mechanical sense) by the $\dot{P}_0 P_0$ piece of the Kato Hamiltonian $K$. The second is that adiabatic evolution realizes parallel transport of $|\psi(s)\rangle$ along a curve in parameter space. As we show now, either framework can be used to determine the adiabatically evolved state and the resulting Berry phase.

We now define a coordinate basis $\{\mathbf{x}_\alpha\}$ for the parameter space $\mathsf{M}$. In other words,

$$\partial_t = \frac{1}{T}\partial_s = \frac{1}{T}\sum_\alpha \dot{\mathbf{x}}_\alpha \partial_\alpha, \quad (E14)$$

where $\partial_s$ is the derivative along the path, $\partial_\alpha \equiv \partial/\partial\mathbf{x}_\alpha$ are derivatives in various directions in parameter space, and $\dot{\mathbf{x}}_\alpha \equiv \frac{d\mathbf{x}_\alpha}{ds}$ are (unitless) parameter velocities. Combining Eqs. (E2) and (E14) with the parallel transport condition Eq. (E12) yields

$$0 = P_0\partial_s|\psi\rangle = i\sum_\alpha \dot{\mathbf{x}}_\alpha(\partial_\alpha\theta - A_{\alpha,00})|\psi\rangle, \quad (E15)$$

where the adiabatic (Berry) connection $A_{\alpha,00} = i\langle\psi_0|\partial_\alpha\psi_0\rangle$ is a vector (gauge) potential in parameter space. The reason we can think of $A_{\alpha,00}$ as a gauge potential is because it transforms as one under gauge transformations $|\psi_0\rangle \to e^{i\vartheta}|\psi_0\rangle$, where $\vartheta \in \mathbb{R}$:

$$A_{\alpha,00} \to A_{\alpha,00} - \partial_\alpha\vartheta. \quad (E16)$$

These structures arise because the adiabatic theorem has furnished for us a vector bundle over the parameter-space manifold M [188,189]. More formally, given the trivial bundle M × H (where at each point in M we have a copy of the full Hilbert space H), the projection $P_0$ defines a (possibly nontrivial) sub-bundle of M × H (in this case, a line bundle, since $P_0$ is rank one). The trivial bundle has a covariant derivative $\nabla_\alpha \equiv \partial_\alpha$ with an associated connection that can be taken to vanish. The Berry connection $A_{\alpha,00}$ is then simply the connection associated with the covariant derivative $P_0 \nabla_\alpha$ induced on the sub-bundle defined by $P_0$.

The Berry connection describes what happens to the initial state vector as it is parallel transported. It may happen that the vector does not return to itself after transport around a closed path in parameter space (due to, e.g., curvature or nonsimple connectedness of M). Given an initial condition $\theta(0) = 0$, the parallel transport condition Eq. (E15) uniquely determines how $\theta$ will change during adiabatic traversal of a path $C$ parametrized by $s \in [0, 1]$, i.e., from a point $\mathbf{x}_\alpha^{(s=0)} \in M$ to $\mathbf{x}_\alpha^{(1)}$. For a closed path ($\mathbf{x}_\alpha^{(1)} = \mathbf{x}_\alpha^{(0)}$) and assuming $A_{\alpha,00}$ is defined uniquely for the whole path [146], the state transforms as $|\psi(0)\rangle \to B|\psi(0)\rangle$, with resulting gauge-invariant holonomy (here, Berry phase)

$$B \equiv \exp\left(i \sum_\alpha \oint_C A_{\alpha,00} d\mathbf{x}_\alpha\right). \quad (E17)$$

Alternatively, we can use Eq. (E14) and the Schrödinger equation (E9): $|\psi(0)\rangle \to U|\psi(0)\rangle$, with holonomy

$$U \equiv \mathbb{P} \exp\left(\sum_\alpha \oint_C \partial_\alpha P_0 P_0 d\mathbf{x}_\alpha\right). \quad (E18)$$

Since the geometric and Kato Hamiltonian formulations of adiabatic evolution are equivalent, Eqs. (E17) and (E18) offer two ways to get to the same answer. They reveal two representations of the Berry connection and holonomy: the coordinate representation $\{iA_{\alpha,00}, B\}$, which determines evolution of $\theta$ from Eq. (E2), and the operator representation $\{\partial_\alpha P_0 P_0, U\}$, which determines evolution of $|\psi_0\rangle$ [see Proposition 1.2 of Ref. [190] and Eq. (5) of Ref. [111]]. Despite the latter being a path-ordered product of matrices, it simplifies to the Berry phase in the case of closed paths.

For completeness, we also state an alternative form for each holonomy representation [Eqs. (E17) and (E18)]. If there are two or more parameters, then the coordinate representation can be expressed in terms of the (here, Abelian) Berry curvature $F_{\alpha\beta,00} \equiv \partial_\alpha A_{\beta,00} - \partial_\beta A_{\alpha,00}$ using Stokes's theorem:

$$B = \exp\left(\frac{i}{2} \sum_{\alpha,\beta} \iint_S F_{\alpha\beta,00} d\mathbf{x}_\alpha d\mathbf{x}_\beta\right), \quad (E19)$$

where $S$ is a surface whose boundary is the contour $C$. The operator representation can also be written as a product of the path-dependent projections $P_0$:

$$U = \mathbb{P} \prod_{s \in C} P_0^{(s)}, \quad (E20)$$

where $\mathbb{P} \prod$ denotes a continuous product ordered from right to left along the path $C$ [see Eq. (47) of Ref. [72] and Proposition 1 of Ref. [60]]. This form of the holonomy should be reminiscent of the Pancharatnam phase [183,187] and, more generally, of a dynamical quantum Zeno effect (Refs. [63,64,66]; see also Refs. [97,98]).

### 2. DFS case

We briefly provide, in addition to Eq. (5.17), another proof of unitarity of the holonomy for the DFS case. Here, we do not need the reference basis of Sec. V B, so we let $|\Psi_\mu^{\text{DFS}}\rangle\rangle \equiv \mathcal{S}(s)|\bar{\Psi}_\mu^{\text{DFS}}\rangle\rangle$ and the same for $\langle\langle J^\mu| = \langle\langle \Psi_\mu^{\text{DFS}}|$. Now $\mathcal{A}_\alpha$ from Eq. (5.13) reduces to the coordinate form of the DFS connection:

$$\mathcal{A}_{\alpha,\mu\nu}^{\text{DFS}} = \langle\langle\Psi_\mu^{\text{DFS}}|\partial_\alpha \Psi_\nu^{\text{DFS}}\rangle\rangle = \text{Tr}\{\Psi_\mu^{\text{DFS}} \partial_\alpha \Psi_\nu^{\text{DFS}}\}. \quad (E21)$$

Although this can be equivalently expressed using the Wilczek-Zee adiabatic connection [186] $A_{kl}^{\text{DFS}} = i\langle\psi_k|\partial_\alpha\psi_l\rangle$, we briefly examine the superoperator counterpart. Sticking with the convention that

$$\Psi_0^{\text{DFS}} \equiv \frac{1}{\sqrt{d}} P_{\text{DFS}} = \frac{1}{\sqrt{d}} \sum_{k=0}^{d-1} |\psi_k\rangle\langle\psi_k|$$

is the only traceful element and using Eq. (E6), $\Psi_0^{\text{DFS}} \partial_\alpha \Psi_0^{\text{DFS}} \Psi_0^{\text{DFS}} = 0$, we see that $\mathcal{A}_{\alpha,\mu 0}^{\text{DFS}} = \mathcal{A}_{\alpha,0\mu}^{\text{DFS}} = 0$ for all $\mu$. Thus, $\mathcal{A}_\alpha^{\text{DFS}}$ consists of a direct sum of zero with a $(d^2 - 1)$-dimensional antisymmetric matrix acting on the Bloch vector components $\{|\Psi_{\mu \neq 0}^{\text{DFS}}\rangle\rangle\}$. Since the latter is antisymmetric, the holonomy is unitary.

Formally, letting $\text{Op}(\text{H})^\star$ be the space of traceless $d$-dimensional Hermitian matrices, $\mathcal{P}_{\text{DFS}}$ defines a sub-bundle of the trivial bundle M × Op(H)$^\star$ and $\mathcal{A}_\alpha^{\text{DFS}}$ is the connection associated with the covariant derivative $\mathcal{P}_{\text{DFS}} \partial_\alpha$ induced on that sub-bundle.

## APPENDIX F: HAMILTONIAN QGT

Here, we review the Hamiltonian quantum geometric tensor. Some relevant quantities for the Hamiltonian, degenerate Hamiltonian or DFS, and NS cases are summarized in Table I.

### 1. Hamiltonian case

First, let us review the nondegenerate Hamiltonian case before generalizing to the degenerate Hamiltonians in operator or superoperator form. We recommend Ref. [191] for a more detailed exposition.

Continuing from Sec. E 1, we begin with an instantaneous zero-energy state $|\psi_0\rangle$ and projection $P_0 = |\psi_0\rangle\langle\psi_0|$, which are functions of a vector of control parameters $\{\mathbf{x}_\alpha\}$. The distance between the

projections $P_0^{(s)}$ and $P_0^{(s+\delta s)}$ along a path parametrized by $s \in [0, 1]$ (with parameter vectors $\mathbf{x}_\alpha^{(s)}$ at each $s$) is governed by the QGT:

$$Q_{\alpha\beta,00} = \langle \psi_0 | \partial_\alpha P_0 \partial_\beta P_0 | \psi_0 \rangle \tag{F1a}$$

$$= \langle \partial_\alpha \psi_0 | (I - P_0) | \partial_\beta \psi_0 \rangle. \tag{F1b}$$

The latter form can be obtained from the former by explicit differentiation of $P_0$ and $\partial_\alpha P_0 \partial_\beta P_0 = (\partial_\alpha P_0)(\partial_\beta P_0)$ by convention. The $I - P_0$ term makes $Q_{\alpha\beta,00}$ invariant upon the gauge transformations $|\psi_0\rangle \to e^{i\vartheta}|\psi_0\rangle$. The tensor can be split into symmetric and antisymmetric parts,

$$2Q_{\alpha\beta,00} = M_{\alpha\beta,00} - iF_{\alpha\beta,00}, \tag{F2}$$

which coincide with its real and imaginary parts. The antisymmetric part is none other than the adiabatic or Berry curvature from Eq. (E19). The symmetric part is the quantum Fubini-Study metric tensor [74],

$$M_{\alpha\beta,00} = \text{Tr}\{P_0 \partial_{(\alpha} P_0 \partial_{\beta)} P_0\} = \text{Tr}\{\partial_\alpha P_0 \partial_\beta P_0\}, \tag{F3}$$

where $A_{(\alpha}B_{\beta)} = A_\alpha B_\beta + A_\beta B_\alpha$, and the latter form can be obtained using $P_0 \partial_\alpha P_0 P_0 = 0$. This quantity is manifestly symmetric in $\alpha$, $\beta$ and real; it is also non-negative when evaluated in parameter space (see Ref. [192], Appendix D).

### 2. DFS case

For degenerate Hamiltonian systems [192,193] and in the DFS case, the QGT $Q^{\text{DFS}}$ is a tensor in both parameter ($\alpha$, $\beta$) and state ($k$, $l$) indices and can be written as

$$Q^{\text{DFS}}_{\alpha\beta,kl} = \langle \psi_k | \partial_\alpha P_{\text{DFS}} \partial_\beta P_{\text{DFS}} | \psi_l \rangle \tag{F4a}$$

$$= \langle \partial_\alpha \psi_k | (I - P_{\text{DFS}}) | \partial_\beta \psi_l \rangle, \tag{F4b}$$

where $P_{\text{DFS}} = \sum_{k=0}^{d-1} |\psi_k\rangle\langle\psi_k|$ is the projection onto the degenerate zero eigenspace of $H(s)$. Since projections are invariant under changes of basis of their constituents, it is easy to see that $Q^{\text{DFS}}_{\alpha\beta} \to R^\dagger Q^{\text{DFS}}_{\alpha\beta} R$ under DFS changes of basis $|\psi_k\rangle \to |\psi_l\rangle R_{lk}$ for $R \in U(d)$. Notice that the QGT in Eq. (F4b) consists of overlaps between states outside of the zero eigenspace. For our applications, we write the QGT in a third way such that it consists of overlaps within the zero eigenspace only:

$$Q^{\text{DFS}}_{\alpha\beta,kl} = -i\partial_\alpha A^{\text{DFS}}_{\beta,kl} - (A^{\text{DFS}}_\alpha A^{\text{DFS}}_\beta)_{kl} - \langle \psi_k | \partial_\alpha \partial_\beta \psi_l \rangle, \tag{F4c}$$

where $A^{\text{DFS}}_\alpha$ is the DFS Berry connection, and we use

$$0 = \partial_\beta \langle \psi_k | \psi_l \rangle = \langle \partial_\beta \psi_k | \psi_l \rangle + \langle \psi_k | \partial_\beta \psi_l \rangle,$$
$$\partial_\alpha \langle \psi_k | \partial_\beta \psi_l \rangle = \langle \partial_\alpha \psi_k | \partial_\beta \psi_l \rangle + \langle \psi_k | \partial_\alpha \partial_\beta \psi_l \rangle. \tag{F5}$$

The Berry curvature is the part of the QGT antisymmetric in $\alpha$, $\beta$ (here, also the imaginary part of the QGT): $F^{\text{DFS}}_{\alpha\beta} = iQ^{\text{DFS}}_{[\alpha\beta]}$. From Eq. (F4c), we recover the form of the DFS Berry curvature.

The symmetric part of the QGT appears in the infinitesimal distance between nearby parallel transported rays (i.e., states of arbitrary phase) $\psi(s)$ and $\psi(s + \delta s)$ in the degenerate subspace:

$$\langle \partial_s \psi | \partial_s \psi \rangle = \langle \partial_s \psi | (I - P_{\text{DFS}}) | \partial_s \psi \rangle, \tag{F6}$$

where we use the parallel transport condition $P_{\text{DFS}} | \partial_s \psi \rangle = 0$. Expanding $\partial_s$ into parameter derivatives using Eq. (E14) and writing out $|\psi\rangle = \sum_{k=0}^{d-1} c_k |\psi_k\rangle$ yields

$$\langle \partial_s \psi | \partial_s \psi \rangle = \frac{1}{2} \sum_{\alpha,\beta} \sum_{k,l=0}^{d-1} Q^{\text{DFS}}_{(\alpha\beta),kl} \dot{\mathbf{x}}_\alpha \dot{\mathbf{x}}_\beta c_k^\star c_l. \tag{F7}$$

The corresponding Fubini-Study metric on the parameter space M is $Q^{\text{DFS}}_{(\alpha\beta)}$ traced over the degenerate subspace:

$$M^{\text{DFS}}_{\alpha\beta} \equiv \sum_{k=0}^{d-1} Q^{\text{DFS}}_{(\alpha\beta),kk} = \langle\!\langle P_{\text{DFS}} | \partial_{(\alpha} P_{\text{DFS}} \partial_{\beta)} P_{\text{DFS}} \rangle\!\rangle. \tag{F8}$$

All of this reasoning easily extends to the superoperator formalism ($|\psi_k\rangle \to |\Psi^{\text{DFS}}_\mu\rangle\!\rangle$). The superoperator QGT corresponding to $Q^{\text{DFS}}$ can be written as

$$\mathcal{Q}^{\text{DFS}}_{\alpha\beta,\mu\nu} = \langle\!\langle \Psi^{\text{DFS}}_\mu | \partial_\alpha \mathcal{P}_{\text{DFS}} \partial_\beta \mathcal{P}_{\text{DFS}} | \Psi^{\text{DFS}}_\nu \rangle\!\rangle$$
$$= \partial_\alpha \mathcal{A}^{\text{DFS}}_{\beta,\mu\nu} + (\mathcal{A}^{\text{DFS}}_\alpha \mathcal{A}^{\text{DFS}}_\beta)_{\mu\nu} - \langle\!\langle \Psi^{\text{DFS}}_\mu | \partial_\alpha \partial_\beta \Psi^{\text{DFS}}_\nu \rangle\!\rangle, \tag{F9}$$

where $\mathcal{A}^{\text{DFS}}_\alpha$ is the adiabatic connection Eq. (E21). The QGT is a real matrix (since $\mathcal{A}^{\text{DFS}}_\alpha$ is real) and consists of parts symmetric ($\mathcal{Q}^{\text{DFS}}_{(\alpha\beta)}$) and antisymmetric ($\mathcal{Q}^{\text{DFS}}_{[\alpha\beta]}$) in $\alpha$, $\beta$. Observing the second line of Eq. (F9), it should be easy to see that the Berry curvature $\mathcal{F}^{\text{DFS}}_{\alpha\beta} = \mathcal{Q}^{\text{DFS}}_{[\alpha\beta]}$. The symmetric part of the superoperator QGT appears in the infinitesimal Hilbert-Schmidt distance (Ref. [174], Sec. 14.3) between nearby parallel transported DFS states $\rho(s)$ and $\rho(s + \delta s)$:

$$\langle\!\langle \partial_s \rho | \partial_s \rho \rangle\!\rangle = \langle\!\langle \partial_s \rho | (\mathcal{I} - \mathcal{P}_{\text{DFS}}) | \partial_s \rho \rangle\!\rangle, \tag{F10}$$

where we use the parallel transport condition $\mathcal{P}_{\text{DFS}} | \partial_s \rho \rangle\!\rangle = 0$. Similar manipulations as with the operator QGT, including the expansion $|\rho\rangle\!\rangle = \sum_{\mu=0}^{d^2-1} c_\mu | \Psi^{\text{DFS}}_\nu \rangle\!\rangle$, yield

$$\langle\langle\partial_s\rho_\infty|\partial_s\rho_\infty\rangle\rangle = \frac{1}{2}\sum_{\alpha,\beta}\sum_{\mu,\nu=0}^{d^2-1} \mathcal{Q}^{\text{DFS}}_{(\alpha\beta),\mu\nu}\dot{\mathbf{x}}_\alpha\dot{\mathbf{x}}_\beta c_\mu c_\nu. \quad (F11)$$

The corresponding superoperator metric

$$\mathcal{M}^{\text{DFS}}_{\alpha\beta} \equiv \text{TR}\{\mathcal{P}_{\text{DFS}}\partial_{(\alpha}\mathcal{P}_{\text{DFS}}\partial_{\beta)}\mathcal{P}_{\text{DFS}}\}, \quad (F12)$$

where TR is the trace in superoperator space, is the symmetric part of the superoperator QGT traced over the degenerate subspace. Since $\text{Op}(\mathsf{H}) = \mathsf{H}\otimes\mathsf{H}^\star$, it is not surprising that $\mathcal{M}^{\text{DFS}}_{\alpha\beta}$ is proportional to the operator metric $M^{\text{DFS}}_{\alpha\beta}$:

$$\mathcal{M}^{\text{DFS}}_{\alpha\beta} = \sum_{\mu=0}^{d^2-1}\mathcal{Q}^{\text{DFS}}_{(\alpha\beta),\mu\mu} = 2dM^{\text{DFS}}_{\alpha\beta}. \quad (F13)$$

## APPENDIX G: OTHER GEOMETRIC TENSORS

In Sec. VI, we show that the antisymmetric part of the QGT

$$\mathcal{Q} = \mathcal{P}_\Psi\partial\mathcal{P}_\Psi\partial\mathcal{P}_\Psi\mathcal{P}_\Psi,$$

corresponds to the curvature $\mathcal{F}$ associated with the adiabatic connection $\mathcal{A}$ from Sec. V. We thus postulate that this QGT and its corresponding symmetric part should be relevant in determining distances between adiabatically connected Lindbladian steady states. However, the story does not end there as there are two more tensorial quantities that can be defined using the steady-state subspace. The first is an extension of the Fubini-Study metric to non- or pseudo-Hermitian Hamiltonians [151,152,194,195] (*different* from Ref. [150]) that can also be generalized to Lindblad systems; we do not further comment on it here. The second is the alternative geometric tensor,

$$\mathcal{Q}^{\text{alt}} \equiv \mathcal{P}^{\ddagger}_\Psi\partial\mathcal{P}^{\ddagger}_\Psi\partial\mathcal{P}_\Psi\mathcal{P}_\Psi. \quad (G1)$$

We show that $\mathcal{Q}^{\text{alt}}$ appears in a bound on the adiabatic path length for Lindbladian systems, which has traditionally been used to determine the shortest possible distance between states in a parameter space $\mathsf{M}$. Here, we introduce the adiabatic path length, generalize it to Lindbladians, and comment on $\mathcal{Q}^{\text{alt}}$.

The adiabatic path length for Hamiltonian systems quantifies the distance between two adiabatically connected states $|\psi_0^{(s=0)}\rangle$ and $|\psi_0^{(1)}\rangle$. The adiabatic evolution operator (derived in Sec. E 1) for an arbitrary path $s\in[0,1]$ and for initial zero-energy state $|\psi_0^{(0)}\rangle$ is

$$U^{(1)} = \mathbb{P}\exp\left(\int_0^1 \dot{P}_0 P_0 ds\right). \quad (G2)$$

Consider the Frobenius norm Eq. (A1) of $U^{(1)}$. By expanding the definition of the path-ordered exponential, one can show that $\|U^{(1)}\|\leq\exp(S)$ with path length

$$L_0 \equiv \int_0^1 \|\dot{P}_0 P_0\| ds. \quad (G3)$$

Remembering that $\|A\| = \sqrt{\text{Tr}\{A^\dagger A\}}$ and writing $\partial_s$ in terms of parameter derivatives, we see that the Fubini-Study metric appears in the path length:

$$\|\dot{P}_0 P_0\|^2 = \frac{1}{2}\sum_{\alpha,\beta}M_{\alpha\beta,00}\dot{\mathbf{x}}_\alpha\dot{\mathbf{x}}_\beta. \quad (G4)$$

Therefore, the shortest path between states in Hilbert space projects to a geodesic in parameter space satisfying the Euler-Lagrange equations associated with the metric $M_{\alpha\beta,00}$ and minimizing the path length [see, e.g., Ref. [148], Eq. (7.58)] (with sum implied)

$$L_0 = \int_0^1 \sqrt{\frac{1}{2}G_{\alpha\beta,00}\dot{\mathbf{x}}_\alpha\dot{\mathbf{x}}_\beta} ds. \quad (G5)$$

In Hamiltonian systems, the adiabatic path length appears in bounds on corrections to adiabatic evolution (Ref. [196], Theorem 3; see also Ref. [192]). This path length is also applicable when one wants to simulate adiabatic evolution in a much shorter time (counterdiabatic or superadiabatic dynamics [197–199] or shortcuts to adiabaticity [200,201]) by explicitly engineering the Kato Hamiltonian $i[\dot{P}_0, P_0]$ from Eq. (E5).

The tensor $\mathcal{Q}^{\text{alt}}_{\alpha\beta}$ arises in the computation of the corresponding Lindbladian adiabatic path length,

$$L \equiv \int_0^1 \|\dot{\mathcal{P}}_\Psi\mathcal{P}_\Psi\| ds, \quad (G6)$$

where the superoperator norm of $\dot{\mathcal{P}}_\Psi\mathcal{P}_\Psi$ is the analogue of the operator Frobenius norm from Eq. (A1): $\|\mathcal{O}\| \equiv \sqrt{\text{TR}\{\mathcal{O}^\ddagger\mathcal{O}\}}$, where $\mathcal{O}$ is a superoperator. This path length provides an upper bound on the norm of the Lindblad adiabatic evolution superoperator Eq. (5.6):

$$\mathcal{U}^{(1,0)} = \mathbb{P}\exp\left(\int_0^1 \dot{\mathcal{P}}_\Psi\mathcal{P}_\Psi ds\right). \quad (G7)$$

Using properties of norms and assuming one NS block, it is straightforward to show that

$$\|\mathcal{U}^{(1,0)}\| \leq \exp(L) \quad \text{with} \quad L = \int_0^1 \sqrt{\frac{1}{2}d_{\text{ax}}\mathcal{M}^{\text{alt}}_{\alpha\beta}\dot{\mathbf{x}}_\alpha\dot{\mathbf{x}}_\beta} ds \quad (G8)$$

(with sum over $\alpha$, $\beta$ implied). The metric governing this path length turns out to be

$$\mathcal{M}^{\text{alt}}_{\alpha\beta} = \langle\!\langle \varrho_{\text{ax}} | \varrho_{\text{ax}} \rangle\!\rangle \sum_{\alpha,\beta} \mathcal{Q}^{\text{alt}}_{(\alpha\beta),\mu\mu}. \tag{G9}$$

For a unique steady state $\varrho$, this alternative metric reduces to the Hilbert-Schmidt metric

$$\mathcal{M}^{\text{alt}}_{\alpha\beta} = \langle\!\langle \partial_{(\alpha}\varrho | \partial_{\beta)}\varrho \rangle\!\rangle. \tag{G10}$$

Note the subtle difference between this metric and the QGT metric [Eq. (6.7)]

$$\mathcal{M}_{\alpha\beta} = \langle\!\langle \partial_{(\alpha}P | \partial_{\beta)}\varrho \rangle\!\rangle. \tag{G11}$$

This difference is precisely due to the absence of $\varrho$ in the left eigenmatrices $J_{\boxminus}$. For the QGT metric, $\varrho$ is never in the same trace twice, while for the alternative metric, the presence of $\mathcal{P}^{\ddagger}_{\Psi}$ yields such terms. We note that for a pure steady state $\varrho = P$ (with $P$ being rank one), both metric tensors reduce to the Fubini-Study metric:

$$\mathcal{M}_{\alpha\beta} = \mathcal{M}^{\text{alt}}_{\alpha\beta} = \langle\!\langle \partial_{(\alpha}P | \partial_{\beta)}P \rangle\!\rangle. \tag{G12}$$

Another notable example is the DFS case ($\varrho_{\text{ax}} = 1$). In that case, $J^{\mu}_{\boxminus} = \Psi_{\mu}$—the QGT and alternative tensor become equal ($\mathcal{Q}^{\text{alt}} = \mathcal{Q}$). Therefore, it is the presence of $\varrho_{\text{ax}}$ that allows for two different metrics $\mathcal{M}_{\alpha\beta}$ and $\mathcal{M}^{\text{alt}}_{\alpha\beta}$. However, for the NS case, the "alternative" curvature $\mathcal{Q}^{\text{alt}}_{[\alpha\beta],\mu\nu}$ does not reduce to the adiabatic curvature $\mathcal{F}_{\alpha\beta,\mu\nu}$ associated with the connection $\mathcal{A}_{\alpha}$ (unlike the QGT curvature). How this subtle difference between $\mathcal{Q}_{\alpha\beta}$ and $\mathcal{Q}^{\text{alt}}_{\alpha\beta}$ for the NS and unique steady-state cases is relevant in determining distances between adiabatic steady states of Lindbladians should be a subject of future investigation.

---